\documentclass[a4paper,11pt]{article}
\usepackage{graphicx}
\usepackage{amsmath}
\usepackage{amssymb}
\usepackage{feynarts}
\usepackage{cite}
\usepackage{mcite}

\begin{document}
\newcommand{\beq}{\begin{eqnarray}}
\newcommand{\eeq}{\end{eqnarray}}
\newcommand{\non}{\nonumber}
\newcommand{\mc}{\mathcal}
\newcommand{\ra}{\rightarrow}

\title{\ \\ \ \\ 
\LARGE{\bf{Radiative corrections to $h^0\to WW^*/ZZ^*\to 4$\,leptons in the MSSM}}}
\author{\Large{Wolfgang Hollik${}^{\,a}$ and {Jian-Hui Zhang${}^{\,a,b}$}}}
\date{${}^{a}$\it{Max-Planck-Institut f\"ur Physik, \\
F\"ohringer Ring 6, 80805 M\"unchen, Germany}
\vskip .3em
      ${}^{b}$Center for High Energy Physics, \\
      Peking University, Beijing, China}

\maketitle

\begin{abstract}
The electroweak $\mc O(\alpha)$ radiative corrections to the decay of the lightest MSSM Higgs boson to four leptons are presented, improved by the two-loop corrections provided by the program package FeynHiggs. We also analyze the results in the decoupling limit and investigate the numeric impact of contributions from the genuine supersymmetric particle spectrum.
\end{abstract}

\section{Introduction}
\label{intro}
\hspace*{2ex} Deciphering the mechanism that breaks the electroweak symmetry and generates the masses of fundamental particles is one of the central tasks of Tevatron and LHC. In the standard model (SM), the electroweak symmetry breaking is realized through the Higgs mechanism where the neutral component of an SU(2) complex scalar doublet acquires a non-zero vacuum expectation value. While in the minimal supersymmetric extension of the standard model (MSSM), two Higgs doublets are required, resulting in five physical Higgs bosons. Two of them, $h^0$ and $H^0$,  are CP-even, one is CP-odd $A^0$, and the other two, $H^\pm$, are charged. Among them the lightest CP-even Higgs boson mass is bounded from above by $M_{h^0}\lesssim 135\,\mbox{GeV}$, including radiative corrections up to two-loop order~\cite{Heinemeyer:1998jw,*Heinemeyer:1998kz,*Heinemeyer:1998np,Degrassi:2002fi,Allanach:2004rh}. In this mass range, the Higgs boson decays dominantly to $b\bar b$ pair. However, this is not a promising channel for the discovery of the Higgs boson at hadron colliders due to the large QCD background~\cite{Ball:2007zza}. Detailed investigations of other decay modes of the Higgs boson are thus necessary, and such investigations also have further implications. At the LHC at least one MSSM Higgs boson can be discovered over all of the MSSM parameter space. In the region where $\tan\beta$ and $M_{A^0}$ take on moderate values and the region with large $M_{A^0}$ values, only the lightest Higgs boson would be observable~\cite{Atlas:1999fr}. In this case, precision measurements of the Higgs decay properties would indicate if the Higgs boson originates from the SM or the MSSM and for the latter case allow to derive indirect bounds on other MSSM parameters, e.g. on the mass of the CP-odd Higgs boson~\cite{Dedes:2003cg}. Over a large fraction of the parameter space more than one Higgs boson would be accessible. Then precision measurements of the lightest Higgs boson properties at the linear collider can help to distinguish between different soft SUSY-breaking scenarios~\cite{Dedes:2003cg}.

The decay properties of the SM Higgs boson have been studied intensively in the literature. In refs.~\cite{Bredenstein:2006rh,Bredenstein:2006ha} the decay of the SM Higgs boson to four fermions via a gauge boson pair was investigated, where the complete $\mc O(\alpha)$ electroweak corrections were presented for leptonic final states, and for semi-leptonic and hadronic final states also complete $\mc O(\alpha_s)$ QCD corrections were available. The results were further improved by corrections beyond $\mc O(\alpha)$ originating from heavy-Higgs effects and final state radiation. 

In the present paper we consider similar processes in the CP-conserving MSSM with real parameters and compute the $\mc O(\alpha)$ electroweak corrections to the decay $h^0\to WW^*/ZZ^*\to 4$\,leptons. In contrast to the SM, in the real MSSM the two CP-even Higgs bosons, $h^0$ and $H^0$, can mix beyond the lowest order. The resulting Higgs propagator corrections are numerically important, we use an effective decay amplitude to account for such corrections. In the presence of the mixing between Higgs bosons, the radiative corrections to the coupling of the heavy CP-even Higgs boson to gauge bosons can be numerically relevant due to heavy fermions and sfermions in the loop and thus should be taken into account. The numerical results are analyzed in the benchmark scenarios suggested in~\cite{Carena:1999xa,*Carena:2002qg}. We also investigate the results in the decoupling limit~\cite{Gunion:2002zf,*Carena:2002es}. 

This paper is organized as follows. In section~\ref{mssmHiggssec} we describe the renormalization of the MSSM Higgs sector. The strategy of our computation is outlined in section~\ref{LOresults} to section~\ref{HOfsr}, and the numerical results are discussed in section~\ref{numeresults}. We draw our conclusions in section~\ref{concl}. In Appendix~\ref{appenintegrals} we list the analytical results for the scalar integrals that are relevant for our computation. The methods used to deal with the soft and collinear photon emission are briefly summarized in Appendix~\ref{appensoftandcoll}.

\section{Renormalization of the MSSM Higgs sector}
\label{mssmHiggssec}
The Higgs sector of the MSSM consists of two complex scalar doublets with opposite hypercharges, which give masses to up- and down-type fermions, respectively. The two doublets can be decomposed as
\begin{align}\label{Higgsdecomp}
H_1&=\left(\begin{array}{c} v_1+\frac{1}{\sqrt 2}(\phi_1+i\chi_1) \\ H_1^- \end{array} \right)=\left(\begin{array}{c} v\cos\beta +\frac{1}{\sqrt 2}(\phi_1+i\chi_1) \\ H_1^- \end{array} \right)\ , \non \\
H_2&=\left(\begin{array}{c} H_2^+ \\ v_2+\frac{1}{\sqrt 2}(\phi_2+i\chi_2) \end{array} \right)=\left(\begin{array}{c} H_2^+ \\ v\sin\beta+\frac{1}{\sqrt 2}(\phi_2+i\chi_2) \end{array} \right)\ ,
\end{align}
where $v=\sqrt{v_1^2+v_2^2}=174\,\mbox{GeV}$ and $\tan\beta=v_2/v_1$ with $0<\beta<\pi/2$. 

The tree-level masses of Higgs bosons are determined by the bilinear terms of Higgs fields in the tree-level scalar potential. In the $\phi_1$ and $\phi_2$ basis, such bilinear terms give rise to a non-diagonal tree-level mass matrix of the neutral CP-even Higgs bosons. It can be diagonalized by transforming into the mass eigenstate basis $h^0$ and $H^0$, leading to
\beq
m^2_{h^0,H^0}=\left(\begin{array}{cc}m^2_{h^0}&0\\ 0&m^2_{H^0}\end{array}\right)\ ,
\eeq
where the mass eigenstates $h^0$ and $H^0$ are given by the following rotation
\beq\label{Higgsrotation}
\left(\begin{array}{c} h^0 \\ H^0 \end{array}\right) &=\left( \begin{array}{cc}-\sin\alpha & \cos\alpha \\ \cos\alpha &  \sin\alpha \end{array}\right) \left(\begin{array}{c} \phi_1 \\ \phi_2 \end{array}\right)\ .
\eeq
The mixing angle $\alpha$ satisfies
\beq\label{treealpha}
\alpha=\frac{1}{2}\arctan\Big[\tan 2\beta \frac{M_{A^0}^2+M_Z^2}{M_{A^0}^2-M_Z^2}\Big]\ , \hspace{1cm} -\frac{\pi}{2}<\alpha<0\ 
\eeq
with $M_{A^0}$ the mass of the neutral CP-odd Higgs boson. This relation follows from the requirement that the lowest order tadpoles and the non-diagonal entries of the CP-even Higgs mass matrix in the $h^0,H^0$ basis vanish. The mixing angles for the CP-odd and charged Higgs bosons can be determined analogously. The tree-level mass of the light CP-even Higgs boson, $m_{h^0}$, has an upper bound of $M_Z$, as a consequence of the fact that Higgs self couplings are determined by gauge couplings in the MSSM. 

The evaluation of higher order corrections requires the renormalization of the Higgs sector. In the following we concentrate on the renormalization that is needed for the present work. The Higgs tadpole counter terms are introduced via 
\beq
T_{h^0}\to T_{h^0}+\delta T_{h^0}\ , \hspace{1cm} T_{H^0}\to T_{H^0}+\delta T_{H^0}\ .
\eeq
They are fixed by requiring that the renormalized tadpoles vanish, which leads to
\beq
\delta T_{h^0}=-T_{h^0}\ , \hspace{1cm} \delta T_{H^0}=-T_{H^0}\ .
\eeq
The mass counter terms $\delta M_Z^2$ and $\delta M_{A^0}^2$ follow from the on-shell conditions
\beq
\mbox{Re}\,\hat\Sigma_{Z}^T(M_Z^2)=0\ , \hspace{1cm} \mbox{Re}\,\hat\Sigma_{A^0}(M_{A^0}^2)=0\ ,
\eeq
where the superscript $T$ denotes the transverse part of the gauge boson self energy. These conditions yield
\beq
\delta M_Z^2=\mbox{Re}\,\Sigma_{Z}^T(M_Z^2)\ , \hspace{1cm}
\delta M_{A^0}^2=\mbox{Re}\,\Sigma_{A^0}(M_{A^0}^2)\ .
\eeq
In the mass eigenstate basis, the mass counter terms for the CP-even Higgs bosons can be written as 
\beq
\delta m^2_{h^0,H^0}=\left(\begin{array}{cc}\delta m_{h^0}^2&\delta m_{h^0H^0}^2\\ \delta m_{h^0H^0}^2&\delta m_{H^0}^2\end{array}\right)\ .
\eeq
Note that the mixing angles are not renormalized, these mass counter terms follow from their tree-level expressions as~\cite{Frank:2006yh}
\begin{align}\label{hHmassct}
\delta m_{h^0H^0}^2&=-\frac{1}{2}\delta M_Z^2 \sin 2(\alpha+\beta)+\frac{1}{2}\delta M_{A^0}^2 \sin 2(\alpha-\beta) \non \\
&+\frac{e}{2M_Z s_W c_W}[\delta T_{H^0}\sin^3(\alpha-\beta)-\delta T_{h^0}\cos^3(\alpha-\beta)] \non \\
&-\delta\tan\beta \cos^2\beta[M_Z^2\cos 2(\alpha+\beta)+M_{A^0}^2\cos 2(\alpha-\beta)]\ , \non\\
\delta m_{h^0}^2&=\delta M_Z^2 \sin^2(\alpha+\beta)+\delta M_{A^0}^2\cos^2(\alpha-\beta)\non \\
&+\frac{e}{2M_Z s_W c_W}[\delta T_{H^0}\cos(\alpha-\beta)\sin^2 (\alpha-\beta)+\delta T_{h^0}\sin(\alpha-\beta)(1+\cos^2(\alpha-\beta))]\non \\
&+\delta\tan\beta \cos^2\beta[M_Z^2 \sin 2(\alpha+\beta)+M_{A^0}^2 \sin 2(\alpha-\beta)] \ ,\non \\
\delta m_{H^0}^2&=\delta M_Z^2 \cos^2(\alpha+\beta)+\delta M_{A^0}^2 \sin^2(\alpha-\beta)\non \\
&-\frac{e}{2M_Z s_W c_W}[\delta T_{H^0}\cos(\alpha-\beta)(1+\sin^2(\alpha-\beta))+\delta T_{h^0}\sin(\alpha-\beta)\cos^2(\alpha-\beta)]\non \\
&-\delta\tan\beta \cos^2\beta[M_Z^2 \sin 2(\alpha+\beta)+M_{A^0}^2\sin 2(\alpha-\beta)]\ ,
\end{align}
where the counter term $\delta\tan\beta$ is introduced via $\delta\tan\beta \to \tan\beta+\delta\tan\beta$, its renormalization condition will be given below.

In order to have finite Green functions, the Higgs fields have to be renormalized as well. For the renormalization of the neutral CP-even Higgs fields, we can choose to renormalize either the fields $h^0$ and $H^0$ or $\phi_1$ and $\phi_2$. This is in analogy to the renormalization of gauge boson fields in the SM, where one can renormalize either $W^3$ and $B$ bosons or alternatively their mixtures, the $\gamma$ and $Z$ bosons. In the SM the weak mixing angle is defined by $\sin^2 \theta_W=1-M_W^2/M_Z^2$ in the on-shell scheme~\cite{Denner:1991kt}. This defining relation is valid to all orders in perturbation theory. The weak mixing angle thus receives renormalization due to the renormalization of $M_W$ and $M_Z$. In the Higgs sector of the MSSM, the relations between the mixing angles and input parameters hold only at tree-level, the mixing angles can be kept unrenormalized. In this paper we renormalize the Higgs fields following~\cite{Heinemeyer:2004ms,Frank:2006yh}, i.e. we introduce a renormalization constant for each Higgs doublet,
\beq\label{Hdoubletrenconst}
H_1\ra (1+\frac{1}{2}\delta Z_{H_1})H_1, \hspace{1cm} H_2\ra (1+\frac{1}{2}\delta Z_{H_2})H_2\ .
\eeq
Their vacuum expectation values then renormalize as follows
\begin{align}\label{vevrenconst}
v_1&\ra (1+\frac{1}{2}\delta Z_{H_1})(v_1-\delta v_1)=v_1(1+\frac{1}{2}\delta Z_{H_1}-\frac{\delta v_1}{v_1}) \ ,\non \\
 v_2&\ra (1+\frac{1}{2}\delta Z_{H_2})(v_2-\delta v_2)=v_2(1+\frac{1}{2}\delta Z_{H_2}-\frac{\delta v_2}{v_2})\ .
\end{align}
The freedom of field renormalization allows us to impose the condition $\frac{\delta v_1}{v_1}=\frac{\delta v_2}{v_2}$. The renormalization of $\tan\beta$ then follows from the renormalization of the vacuum expectation values of the two Higgs doublets as
\beq\label{tbrenconst}
\frac{\delta\tan\beta}{\tan\beta}=\frac{1}{2}(\delta Z_{H_2}-\delta Z_{H_1})\ .
\eeq
We can write the renormalized self energies of the CP-even Higgs bosons in terms of the unrenormalized ones, the field renormalization constants and the mass counter terms
\begin{align}\label{renSE}
\hat\Sigma_{h^0}(k^2)&=\Sigma_{h^0}(k^2)+(\sin^2\alpha\,\delta Z_{H_1}+\cos^2\alpha\,\delta Z_{H_2})(k^2-m_{h^0}^2)-\delta m_{h^0}^2\ , \non \\
\hat\Sigma_{H^0}(k^2)&=\Sigma_{H^0}(k^2)+(\cos^2\alpha\,\delta Z_{H_1}+\sin^2\alpha\,\delta Z_{H_2})(k^2-m_{H^0}^2)-\delta m_{H^0}^2 \ ,\non \\
\hat\Sigma_{h^0H^0}(k^2)&=\Sigma_{h^0H^0}(k^2)+\sin\alpha\cos\alpha(\delta Z_{H_2}-\delta Z_{H_1})(k^2-\frac{1}{2}(m_{h^0}^2+m_{H^0}^2))-\delta m_{h^0H^0}^2\ .
\end{align}
The field renormalization constants $\delta Z_{H_1}$ and $\delta Z_{H_2}$ are determined in the $\overline{DR}$ scheme. From Eq.~(\ref{renSE}) one finds
\begin{align}\label{DRbarfieldrenconst}
\delta Z_{H_1}^{\overline{DR}}&=-[\,\mbox{Re}\Sigma'_{H^0}(m_{H^0}^2)|_{\alpha=0}\,]^{\mbox{div}}\ , \non\\
\delta Z_{H_2}^{\overline{DR}}&=-[\,\mbox{Re}\Sigma'_{h^0}(m_{h^0}^2)|_{\alpha=0}\,]^{\mbox{div}}\ .
\end{align}
In the $\overline{DR}$ scheme $\delta\tan\beta$ is fixed by Eq.~(\ref{tbrenconst}) and (\ref{DRbarfieldrenconst}). At one-loop level, this $\overline{DR}$ renormalization of $\tan\beta$ yields gauge independent results within the class of $R_\xi$ gauges (the gauge dependence arises at two-loop level even within $R_\xi$ gauges)~\cite{Freitas:2002um}. Hence the $\overline{DR}$ scheme is a convenient choice for the evaluation of one-loop corrections. Moreover, this scheme has stable numerical behavior~\cite{Freitas:2002um,Brignole:1992uf,Frank:2002qf}. In this work we use the $\overline{DR}$ renormalization of $\tan\beta$ and choose the renormalization scale as $\mu^{\overline{DR}}=1.5M_W$, which is the scale of the physical mass of the lightest CP-even Higgs boson for moderate $\tan\beta$ and $M_{A^0}$ values. Other SM parameters are renormalized as in the on-shell scheme in~\cite{Denner:1991kt}. 

Beyond the lowest order, the mass of the light CP-even Higgs boson receives large radiative corrections, predominantly from the top/stop loops, and also from bottom/sbottom loops for large $\tan\beta$. In addition, the two CP-even Higgs bosons can mix with each other. These lead to finite wave function normalization factors for the external Higgs boson in our process, which have to be taken into account so that a correct normalization of the S-matrix is ensured. As a consequence of these effects, the decay amplitude of $h^0\to 4$\,leptons can be written as 
\beq
\mc M(h^0\ra 4l)=\sqrt{Z_{h^0}}(\mc M_{h^0}+Z_{h^0H^0}\mc M_{H^0})\ ,
\eeq
where the wave function normalization factors $Z_{h^0}$ and $Z_{h^0H^0}$ are given by~\cite{Heinemeyer:2000fa}
\begin{align}
Z_{h^0}&=\frac{1}{1+\mbox{Re}\hat\Sigma_{h^0}'(k^2)-\mbox{Re}\left(\frac{\hat\Sigma^2_{h^0H^0}(k^2)}{k^2-m_{H^0}^2+\hat\Sigma_{H^0}(k^2)}\right)'}\Bigg|_{k^2=M^2_{h^0}}\ ,\non\\
Z_{h^0H^0}&=-\frac{\hat\Sigma_{h^0H^0}(M^2_{h^0})}{M^2_{h^0}-m_{H^0}^2+\hat\Sigma_{H^0}(M^2_{h^0})}\ .
\end{align}
$M_{h^0}$ denotes the physical mass of $h^0$. These finite wave function normalization factors, as well as the physical masses of the Higgs bosons can be computed by the program package FeynHiggs~\cite{Heinemeyer:1998yj,*Hahn:2005cu,*Hahn:2006np}, in which also the dominant two-loop corrections to Higgs boson self energies are taken into account. 

\section{Lowest order results}
\label{LOresults}
We consider the following leptonic decay processes
\beq\label{procWW}
h^0(k_1)\to WW^*\to e^-(k_2)+\bar\nu_e(k_3)+\mu^+(k_4)+\nu_\mu(k_5)
\eeq
and 
\beq\label{procZZ}
h^0(k_1)\to ZZ^*\to e^-(k_2)+e^+(k_3)+\mu^+(k_4)+\mu^-(k_5)\ ,
\eeq
where the particle momenta are given in the parentheses, the helicity indices are suppressed. Throughout this work the masses of the final state leptons are neglected whenever possible, i.e. we keep them only as regulators for the collinear singularities. In these decay processes, one of the intermediate gauge bosons can become resonant due to the upper bound of the $h^0$ mass. The finite width has to be incorporated for the resonant gauge boson in order to avoid the occurrence of singularities. 

In this work the Feynman diagrams are generated by FeynArts~\cite{Kublbeck:1990xc,*Hahn:2000kx,*Kublbeck:1992mt}.  FormCalc~\cite{Hahn:2001rv,*Hahn:2006qw,*Hahn:2007px} and LoopTools~\cite{Hahn:2001rv,Hahn:1999wr} are then used to algebraically simplify the amplitudes and evaluate the one-loop scalar and tensor integrals that do not require the finite gauge boson width as regulators. The computation is carried out in the 't Hooft-Feynman gauge.

\subsection{Implementation of gauge boson width}
The finite width has to be included for the intermediate gauge boson when it becomes resonant. However, in perturbation theory the description of resonances requires a Dyson resummation of self energy insertions. This mixes different perturbative orders as only partial higher order corrections are taken into account. As a consequence, gauge invariance might be spoiled, since it is preserved order by order in perturbation theory. There have been proposals for a consistent implementation of the width of gauge boson. One of them is the complex mass scheme~\cite{Denner:1999gp,*Denner:2005fg}, in which the gauge boson masses are taken as complex quantities that are defined by the poles of the propagators. This scheme has been successfully applied to tree-level computations as well as the evaluation of radiative corrections in the SM. The implementation of the complex mass scheme would require independent mass parameters, which is clearly not the case in the MSSM, since, for example, the CP-even Higgs boson masses depend on the gauge boson masses. Another proposal is the pole scheme~\cite{Stuart:1991xk,Aeppli:1993cb,Aeppli:1993rs}, in which one performs an expansion of the amplitude according to its pole structure and includes the finite width only in the pole term. The drawback of this scheme is that it is not applicable near and below the threshold region. In this work the width of the resonant gauge boson is incorporated according to the factorization scheme~\cite{Baur:1991pp,Kurihara:1994fz,Denner:2003iy}, which yields a simple rescaling at tree-level (for convenience, we use a subscript $V$ to denote the vector gauge boson, it should be replaced by $W$ or $Z$ accordingly for the above processes)
\beq\label{factschemetree}
\mc M_{born}=\frac{k_V^2-M_V^2}{k_V^2-M_V^2+iM_V\Gamma_V}\mc M_{born}(\Gamma_V=0)\ ,
\eeq
where $k_V$ is the four-momentum of the resonant gauge boson, $\mc M_{born}(\Gamma_V=0)$ denotes the decay amplitude before Dyson resummation. At one-loop level, the width of resonant gauge boson can be incorporated as follows
\begin{align}\label{factschemeloop}
\mc M_{loop}&=\frac{k_V^2-M_V^2}{k_V^2-M_V^2+iM_V\Gamma_V}\mc M_{loop}(\Gamma_V=0)+\frac{i\,\mbox{Im}\Sigma_V^T(M_V^2)}{k_V^2-M_V^2}\mc M_{born} \non\\
&=\frac{k_V^2-M_V^2}{k_V^2-M_V^2+iM_V\Gamma_V}\mc M_{loop,\, no\ r.s.}(\Gamma_V=0) \non \\
&-\left(\frac{\Sigma_V^T(k_V^2)-\Sigma_V^T(M_V^2)}{k_V^2-M_V^2}+\delta Z_V\right)\mc M_{born}\ ,
\end{align}
where the $\mc M_{loop,\, no\ r.s.}$ term in the second row denotes the one-loop corrections excluding the self energy corrections to the resonant gauge boson. The last term in the first row is required to avoid double-counting from the inclusion of finite width in the lowest order amplitude. This term is absorbed into the self energy corrections, yielding the last term in the above equation. In the factorization scheme, the non-resonant terms are treated incorrectly, since they are simply put to zero on the resonance. The resulting error is, however, of higher order.

\subsection{The Effective Born amplitude}
\begin{figure}[htbp]
\vspace*{-0.5cm}
\centering
\input{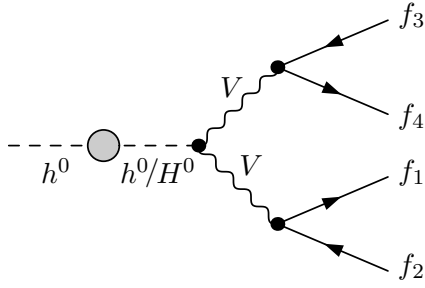}
\vspace*{-1cm}
\caption{Lowest order diagram including Higgs propagator corrections.}
\label{LOdiagram}
\end{figure}

The finite wave function normalization factors induce numerically important corrections. To account for these corrections, we use an effective Born amplitude
\begin{align}\label{effbornamp}
\mc M_{born}(\Gamma_V=0)&=\sqrt{Z_{h^0}}\left(\mc M^0_{h^0}(\Gamma_V=0)+Z_{h^0H^0}\mc M_{H^0}^0(\Gamma_V=0)\right) \non\\
&=\sqrt{Z_{h^0}}\mc M_{h^0}^0(\Gamma_V=0)\left(1+\cot(\beta-\alpha)Z_{h^0H^0}\right)\ ,
\end{align}
where $\mc M_{h^0}^0(\Gamma_V=0)$ and $\mc M_{H^0}^0(\Gamma_V=0)$ denote the respective tree-level decay amplitude of $h^0$ and $H^0$ before Dyson resummation. In the second row we have written the amplitude $\mc M_{H^0}^0$ in terms of $\mc M_{h^0}^0$. The lowest order Feynman diagram including the Higgs propagator corrections is shown in Fig.~\ref{LOdiagram}.

For the process in Eq.~(\ref{procWW}), the tree-level amplitude $\mc M_{h^0}^0(\Gamma_W=0)$ is given by
\begin{align}
\mc M_{h^0}^0(\Gamma_W=0)&=\frac{2\pi\alpha\, e M_W\sin(\beta-\alpha)}{s_W^3}\frac{1}{k_+^2-M_W^2}\non \\
&\times\frac{1}{k_-^2-M_W^2}[\bar u_{e^-}(k_2)\gamma_\rho\omega_- v_{\bar\nu_e}(k_3)][\bar u_{\nu_\mu}(k_5)\gamma^\rho\omega_- v_{\mu^+}(k_4)]\ ,
\end{align}
while for the process in Eq.~(\ref{procZZ}) the tree-level amplitude $\mc M_{h^0}^0(\Gamma_Z=0)$ reads
\begin{align}
\mc M_{h^0}^0(\Gamma_Z=0)&=\frac{\pi\alpha\,e M_W\sin(\beta-\alpha)}{c_W^4 s_W^3}\frac{1}{k_+^2-M_Z^2}\frac{1}{k_-^2-M_Z^2}\non \\
&\times [(-1+2s_W^2)\bar u_{e^-}(k_2)\gamma_\rho\omega_- v_{e^+}(k_3)+2s_W^2\bar u_{e^-}(k_2)\gamma_\rho\omega_+ v_{e^+}(k_3)]\non \\
&\times [(-1+2s_W^2)\bar u_{\mu^-}(k_5)\gamma^\rho\omega_- v_{\mu^+}(k_4)+2s_W^2\bar u_{\mu^-}(k_5)\gamma^\rho\omega_+ v_{\mu^+}(k_4)]\ ,
\end{align}
where we have introduced the variables $k_\pm$ with $k_+=k_2+k_3$ and $k_-=k_4+k_5$, and $\omega_{\pm}=\frac{1}{2}(1\pm\gamma_5)$. These decay amplitudes differ from their SM counterparts only by a factor of $\sin(\beta-\alpha)$.

The effective Born partial decay width is then given by
\beq
\Gamma_{born}=\frac{1}{2M_{h^0}}\int \sum_{pol}|\mc M_{born}|^2 d\Phi\ 
\eeq
with the squared matrix element summed over the final state polarizations and the phase space factor
\beq
d\Phi= \left(\prod_{i=2}^5\frac{d^3\mathbf k_i}{(2\pi)^3\;2k_i^0}\right)(2\pi)^4\delta^{(4)}(k_1-\sum_{i=2}^5 k_i)\ .
\eeq

\section{Virtual corrections}
\label{Virtcorr}
The evaluation of virtual corrections involves several additional issues. As in the lowest order amplitude, the wave function normalization factors for the external Higgs boson are taken into account by using an effective amplitude, in which we also include the one-loop corrections to the coupling of $H^0$ to gauge bosons from the fermionic and sfermionic sector, since in the presence of mixing between the two CP-even Higgs bosons, such corrections (especially the corrections from the third generation fermions and sfermions) may yield sizeable contributions as they involve potentially large Yukawa couplings. In addition, the photonic one-loop diagrams may involve not only infrared singularities, but also on-shell singularities. The on-shell singularities are closely related to the presence of resonant gauge boson in the loop and have to be cured by including the finite width of gauge boson. For this purpose, the one-loop integrals that contribute to the on-shell singularities are computed analytically. As before, the final state fermion masses are neglected whenever possible in the evaluation of these integrals. If the contribution of real photon emission process is taken into account, the infrared singularities will cancel out.

\subsection{Virtual corrections to $h^0\to WW^*\to 4l$}
\begin{figure}[t]
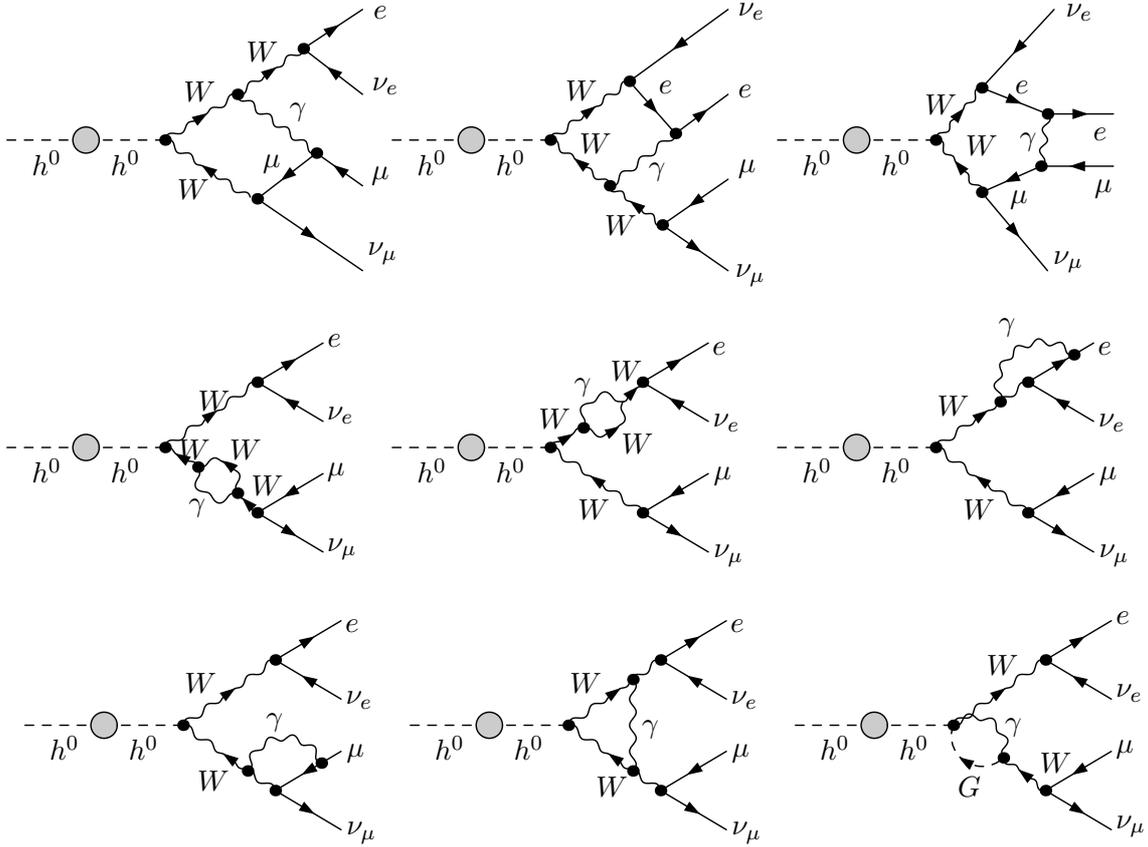

\vspace*{-0.5cm}
\centering
\input{./figtex/boxWWphot.tex}
\vspace*{-0.4cm}
\input{./figtex/selfvertWWphot.tex}
\vspace*{-0.8cm}
\caption{Examples of photonic one-loop diagrams for the process $h^0\to WW^*\to 4l$.}
\label{PhotondiagramWW}
\end{figure}

\begin{figure}[t]
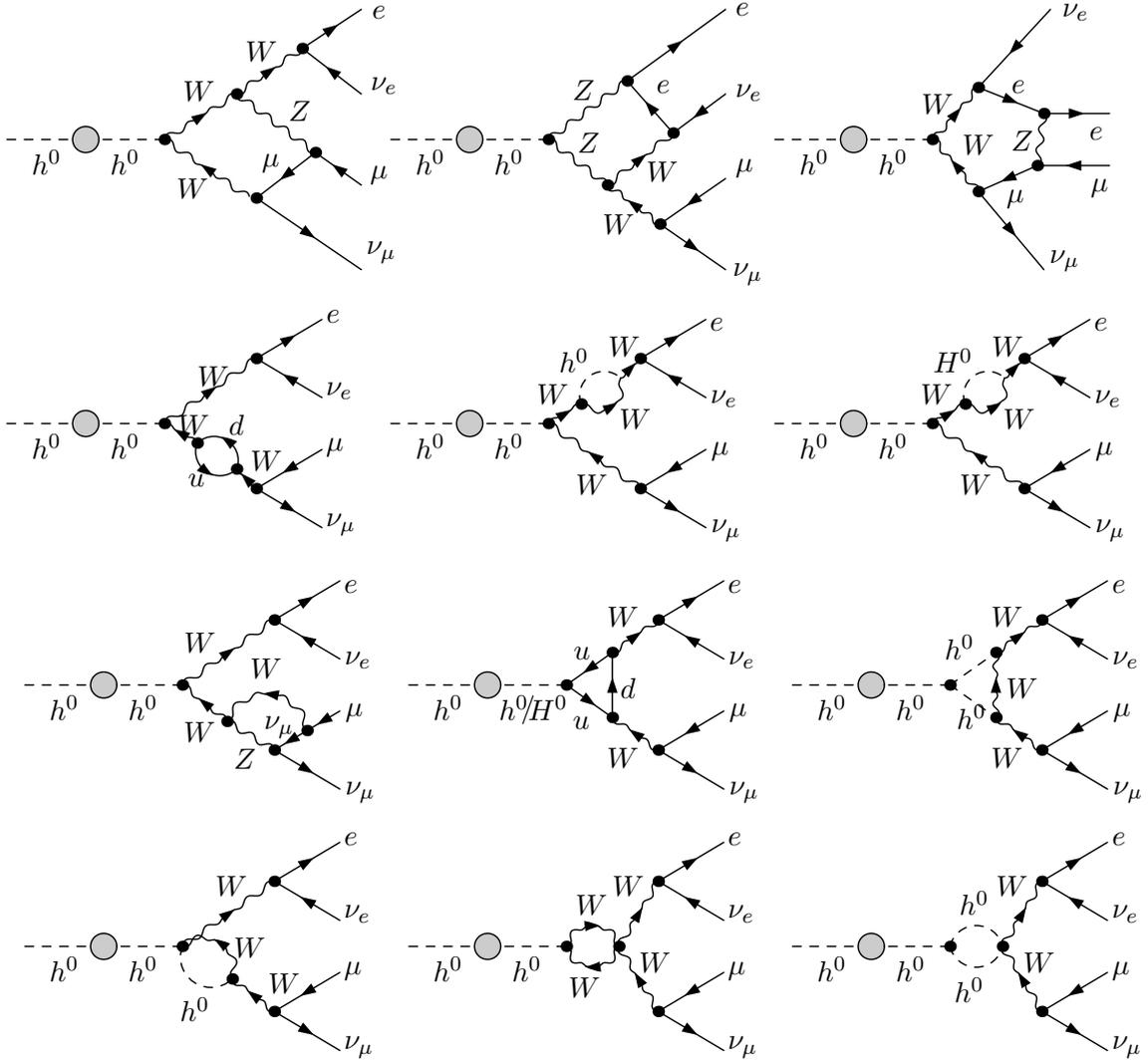

\vspace*{-0.5cm}
\centering
\input{./figtex/boxWWSMlike.tex}
\vspace*{-0.4cm}
\input{./figtex/selfvertWWSMlike.tex}
\vspace*{-0.9cm}
\caption{Examples of one-loop diagrams for the process $h^0\to WW^*\to 4l$ involving SM particles other than photon and the MSSM Higgs bosons.}
\label{SMlikediagramWW}
\vspace{-0.1cm}
\end{figure}

\begin{figure}[htbp]
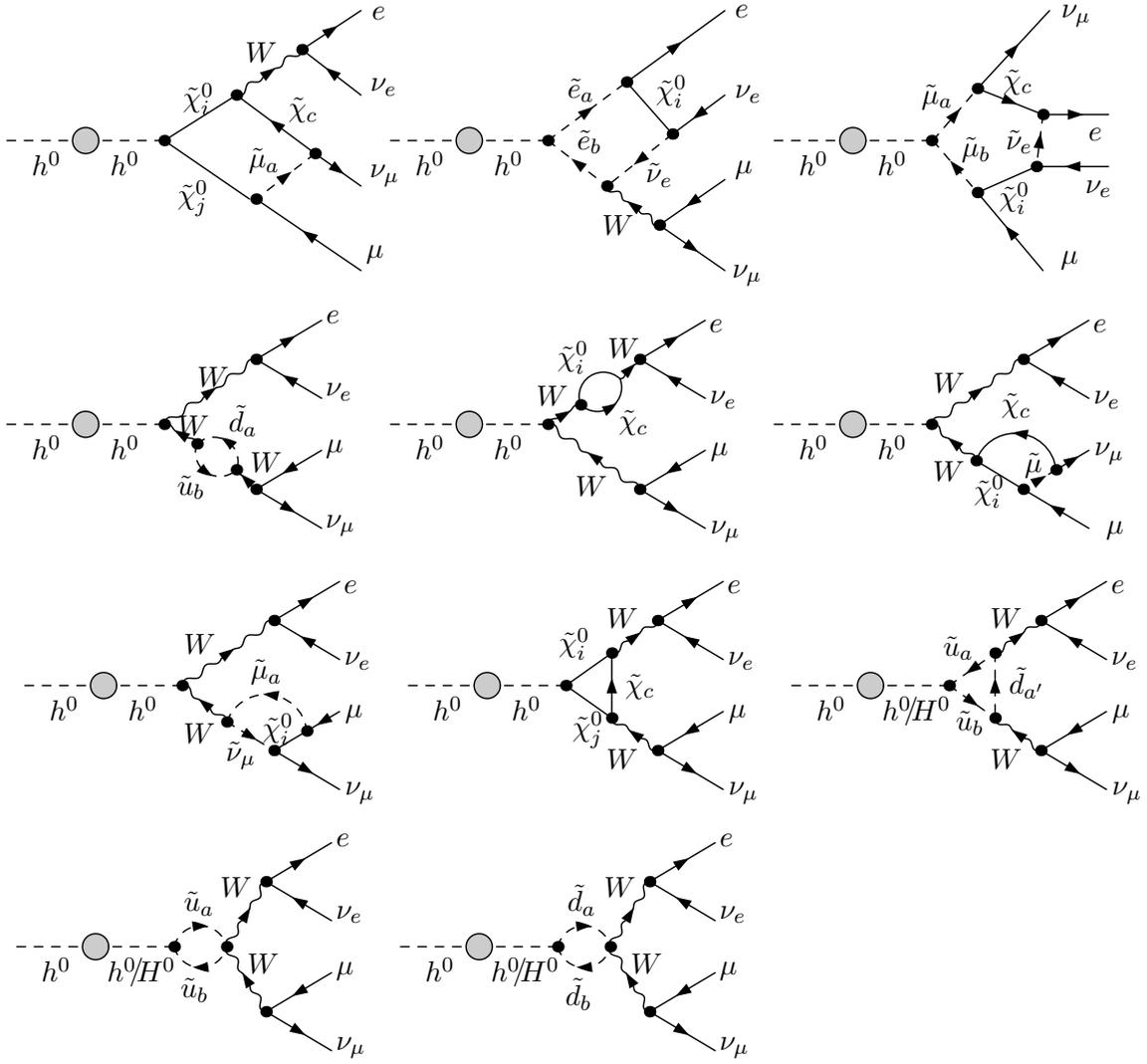

\vspace*{-0.5cm}
\centering
\input{./figtex/boxWWSUSY.tex}
\vspace*{-0.4cm}
\input{./figtex/selfvertWWSUSY.tex}
\vspace*{-0.8cm}
\caption{Some representative genuine SUSY one-loop diagrams for the process $h^0\to WW^*\to 4l$.}
\label{SUSYdiagramWW}
\end{figure}

\begin{figure}[htbp]
\vspace*{-0.5cm}
\centering
\input{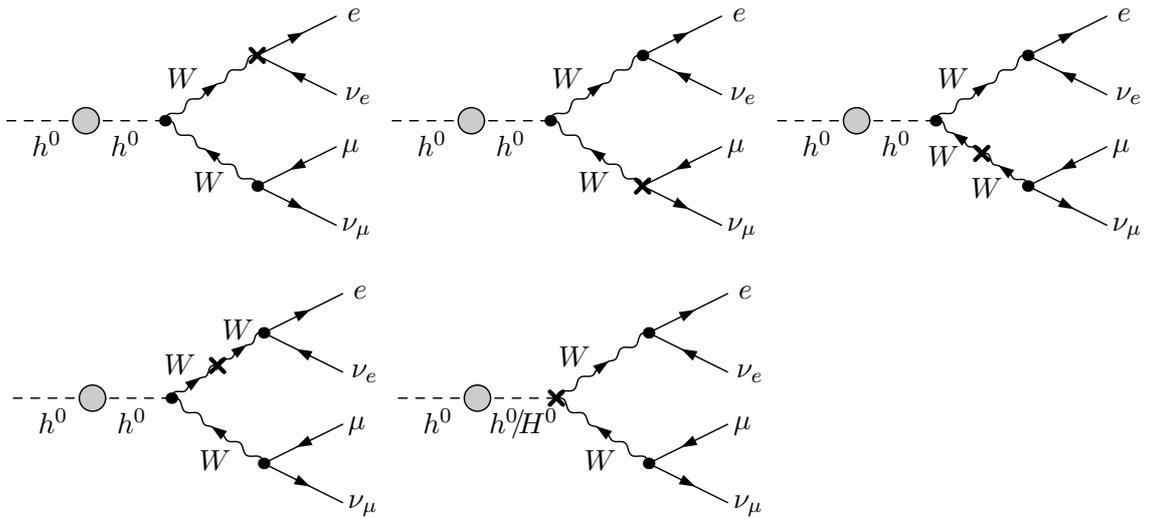}
\vspace*{-0.8cm}
\caption{Counter term diagrams for the process $h^0\to WW^*\to 4l$.}
\label{CTdiagramWW}
\end{figure}

In this subsection we describe the computation of virtual corrections to $h^0\to WW^*\to 4l$. It is convenient to classify the one-loop diagrams as: photonic diagrams, SM-like diagrams and genuine SUSY diagrams. The photonic diagrams are the same as in the SM, in Fig.~\ref{PhotondiagramWW} we show some examples of them. Infrared and on-shell singularities can arise only from the photonic diagrams. Note that the diagram with a photon exchanged between the two intermediate $W$ bosons does not contribute to on-shell singularities, since only one of these $W$ bosons can be resonant. In the factorization scheme, power counting tells us that only scalar integrals resulting from the virtual photonic diagrams can contribute to on-shell singularities. In addition, soft singularities occur only in these scalar integrals as well. We will evaluate these scalar integrals analytically. The SM-like diagrams consist of diagrams involving the SM particles other than photon and the MSSM Higgs bosons in the loop. Examples of these diagrams are depicted in Fig.~\ref{SMlikediagramWW}. In the decoupling limit $M_{A^0}\gg M_Z$, the lightest MSSM Higgs boson behaves like a SM Higgs boson and all other heavy Higgs bosons decouple, the contribution of these SM-like diagrams is expected to approach the corresponding SM contribution in this limit. The one-loop diagrams involving all other SUSY particles constitute the genuine SUSY diagrams, some representative of them are shown in Fig.~\ref{SUSYdiagramWW}. The counter term diagrams are depicted in Fig.~\ref{CTdiagramWW}.

The structure of the counter term contribution from the first four diagrams in Fig.~\ref{CTdiagramWW} has the same form as in the SM, while the last diagram yields (see e.g.~\cite{Hahn:2002gm})
\begin{align}\label{ct}
\mc M_{h^0}^{CT}&=\mc M_{h^0}^0\Big[\delta Z_e+\delta Z_W+\frac{1}{2}\frac{\delta M_W^2}{M_W^2}+\frac{\delta s_W}{s_W}+\frac{\cos(\beta-\alpha)}{\sin(\beta-\alpha)}\Big(\cos^2\beta\delta\tan\beta \non\\
&+\frac{1}{2}\sin\alpha\cos\alpha(\delta Z_{H_2}-\delta Z_{H_1})\Big)+\frac{1}{2}(\sin^2\alpha\delta Z_{H_1}+\cos^2\alpha\delta Z_{H_2})\Big]\ , \non\\
\mc M_{H^0}^{CT}&=\mc M_{H^0}^0\Big[\delta Z_e+\delta Z_W+\frac{1}{2}\frac{\delta M_W^2}{M_W^2}+\frac{\delta s_W}{s_W}+\frac{\sin(\beta-\alpha)}{\cos(\beta-\alpha)}\Big(-\cos^2\beta\delta\tan\beta \non\\
&+\frac{1}{2}\sin\alpha\cos\alpha(\delta Z_{H_2}-\delta Z_{H_1})\Big)+\frac{1}{2}(\cos^2\alpha\delta Z_{H_1}+\sin^2\alpha\delta Z_{H_2})\Big]\ 
\end{align}
with
\beq
\mc M_{h^0,H^0}^0=\frac{k_V^2-M_V^2}{k_V^2-M_V^2+iM_V\Gamma_V}\mc M_{h^0,H^0}^0(\Gamma_V=0)\ .
\eeq
The counter terms $\delta Z_{H_1}$, $\delta Z_{H_2}$ and $\delta\tan\beta$ have been given in Eqs.~(\ref{DRbarfieldrenconst}) and (\ref{tbrenconst}), the remaining counter terms are determined in the on-shell scheme and can be found in ref.~\cite{Denner:1991kt}.

In the photonic diagrams, on-shell singularities can arise if the exchanged photon becomes soft. In the following we describe the extraction of these singularities from the virtual photonic contributions.

\subsubsection{On-shell singular virtual contributions}
\label{ossingvirtcorr}
As argued previously, only scalar integrals resulting from the photonic diagrams can contribute to on-shell singularities in the factorization scheme. Here we evaluate the relevant scalar integrals and extract from them the on-shell singular virtual contributions. These scalar integrals are computed for zero gauge boson width. A finite width is inserted afterwards wherever a singularity arises when the resonant gauge boson goes on-shell. 

For the notation of the one-loop integrals, we follow the convention of~\cite{Denner:1997ia,*Denner:2002ii}. Our evaluation of on-shell singular scalar integrals is based on the observation that only one of the intermediate gauge bosons can be resonant due to the upper bound of the mass of $h^0$. Therefore it is possible to make a decomposition of these scalar integrals by extracting from them the non-resonant gauge boson propagator. This decomposition leaves the on-shell singular parts of the integrals intact but simplifies the extraction of such singular parts from the integrals. 

We start from the on-shell singular scalar 5-point integral. Assuming the gauge boson with momentum $k_-$ is resonant, the scalar 5-point integral arising from the photonic pentagon diagram in Fig.~\ref{PhotondiagramWW} can be decomposed as follows 
\begin{align}\label{E0}
&E_0(-k_2,-k_+,k_-,k_4,\lambda,m_e,M_V,M_V,m_{\mu})\non\\
&=\frac{1}{k_+^2-M_V^2}\Big\{D_0(-k_2,k_-,k_4,\lambda,m_e,M_V,m_\mu)-D_0(k_2+k_4,-k_3,k_2+k_-,m_e,m_\mu,M_V,M_V)\non\\
&-S_{23}E_1(-k_2,-k_+,k_-,k_4,0,m_e,M_V,M_V,m_{\mu})\non\\
&+(S_{24}+S_{34})E_2(-k_2,-k_+,k_-,k_4,0,m_e,M_V,M_V,m_{\mu})\non\\
&-2S_{23}E_3(-k_2,-k_+,k_-,k_4,0,m_e,M_V,M_V,m_{\mu})\non\\
&+(S_{24}+S_{25}+S_{34}+S_{35})E_4(-k_2,-k_+,k_-,k_4,0,m_e,M_V,M_V,m_{\mu})\Big\}\ ,
\end{align}
where $S_{ij}=(k_i+k_j)^2$, and a fictitious photon mass $\lambda$ is introduced to regularize the soft singularity. While the second scalar 4-point integral in the curly bracket is finite, the first one contains both the on-shell singular logarithm $\ln(k_-^2-M_V^2+i\epsilon)$ and the resonant factor $1/(k_-^2-M_V^2)$, as well as the soft singularities, as one can see from its analytical expression in  Appendix~\ref{appenintegrals}. The tensor coefficients $E_i$ that arise from the covariant decomposition of the vector 5-point integral (see ref.~\cite{Denner:1997ia,*Denner:2002ii}) do not contain soft singularity. They may contain the singular logarithm $\ln(k_-^2-M_V^2+i\epsilon)$, but not the resonant factor $1/(k_-^2-M_V^2)$. Therefore it follows from Eq.~(\ref{factschemeloop}) that only the first scalar 4-point integral gives rise to a singular logarithm $\ln(k_-^2-M_V^2+i\epsilon)$ when $k_-^2$ approaches $M_V^2$. This singular logarithm has to be replaced by $\ln(k_-^2-M_V^2+iM_V\Gamma_V)$ in the final result, as will be done later on for other scalar integrals involving this singular logarithm. This replacement does not disturb gauge invariance~\cite{Aeppli:1993rs,Dittmaier:2001ay}. 

The photonic box diagrams in Fig.~\ref{PhotondiagramWW} involve scalar 4-point integrals that might contribute to on-shell singularities. These integrals can be decomposed analogously by extracting the non-resonant gauge boson propagator, yielding
\begin{align}\label{D0}
&D_0(-k_4,k_+,-k_-,0,m_\mu,M_V,M_V)\non\\
&=\frac{1}{k_+^2-M_V^2}\Big\{C_0(-k_4,-k_-,0,m_\mu,M_V)-C_0(-k_5,k_4+k_+,m_\mu,M_V,M_V)\non\\
&+(S_{24}+S_{34})D_1(-k_4,k_+,-k_-,0,m_\mu,M_V,M_V)\non\\
&+(S_{24}+S_{34}+S_{25}+S_{35})D_2(-k_4,k_+,-k_-,0,m_\mu,M_V,M_V)\non\\
&-2S_{23}D_3(-k_4,k_+,-k_-,0,m_\mu,M_V,M_V)\Big\}\ ,\non
\end{align}
\begin{align}
&D_0(-k_2,k_-,-k_+,0,m_e,M_V,M_V)\non\\
&=\frac{1}{k_+^2-M_V^2}\Big\{C_0(-k_2,k_-,0,m_e,M_V)-C_0(-k_3,k_2+k_-,m_e,M_V,M_V)\non\\
&-S_{23}D_1(-k_2,k_-,-k_+,0,m_e,M_V,M_V)-2S_{23}D_2(-k_2,k_-,-k_+,0,m_e,M_V,M_V)\non\\
&+(S_{24}+S_{34}+S_{25}+S_{35})D_3(-k_2,k_-,-k_+,0,m_e,M_V,M_V)\Big\}\ .
\end{align}
The analytical expressions for the on-shell singular scalar 3-point functions in Eq.~(\ref{D0}) can be found in Appendix~\ref{appenintegrals}.

With these decompositions, the on-shell singular virtual contributions can be extracted straightforwardly. As only scalar integrals contain soft singularities, the analytical results of the scalar integrals can also be used to extract soft singularities from the virtual contributions and allow an analytical check of their cancellation when combining with the real corrections. The soft and on-shell singular terms arising from the photonic box and pentagon diagrams can be summarized in a correction factor to the tree-level amplitude
\begin{align}\label{bpcorrfac}
{\mc M_{h^0}}_{b,p}^{sing}&=\mc M_{h^0}^0\delta_{b,p}^{sing}=\mc M_{h^0}^0\Big\{-\frac{\alpha}{2\pi}\Big(\frac{(S_{24}+S_{34})(k_-^2-M_V^2)}{k_+^2-M_V^2}C_0(-k_4,-k_-,0,m_\mu,M_V) \non\\
&+(S_{24}+S_{25})C_0(-k_2,k_-,0,m_e,M_V) \non\\
&+S_{24}(k_-^2-M_V^2)D_0(-k_2,k_-,k_4,\lambda,m_e,M_V,m_\mu)\Big)
\Big\}\ .
\end{align}

The photonic vertex and self energy diagrams also contribute to soft and on-shell singularities. These contributions originate from the photonic corrections to the $Wff'$ vertex, the $W$ boson self energy and the field renormalization constants of the external charged fermions. The field renormalization constants of $W$ boson give rise to soft singular contributions as well. However, they only appear in intermediate stages and cancel out in the full matrix element. Owing to the presence of one additional resonant propagator, the photonic $W$ boson self energy insertion gives rise to a correction factor to the lowest order amplitude of the form $B_0(k_-^2,0,M_V^2)/(k_-^2-M_V^2)$. After the $W$ boson mass renormalization, the correction factor is modified to be proportional to
\beq
\frac{B_0(k_-^2,0,M_V^2)-B_0(M_V^2,0,M_V^2)}{k_-^2-M_V^2}=-\frac{1}{k_-^2}\ln\left(1-\frac{k_-^2}{M_V^2}-i\epsilon\right)\ ,
\eeq
which is on-shell singular and has to be regularized by the width of $W$ boson. The photonic vertex correction gives rise to a correction factor involving the on-shell singular scalar 3-point integral. Putting all these together, one finds the following correction factor that contains the soft and on-shell singularities from the photonic vertex and self energy diagrams
\begin{align}\label{vscorrfac}
{\mc M_{h^0}}_{v,s}^{sing}&=\mc M_{h^0}^0\delta_{v,s}^{sing}=\mc M_{h^0}^0\Big\{\frac{\alpha}{2\pi}\Big(k_+^2 C_0(-k_2,-k_+,0,m_e,M_V) \non\\
&+k_-^2 C_0(-k_4,-k_-,0,m_\mu,M_V)\Big)+\frac{1}{2}(\delta Z_e^L+\delta Z_\mu^L)_{\mbox{\tiny{IR}}} \non\\
&-\frac{\alpha}{4\pi}(5-M_V^2/k_-^2)\ln\left(1-\frac{k_-^2}{M_V^2}-i\epsilon\right)\Big\}\ ,
\end{align}
where the subscript $\mbox{IR}$ denotes the infrared singular part of the field renormalization costants, whose expression can be found, e.g. in~\cite{Denner:1991kt}. The correction factors defined in Eqs.~(\ref{bpcorrfac}) and (\ref{vscorrfac}) include all the soft singularities from virtual photonic diagrams. The collinear singularities, however, are not fully contained in these correction factors, since the tensor integrals that are not accounted for in these factors, contain collinear singularities as well.

In the case that the other gauge boson becomes resonant, the correction factor resulting from the box and pentagon diagrams becomes
\begin{align}\label{bpcorrfac1}
{\mc M_{h^0}}_{b,p}^{sing}&=\mc M_{h^0}^0\delta_{b,p}^{sing}=\mc M_{h^0}^0\Big\{-\frac{\alpha}{2\pi}\Big(\frac{(S_{24}+S_{25})(k_+^2-M_V^2)}{k_-^2-M_V^2}C_0(-k_2,-k_+,0,m_e,M_V) \non\\
&+(S_{24}+S_{34})C_0(-k_4,k_+,0,m_\mu,M_V) \non\\
&+S_{24}(k_+^2-M_V^2)D_0(-k_2,-k_+,k_4,\lambda,m_e,M_V,m_\mu)\Big)
\Big\}\ .
\end{align}
The correction factor from the self energy and vertex diagrams can be obtained from Eq.~(\ref{vscorrfac}) by replacing $k_-^2$ with $k_+^2$ in the last term.

Now we are able to extract the on-shell singular virtual contributions from Eq.~(\ref{bpcorrfac}), (\ref{vscorrfac}) and the results of the scalar integrals in Appendix~\ref{appenintegrals}. This gives rise to the following correction factor
\begin{align}\label{onshellsin}
{\mc M_{h^0}}^{\!\!\!\!\!\! on-shell, sing}&=\mc M_{h^0}^0\frac{\alpha}{2\pi}\Big\{\ln \Big[\frac{k_-^2}{M_V^2}\Big(\frac{S_{24}+S_{25}}{S_{24}}\Big)^2\Big]+\frac{M_V^2}{2k_-^2}-\frac{5}{2}\Big\}\ln\Big(1-\frac{k_-^2}{M_V^2}-i\epsilon\Big)\ .
\end{align} 
In the case that the other gauge boson becomes resonant, the correction factor describing on-shell singular virtual contributions can be obtained by making the following replacement in the above equation \beq
S_{25}\leftrightarrow S_{34}, \hspace{1cm} k_-^2\leftrightarrow k_+^2\ .
\eeq

\subsubsection{Soft and collinear singular virtual contributions}
For the process $h^0\to WW^*\to 4l$, the photonic diagrams do not build a gauge invariant subset by themselves and their contributions are UV divergent. One can extract the soft and collinear singularities and on-shell logarithms from them and combine with the real photon bremsstrahlung to build the QED-like corrections, which are both IR and UV finite. The soft and collinear singular contributions from the virtual corrections can be extracted by making use of the well-known Kinoshita-Lee-Nauenberg (KLN) theorem~\cite{Kinoshita:1962ur,*Lee:1964is}, according to which the soft and collinear singularities are canceled out between the real and virtual corrections for sufficiently inclusive quantities. In the decay of $h^0$ to leptons, there is no initial state radiation of photons. The soft and collinear singular parts of the virtual corrections are exactly given by the singularities in the final state photon bremsstrahlung, but with opposite sign. The latter can be computed, e.g. with the dipole subtraction approach~\cite{Catani:1996jh,*Catani:1996vz,Dittmaier:1999mb,Roth:1999kk,Denner:2000bj}. Following~\cite{Denner:2000bj}, the soft and collinear singular parts of the virtual contributions can be defined as
\beq\label{virtsin}
d\Gamma_{virt,sing}=d\Gamma_{born}\frac{\alpha}{2\pi}\sum_{i=2}^5\sum_{j=i+1}^5 Q_i\,Q_j\Big(L(S_{ij},m_i^2)+L(S_{ij},m_j^2)-\frac{2\pi^2}{3}+3\Big)\ ,
\eeq
where $d\Gamma_{born}$ is the lowest order decay width, $Q_i$ denote the charge of final state fermions, and the function $L(S_{ij},m_i^2)$ is given by
\beq\label{Lfunction}
L(S_{ij},m_i^2)=\ln\Big(\frac{m_i^2}{S_{ij}}\Big)\ln\Big(\frac{\lambda^2}{S_{ij}}\Big)+\ln\Big(\frac{\lambda^2}{S_{ij}}\Big)-\frac{1}{2}\ln^2\Big(\frac{m_i^2}{S_{ij}}\Big)+\frac{1}{2}\ln\Big(\frac{m_i^2}{S_{ij}}\Big)\ .
\eeq
Note that when computing these contributions, we do not include the soft and collinear singular parts arising from the virtual photonic corrections to the mixed tree-level amplitude, i.e. to the second term in Eq.~(\ref{effbornamp}), since the corresponding photonic loop diagrams are not included in the virtual contribution. The definition of the soft and collinear singular virtual contribution is, of course, not unique. An alternative definition can be found, e.g. in~\cite{Yennie:1961ad}. 

The IR and on-shell singular virtual contributions combined with the contribution of real photon bremsstrahlung yield the QED-like correction. As mentioned, the soft singularities from photonic virtual diagrams are fully contained in the correction factors $\delta_{b,p}^{sin}$ and $\delta_{v,s}^{sin}$ defined in Eqs.~(\ref{bpcorrfac}) and (\ref{vscorrfac}). Subtracting from them the virtual singular factor in Eq.~(\ref{virtsin}) (note that one has to take twice the real part of the correction factor in (\ref{bpcorrfac}) and (\ref{vscorrfac})), the remnant must be free of soft singularities. This provides an analytic check on the cancellation of soft singularities. The cancellation of collinear singularities is checked numerically, since the collinear singularities appear in scalar integrals as well as in tensor ones, the contribution of the latter is not accounted for in the correction factors.

\subsection{Virtual corrections to $h^0\to ZZ^*\to 4l$}
\label{vchzz4f}
\begin{figure}[htbp]
\vspace*{-0.5cm}
\centering
\input{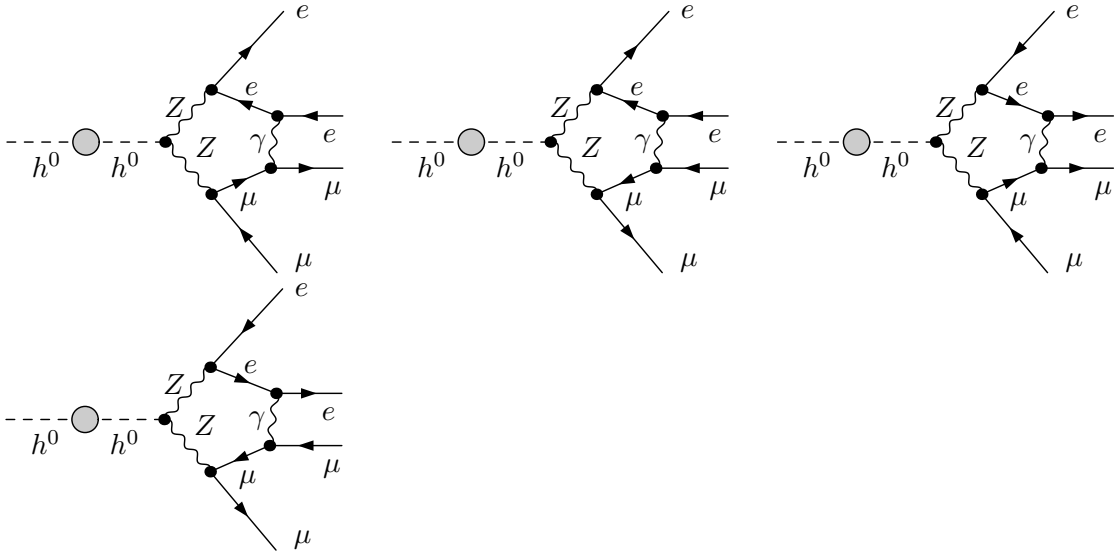}
\vspace*{-0.6cm}
\caption{Photonic one-loop diagrams that contribute to the on-shell singularities.}
\label{PhotondiagramZZ}
\end{figure}

The computation of virtual corrections to the decay of $h^0$ to four leptons via a $Z$ boson pair can be carried out analogously. Consider the process
\beq
h^0(k_1)\to ZZ^*\to e^-(k_2)+e^+(k_3)+\mu^+(k_4)+\mu^-(k_5)\ .
\eeq
We also define $k_+=k_2+k_3$ and $k_-=k_4+k_5$ as before. The virtual one-loop diagrams can be classified as for the process $h^0\to WW^*\to 4l$, and the $h^0 ZZ/H^0 ZZ$ counter term contributions can be easily obtained from Eq.~(\ref{ct}) by simple replacements: $\delta Z_w\to \delta Z_Z$, $\frac{\delta s_W}{s_W}\to \frac{\delta s_W}{s_W}\big(1-2\frac{s_W^2}{c_W^2}\big)$, while the structure of the other counter terms is as in the SM. In Fig.~\ref{PhotondiagramZZ} we show only the photonic diagrams that contribute to the on-shell singularities. 

In this decay process, the intermediate gauge bosons are neutral, thus only pentagon diagrams with a photon exchanged between two external charged fermions can contribute to on-shell singularities. The scalar 5-point integrals can again be decomposed by separating the non-resonant gauge boson propagator. Here we give explicitly the correction factor resulting from the photonic pentagon diagrams that contains the soft and on-shell singularities (assuming the gauge boson with four-momentum $k_-$ is at resonance)
\begin{align}\label{osfactor}
\delta_{p}^{sing}&=-\frac{\alpha}{2\pi}\Big\{\Big[-\mbox{Li}_2\Big(-\frac{S_{25}+k_-^2-M_V^2}{S_{24}}\Big)\non\\
&+2\ln\Big(-\frac{S_{24}}{m_e m_\mu}-i\epsilon\Big)\ln\Big(\frac{M_V^2-k_-^2}{\lambda M_V}-i\epsilon\Big)\non\\
&-\ln^2\Big(\frac{m_\mu}{M_V}\Big)-\ln^2\Big(\frac{M_V^2-k_-^2-S_{24}-S_{25}}{m_e M_V}-i\epsilon\Big)-\frac{\pi^2}{3}\Big]\non\\
&+(2\leftrightarrow3,4\leftrightarrow5)-(2\leftrightarrow3)-(4\leftrightarrow5)\Big\}\ ,
\end{align}
where the expressions in the parentheses are obtained from that in the squared bracket by interchange of indices. In the case that the other gauge boson is resonant, the correction factor can be obtained from Eq.~(\ref{osfactor}) by making the following substitutions
\beq
S_{25}\leftrightarrow S_{34}, \hspace{1cm} k_-^2\to k_+^2, \hspace{1cm} m_e\leftrightarrow m_\mu\ .
\eeq

The on-shell singular virtual contributions can be easily extracted from these correction factors. The photonic corrections to the $Zf\bar f$ vertex and the field renormalization constants of the external charged fermions contribute to soft singularities, but not to on-shell singularities. There is no contribution to these singularities from $Z$ boson self energy corrections.

\subsection{Application to semileptonic and hadronic final states}
\label{otherfs}
In the discussions above, we only consider the decay of $h^0$ to leptonic final states. The procedure outlined there can in principle be applied to semileptonic and hadronic final states as well. For semileptonic and hadronic final states, the number of photonic diagrams involving on-shell singularities would increase, but their evaluation can be carried out in complete analogy to what we did for leptonic final states. With appropriate substitution of momenta and masses, the analytical results for the scalar integrals in Appendix~\ref{appenintegrals} can still be used to extract the on-shell singularities from the virtual contributions. The soft and collinear singular virtual contributions are again given by Eq.~(\ref{virtsin}). The only possible exception in which the procedure used for our computation can not be straightforwardly applied is that the final state involves heavy down-type fermions, e.g. $b$ quarks. In our computation we keep the final state fermion masses only as regulators for collinear singularities and neglect them elsewhere. If the final state involves $b$ quarks, we can not neglect their masses, since their coupling to Higgs bosons can be enhanced at large $\tan\beta$ and give rise to numerically important effects.

\section{Real corrections}
\label{Realcorr}
In order to achieve an IR-finite physical result, the combination of virtual and real corrections is required, as a consequence of the fact that the experimental resolution of soft photons is limited. When considering the corresponding real photon emission for the processes given in Eqs.~(\ref{procWW}) and (\ref{procZZ}), the final state fermion masses are consistently neglected unless they have to be kept as regulators. Soft and collinear real photon emission give rise to singularities, which are regularized by the photon mass $\lambda$ and the fermion mass $m_f$, respectively. These soft and collinear singularities are treated within two different approaches, i.e. the phase space slicing and the dipole subtraction approaches. They will be briefly summarized in Appendix~\ref{appensoftandcoll}.

When evaluating the hard Bremsstrahlung contribution a constant width is introduced for each gauge boson propagator. For the propagator that is not resonant, the corresponding error is of higher order and negligible. If the collinearly emitted photon is not treated inclusively in the evaluation of distributions, the logarithm involving the light fermion masses would survive and become visible.

\section{Higher order final state radiation}
\label{HOfsr}
The emission of photons collinear to final state charged fermions leads to corrections that are enhanced by large logarithm involving the fermion masses. If the collinearly emitted photon is treated inclusively, this logarithm will cancel out as a consequence of the KLN theorem. If this is not the case, for instance, in the evaluation of distributions of final state muons, this logarithm will survive and might yield sizeable effects. Therefore one should take into account the corresponding higher order corrections. This can be done by the structure-function method~\cite{Beenakker:1996kt,Kuraev:1985hb,*Altarelli:1986kq,*Nicrosini:1986sm,*Nicrosini:1987sw,*Berends:1987ab} based on the mass factorization theorem. Here we incorporate the effects of the higher order final state radiation following~\cite{Denner:2000bj,Bredenstein:2006rh} and choose the factorization scale as the physical mass of the light CP-even Higgs boson mass.

\section{Numerical results}
\label{numeresults}
For the numerical evaluation, we use the following inputs for the SM parameters~\cite{Yao:2006px,Arguin:2005cc}
\begin{align}
G_\mu&=1.16637\times 10^{-5}\,\mbox{GeV}^{-2}\ , &\alpha&=1/137.03599968\ , \non\\
M_W&=80.403\,\mbox{GeV}\ , &\Gamma_W&=2.141\,\mbox{GeV} \ ,\non\\
M_Z&=91.1876\,\mbox{GeV}\ , &\Gamma_Z&=2.4952\,\mbox{GeV} \ ,\non\\
m_t&=172.7\,\mbox{GeV}\ , &m_b&=4.2\,\mbox{GeV} \ .\non
\end{align}
The lowest order matrix element is parametrized in such a way that it absorbs the running of the electromagnetic coupling and the leading universal corrections to the $\rho$ parameter, i.e. we use the effective coupling derived from the Fermi constant
\beq
\alpha_{G_\mu}=\frac{\sqrt 2 G_\mu M_W^2 s_W^2}{\pi}
\eeq
for the Born amplitude. In the relative $\mc O(\alpha)$ corrections, we use the coupling $\alpha=\alpha(0)$. 
To avoid double-counting from using $\alpha_{\mbox{\tiny{G}}_\mu}$, the charge renormalization constant in Eq.~(\ref{ct}) is modified to
\beq
\delta\tilde Z_e=\delta Z_e-\frac{1}{2}\Delta r\ ,
\eeq
where $\Delta r$ summarizes the radiative corrections to the muon decay. In the evaluation of distributions, a real photon closer than $5^\circ$ to the emitting charged fermion or with energy less than $1\,\mbox{GeV}$ is combined with the emitting charged fermion in the inclusive treatment.

In this analysis we investigate the results in several suggested benchmark scenarios~\cite{Carena:1999xa,*Carena:2002qg}, which are defined so that the two parameters that govern the tree-level Higgs sector, $M_{A^0}$ and $\tan\beta$, are varied while the other parameters that enter via radiative corrections are fixed. In these scenarios a common soft SUSY-breaking parameter $M_{\mbox{\tiny{SUSY}}}$, as well as the same trilinear coupling for the third generation slepton and squark, is chosen for simplicity. The U(1) gaugino mass parameter is given by the GUT relation $M_1=\frac{5}{3}\frac{s_W^2}{c_W^2}M_2$. Throughout the parameter scan, the experimental mass exclusion limits from direct search of supersymmetric particles and the upper bound on the SUSY corrections to the electroweak $\rho$ parameter~\cite{Yao:2006px} have been taken into account. In the parameter region $50\,\mbox{GeV}<M_{A^0}<1\,\mbox{TeV}$ and $1<\tan\beta<50$, the bound derived from the $\mbox{BR}(B\to X_s\gamma)$ prediction has ruled out the gluophobic scenario~\cite{Brein:2007da}, therefore we will not discuss this scenario here. The investigated scenarios are listed as follows:

\noindent 1. The $m_h^{\mbox{\small{max}}}$ scenario \\
The parameters in this scenario are given by
\begin{align}
M_{\mbox{\tiny{SUSY}}}&=1\,\mbox{TeV}\ , & \mu&=200\,\mbox{GeV}\ , & M_2&=200\,\mbox{GeV} \ ,\non\\
X_t&=2M_{\mbox{\tiny{SUSY}}}\ , & A_b&=A_t=A_{\tau}\ , & m_{\tilde g}&=0.8M_{\mbox{\tiny{SUSY}}}\ .
\end{align}
where $X_t$ is the mixing parameter of the top squark sector and $m_{\tilde g}$ is the gluino mass. This scenario yields a maximal value of the lightest CP-even Higgs boson for given $M_{A^0}$ and $\tan\beta$.

\noindent 2. The no-mixing scenario \\
The only difference of this scenario from the $m_h^{\mbox{\small{max}}}$ scenario is the vanishing mixing in the top squark sector and a higher value of $M_{\mbox{\tiny{SUSY}}}$, where the latter is chosen to avoid the exclusion bounds from the LEP Higgs searches~\cite{Barate:2003sz,Schael:2006cr}. The parameters in this scenario read
\begin{align}
M_{\mbox{\tiny{SUSY}}}&=2\,\mbox{TeV}\ , & \mu&=200\,\mbox{GeV}\ , & M_2&=200\,\mbox{GeV} \ ,\non\\
X_t&=0\ , & A_b&=A_t=A_{\tau}\ , & m_{\tilde g}&=0.8M_{\mbox{\tiny{SUSY}}}\ .
\end{align}

\noindent 3. The small-$\alpha_{\mbox{\small{eff}}}$ scenario \\
In this scenario a suppression of the $h^0b\bar b$ coupling can occur. The parameters are given by
\begin{align}
M_{\mbox{\tiny{SUSY}}}&=800\,\mbox{GeV}\ , & \mu&=2.5M_{\mbox{\tiny{SUSY}}}\ , & M_2&=500\,\mbox{GeV} \ ,\non\\
X_t&=-1100\,\mbox{GeV}\ , & A_b&=A_t=A_{\tau}\ , & m_{\tilde g}&=500\,\mbox{GeV}\ .
\end{align}

\begin{figure}[htbp]
\includegraphics[width=0.46\textwidth]{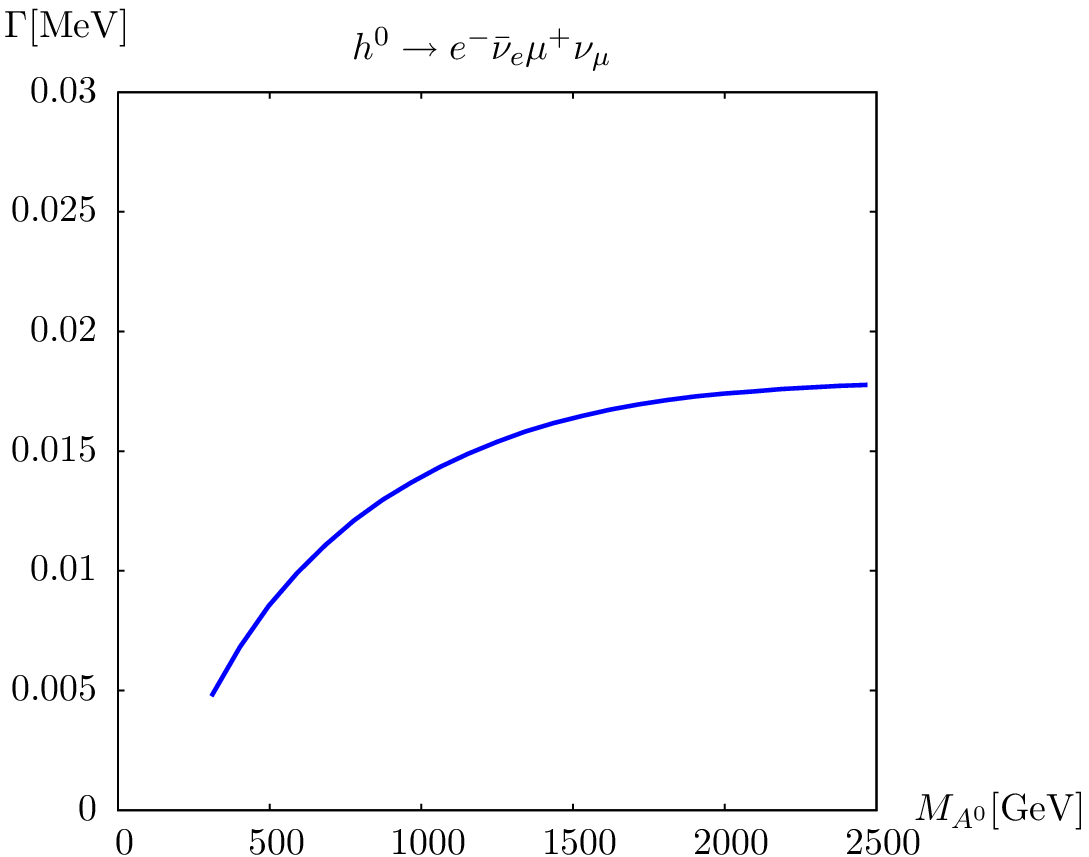}
\hspace{2em}
\includegraphics[width=0.47\textwidth]{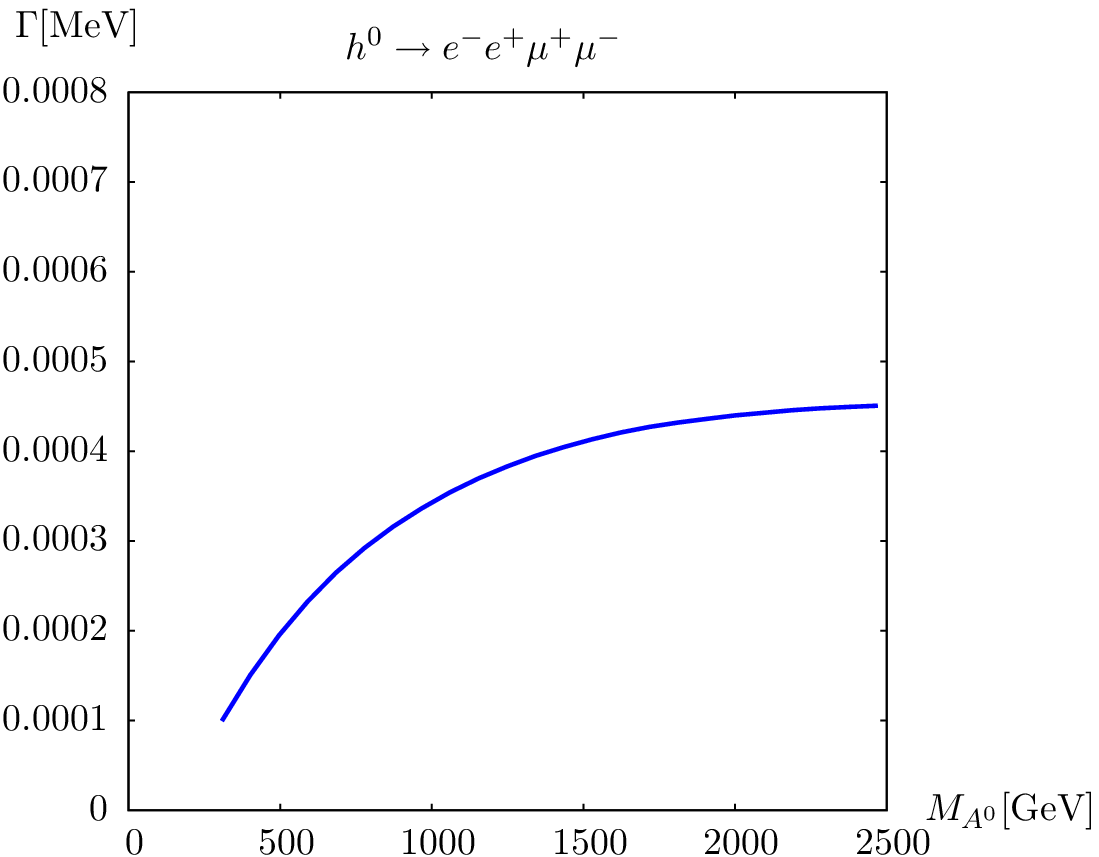}
\caption{Corrected partial decay width of $h^0$ from diagrams excluding genuine SUSY particles as a function of $M_{A^0}$, with $M_{\mbox{\tiny{SUSY}}}=\mu=M_2=M_{A^0}$ and $\tan\beta=6$. The remaining parameters are chosen as in the $m_h^{\mbox{\small{max}}}\;\mbox{scenario}$.}
\label{SMlike}
\end{figure}

In the decay processes considered in the present work, only SM particles are involved in the final state. It is interesting to compare the SM and the MSSM predictions for the partial decay widths. As discussed previously, in the limit that the mass parameter $M_{A^0}$ is much larger than the electroweak scale, all the heavy Higgs bosons will decouple and the contribution to the partial decay width from loop diagrams involving the SM particles and Higgs bosons will approach the SM prediction for a Higgs boson with the same mass. In order to compare the partial decay width of $h^0$ to four leptons in this limiting case with the SM result~\cite{Bredenstein:2006rh}, we also carry out the computation with the input parameters of ref.~\cite{Bredenstein:2006rh}, and choose the SUSY parameters $M_{\mbox{\tiny{SUSY}}}$, $\mu$ and $M_2$ to be  $M_{\mbox{\tiny{SUSY}}}=\mu=M_2=M_{A^0}$, so that the supersymmetric particles decouple when $M_{A^0}$ becomes large. The remaining SUSY parameters are chosen as in the $m_h^{\mbox{\small{max}}}\;\mbox{scenario}$. In Fig.~\ref{SMlike} we show the one-loop corrected partial decay width of $h^0$ to four leptons excluding the contribution of genuine SUSY diagrams as a function of $M_{A^0}$. In the limiting case that $M_{A^0}$ gets large ($M_{A^0}> 1.5\,\mbox{TeV}$), we find an agreement with the SM results.

\begin{figure}[htbp]
\includegraphics[width=0.46\textwidth]{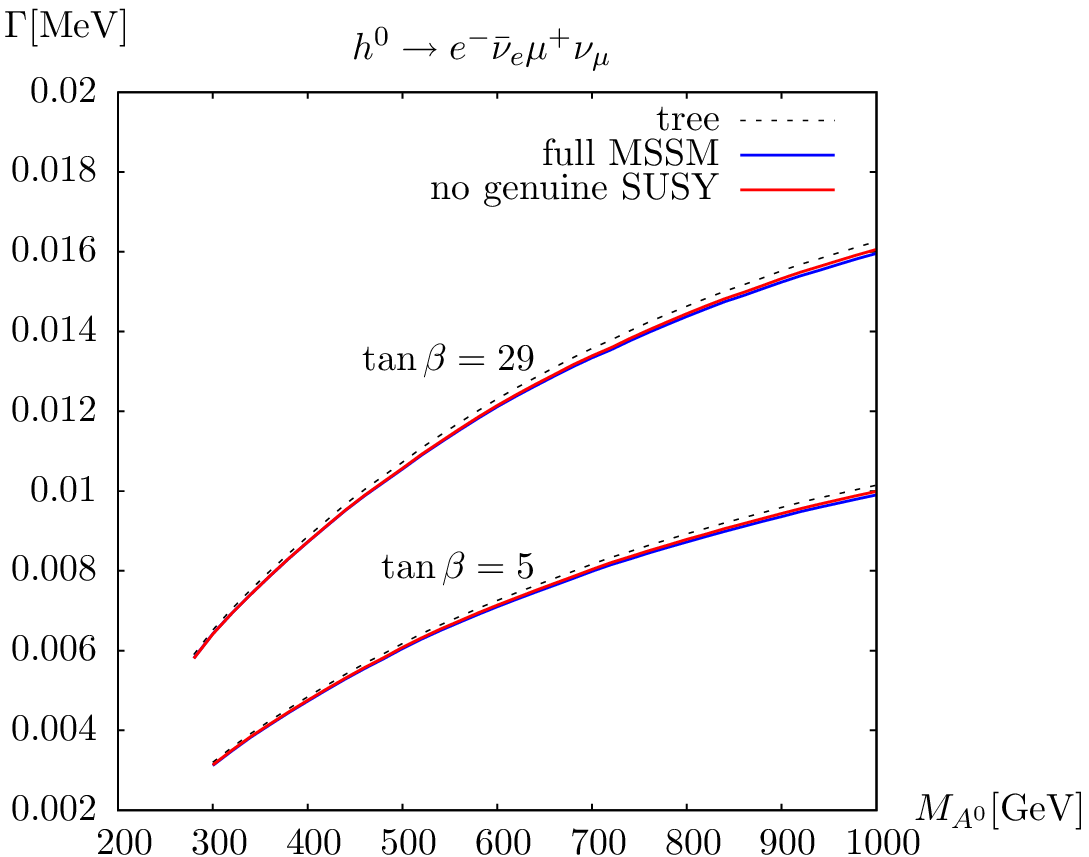}
\hspace{2em}
\includegraphics[width=0.47\textwidth]{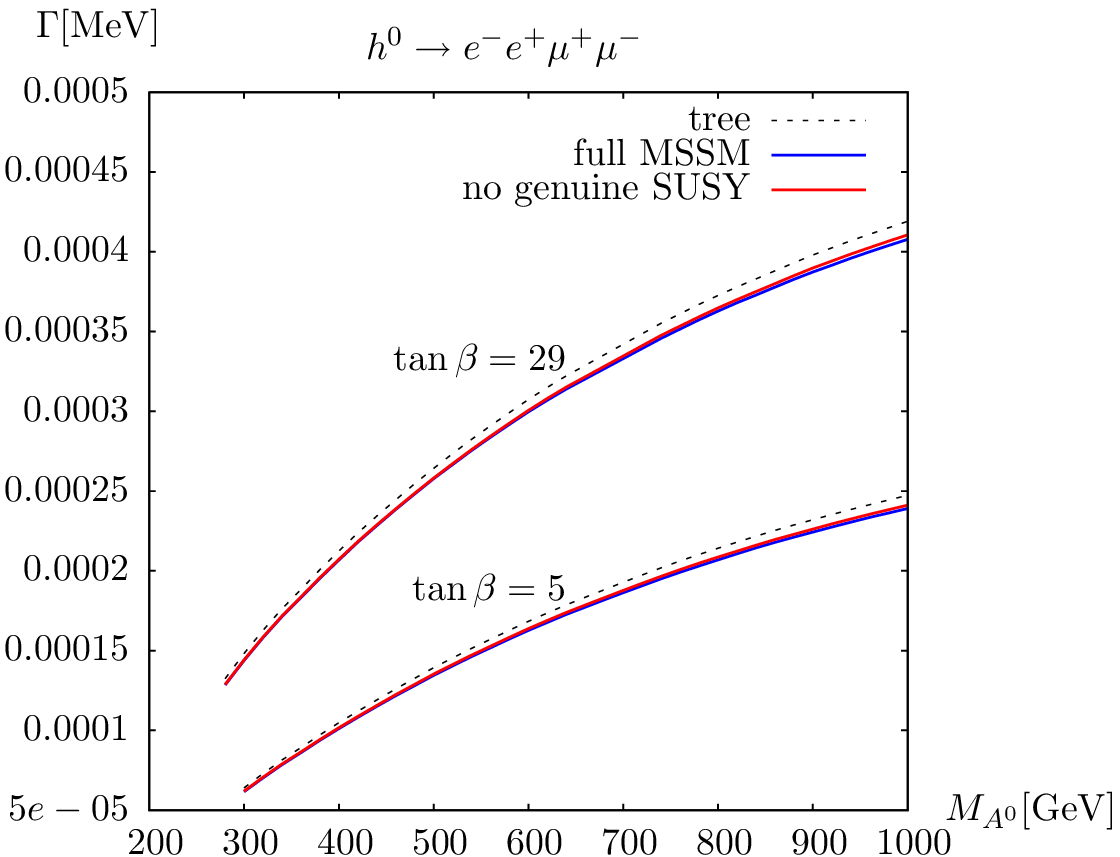}
\caption{Partial decay width of $h^0$ as a function of $M_{A^0}$, with $M_{\mbox{\tiny{SUSY}}}=\mu=M_2=M_{A^0}$ and $\tan\beta=5, 29$. The remaining parameters are chosen as in the $m_h^{\mbox{\small{max}}}\;\mbox{scenario}$. The dashed line denotes the tree-level result. The blue line shows the corrected width with full MSSM corrections, the red line shows the corrected width excluding genuine SUSY contributions.}
\label{compMSSMandSMlike}
\end{figure}

\begin{figure}[t]
\centering
\includegraphics[width=0.46\textwidth]{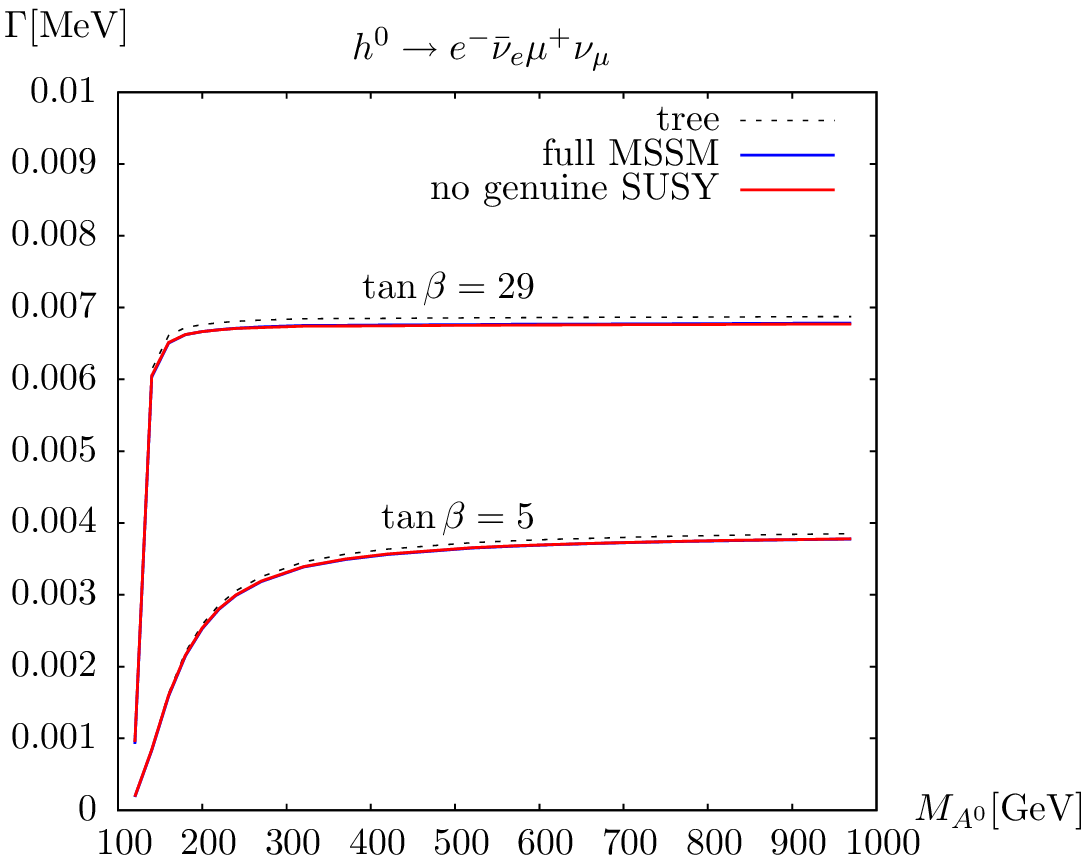}
\hspace{1.5em}
\includegraphics[width=0.49\textwidth]{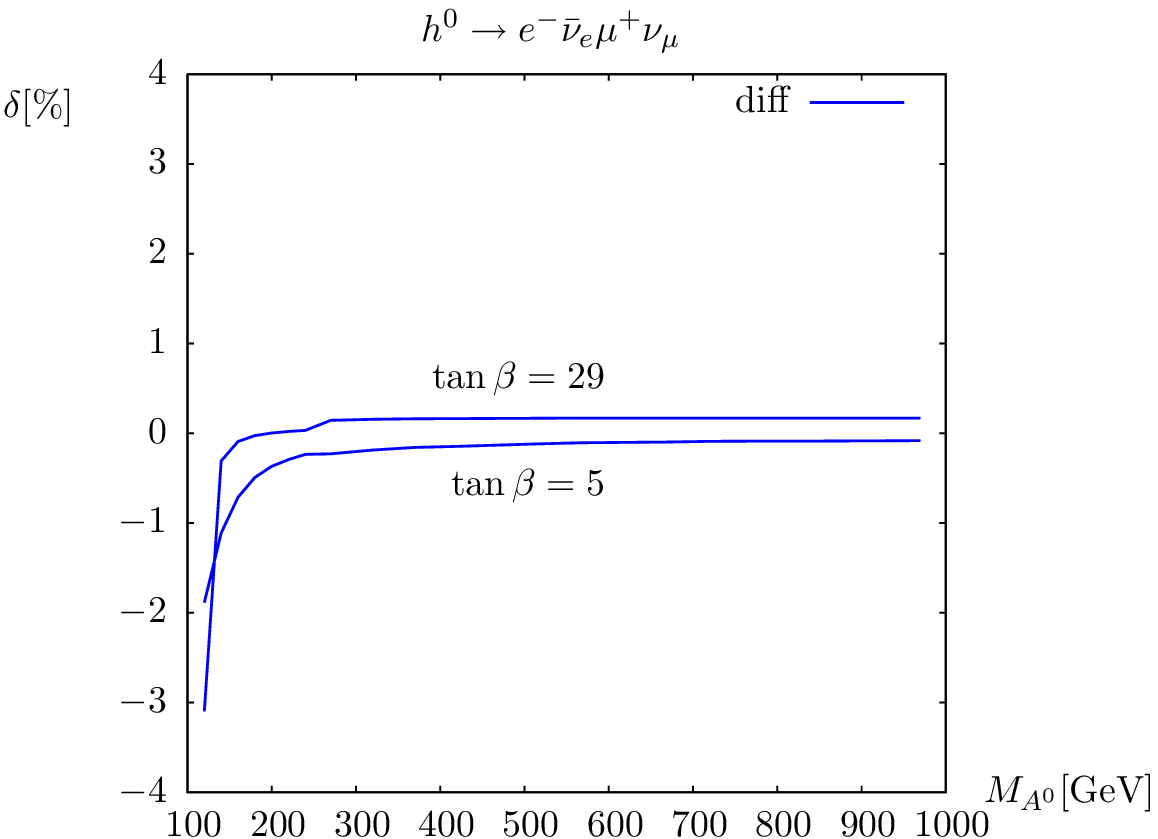}
\caption{Partial decay width of $h^0$ as a function of $M_{A^0}$, with $M_{\mbox{\tiny{SUSY}}}=300\mbox{GeV}$ and $\tan\beta=5, 29$. The remaining parameters are chosen as in the $m_h^{\mbox{\small{max}}}\;\mbox{scenario}$. The right plot shows the effects of the genuine SUSY contributions. }
\label{compMSSMandSMlikeMSusy300}
\end{figure}

If the generic mass scale of SUSY particles $M_{\mbox{\tiny{SUSY}}}$ is not much larger than the electroweak scale, the genuine SUSY particles will not decouple even in the limit $M_{A^0}\gg M_Z$. To investigate the numeric impact of contributions from the genuine SUSY particle spectrum, we compare the one-loop corrected partial decay width of $h^0$ with and without the genuine SUSY loop contributions. In Fig.~\ref{compMSSMandSMlike} the lowest order and corrected partial decay widths of $h^0$ to four leptons are depicted for $M_{\mbox{\tiny{SUSY}}}=\mu=M_2=M_{A^0}$, with the remaining parameters chosen as in the $m_h^{\mbox{\small{max}}}\;\mbox{scenario}$. The numerically most important one-loop corrections have been incorporated into the lowest order result by using the effective amplitude and the effective coupling. For the decay $h^0\ra e^-\bar\nu_e\mu^+\nu_\mu$, the relative size of the full MSSM loop contribution varies between $-2.5\%$ and $-2\%$ for $\tan\beta=5$, while it varies between $-2\%$ and $-1.5\%$ for $\tan\beta=29$. For the process $h^0\ra e^- e^+\mu^+\mu^-$, the relative corrections change from $-3\%$ to $-2\%$ and from $-4\%$ to $-3\%$ for $\tan\beta=5$ and $29$, respectively. As can be seen from the plots, for both processes, at large $M_{A^0}$ the blue curve that includes the genuine SUSY loop contributions and the red curve that does not almost coincide with each other, indicating that the effects of the genuine SUSY loop contributions are negligible in the decoupling limit. In Fig.~\ref{compMSSMandSMlikeMSusy300} we choose a relatively small value $300\,\mbox{GeV}$ for $M_{\mbox{\tiny{SUSY}}}$, so that the genuine SUSY spectrum is not too heavy. The relative corrections vary from $-2.5\%$ to $-1.5\%$ and from $-3\%$ to $-1\%$ for $\tan\beta=5$ and $29$, respectively. However, the genuine SUSY loop contributions again yield negligible effects in the large $M_{A^0}$ limit (while for small value of $M_{A^0}$, their contributions can reach several percent), as one can see from the right plot of Fig.~\ref{compMSSMandSMlikeMSusy300}. This implies that for the processes investigated here the decoupling behavior is essentially dominated by the mass parameter $M_{A^0}$, thus it will be rather difficult, even if one-loop corrections are taken into account, to distinguish the lightest MSSM Higgs boson from the SM Higgs boson if $M_{A^0}$ is large.

We also perform a generic scan over the most relevant parameters of the Higgs sector, $M_{A^0}$ and $\tan\beta$. Fig.~\ref{hWWwidth} shows the results for the one-loop corrected partial decay width of $h^0\ra e^-\bar\nu_e\mu^+\nu_\mu$ including the full MSSM corrections in three different scenarios, where the corrections to the $H^0WW$ vertex from loops involving heavy fermions and sfermions are not included. For $M_{A^0}>500\,\mbox{GeV}$, the results hardly vary with $M_{A^0}$ and therefore are not shown there. For small $M_{A^0}$ values ($M_{A^0}<140\,\mbox{GeV}$), the decay width is rather small due to a cancellation between the two parts of the effective Born amplitude Eq.~(\ref{effbornamp}).  When $M_{A^0}$ and $\tan\beta$ increase, the Higgs boson mass and thus the decay width increases rapidly in all three scenarios and reach a plateau after $\tan\beta>15$ and $M_{A^0}>220\,\mbox{GeV}$. In the small-$\alpha_{\mbox{\small{eff}}}$ scenario there is a slight decrease with $M_{A^0}$ for moderate and large $\tan\beta$ and $M_{A^0}>220\,\mbox{GeV}$. This is basically due to the slight decrease of the light CP-even Higgs boson mass with $M_{A^0}$ in this region. The relative corrections are $-1.5\%\sim1\%$ and $-3\%\sim-2\%$ respectively in the $m_h^{\mbox{\small{max}}}$ and no-mixing scenarios; in the small-$\alpha_{\mbox{\small{eff}}}$ scenario they do not exceed $-4\%$ unless for small $M_{A^0}$ values ($M_{A^0}<140\,\mbox{GeV}$), where the cancellation between the two parts of the effective Born amplitude Eq.~(\ref{effbornamp}) can yield a rather small lowest order result, and the size of the radiative correction is comparable to the lowest order result. This is not shown in Fig.~\ref{hWWwidth} so that the generic size of the relative corrections can be clearly seen. 

Owing to the mixing between the two CP-even Higgs bosons, the coupling of $H^0$ to gauge bosons needs to be taken into account as well. Although at tree-level this coupling is usually suppressed, the one-loop contribution can be numerically relevant, since the fermionic and sfermionic loops involve potentially large Yukawa couplings. In Fig.~\ref{HWWwidth} we show the correction to the partial decay width of $h^0\ra e^-\bar\nu_e\mu^+\nu_\mu$ due to the third generation fermionic and sfermionic loop contribution to the $H^0WW$ coupling. Such correction involves both the Yukawa couplings and the wave function normalization factors resulting from the mixing between the two CP-even Higgs bosons. While the latter can lead to a suppression/enhancement to the correction when $M_{A^0}/\tan\beta$ increase, the former tend to enhance the correction when $\tan\beta$ increases. The combination of these effects may lead to local extremum in the $M_{A^0}-\tan\beta$ plane. For our scan, a maximum appears at $\tan\beta=9,\,M_{A^0}=140\,\mbox{GeV}$ in the $m_h^{\mbox{\small{max}}}$ scenario and at $\tan\beta=19,\,M_{A^0}=120\,\mbox{GeV}$ in the no-mixing scenario. In the small-$\alpha_{\mbox{\small{eff}}}$ scenario, there is a maximum at $\tan\beta=9,\,M_{A^0}=140\,\mbox{GeV}$. The relative corrections are positive and stay below $1.5\%$ in the $m_h^{\mbox{\small{max}}}$ and no-mixing scenarios, and vary from $-0.5\%$ to $1\%$ in the small-$\alpha_{\mbox{\small{eff}}}$ scenario. In all three scenarios the contribution of these fermionic and sfermionic loops decreases very rapidly when $M_{A^0}$ becomes large.

In Fig.~\ref{hZZwidth} and \ref{HZZwidth} we show the same plots for $h^0\ra e^-e^+\mu^+\mu^-$, which exhibit similar features as Fig.~\ref{hWWwidth} and \ref{HWWwidth}. As shown in Fig.~\ref{hZZwidth}, the relative corrections are negative in all three scenarios. In the $m_h^{\mbox{\small{max}}}$ scenario the relative correction varies from $-3\%$ to $-0.5\%$, while it stays between $-5.5\%$ and $-3\%$ in the other two scenarios. In the small-$\alpha_{\mbox{\small{eff}}}$ scenario where the cancellation in the effective Born amplitude can occur, the relative correction can reach $95\%$. We also find that there is a maximum at $\tan\beta=11,\,M_{A^0}=140\,\mbox{GeV}$ in the $m_h^{\mbox{\small{max}}}$ scenario. The maximum occurs at $\tan\beta=23,\,M_{A^0}=120\,\mbox{GeV}$ in the no-mixing scenario and at $\tan\beta=9,\,M_{A^0}=140\,\mbox{GeV}$ in the small-$\alpha_{\mbox{\small{eff}}}$ scenario.

Fig.~\ref{HWWimdmax120and400} shows the invariant mass distribution of the $\mu^+\nu_\mu$ pair in the decay process $h^0\ra e^-\bar\nu_e\mu^+\nu_\mu$ in the $m_h^{\mbox{\small{max}}}$ scenario, where the parameters are chosen as $\tan\beta=30$,  $M_{A^0}=120\,\mbox{GeV}$ and $400\,\mbox{GeV}$. As the mass of $h^0$ stays below the production threshold of the gauge boson pair, only one intermediate gauge boson can be resonant. From the plots it can be seen that in addition to the peak around the W boson mass, there is another broad peak at a small invariant mass. This is the point where the other W boson gets resonant. In the right plot the broad peak is closer to the W resonance peak, as the Higgs boson mass is larger. In Fig.~\ref{HWWimdmax120and400relative1} we show the relative corrections to the invariant mass distribution of the $\mu^+\nu_\mu$ pair in the $m_h^{\mbox{\small{max}}}$ scenario. From the plots one can find an enhancement at low invariant mass due to the emission of photon off the final state fermion. In the case that the collinearly emitted photon is not combined with the emitting fermion, the logarithm involving the light fermion mass would survive and give rise to large corrections. This is shown by the blue curves in  Fig.~\ref{HWWimdmax120and400relative1}, where the red curves show the results with photon combinations, i.e. the collinear photon is combined with the emitting fermion. If the fermion masses are consistently neglected, the invariant mass distribution of the $e^-\bar\nu_e$ pair with collinear photon combination will coincide with the red curves in the plots. In Fig.~\ref{HWWimdmax120and400relative2} we show the relative contributions due to the higher order final state radiation, and due to the corrections of the third generation fermions and sfermions to the $H^0WW$ coupling, where in the right plot the latter is not shown since it is strongly suppressed by the wave function normalization factors and is completely negligible for large $M_{A^0}$ values. For both $M_{A^0}$ values, the higher order final state radiation can lead to corrections less than $2\%$. The corrections of the third generation fermions and sfermions to the $H^0WW$ coupling give rise to a contribution less than $1\%$ for $M_{A^0}=120\,\mbox{GeV}$. In Fig.~\ref{HWWimdnoandsmall400} the invariant mass distributions of the $\mu^+\nu_\mu$ pair without photon combination in the no-mixing and small-$\alpha_{\mbox{\small{eff}}}$ scenarios are shown with $\tan\beta=30$ and $M_{A^0}=400\,\mbox{GeV}$. The relative corrections can reach $\sim 25\%$ in both scenarios. 

Fig.~\ref{HZZimdmax120and400} to \ref{HZZimdnoandsmall400} show the corresponding invariant mass distributions for the process $h^0\ra e^-e^+\mu^+\mu^-$ in three benchmark scenarios. In the left plot of Fig.~\ref{HZZimdmax120and400}, the broad peak at low invariant mass is not clearly visible in the depicted region because of the larger mass of $Z$ boson compared to the $W$ boson mass. As shown in Fig.~\ref{HZZimdmax120and400relative2}, for both $M_{A^0}$ values, the higher order final state radiation gives rise to larger corrections than in the previous process, as the final state now involves two muons. While the correction to the $H^0ZZ$ coupling leads to a contribution of $\sim1\%$ for $M_{A^0}=120\,\mbox{GeV}$ in the $m_h^{\mbox{\small{max}}}$ scenario, it is completely negligible for $M_{A^0}=400\,\mbox{GeV}$. In Fig.~\ref{HZZimdnoandsmall400} the invariant mass distributions in the no-mixing and small-$\alpha_{\mbox{\small{eff}}}$ scenarios are depicted with $\tan\beta=30$ and $M_{A^0}=400\,\mbox{GeV}$ (no photon combination). In both scenarios, the relative corrections vary from $\sim-30\%$ to $\sim10\%$.

\section{Conclusions}
\label{concl}
We have investigated the leptonic decay processes of the light CP-even Higgs boson in the MSSM via a gauge boson pair and computed the corresponding $\mc O(\alpha)$ electroweak corrections, improved by the two-loop corrections provided by FeynHiggs. 

The tree-level coupling of the lightest CP-even MSSM Higgs boson to the SM fermions and gauge bosons approaches that of the SM Higgs boson in the decoupling limit $M_{A^0}\gg M_Z$. At one-loop level, the radiative corrections that arise from loops involving the SM particles and the MSSM Higgs bosons also tend to the SM result in the decoupling limit, since all heavy Higgs bosons decouple in this limit. We computed such corrections in this limit and found an agreement with the SM result. The decoupling of the genuine SUSY particles is governed by their characteristic mass scale. If this mass scale is not much larger than the electroweak scale, these genuine SUSY particles might yield sizeable effects. However, our results show that they only yield negligible effects in the limit that $M_{A^0}\gg M_Z$, even for a relatively light genuine SUSY spectrum. This indicates that in our case the decoupling behavior is essentially dominated by the mass of the CP-odd Higgs boson mass, thus it will be rather difficult, even if one-loop corrections are taken into account, to distinguish the lightest MSSM Higgs boson from the SM Higgs boson if $M_{A^0}$ is large.

For the decay processes considered in this work, the relative corrections to the decay width turn out to be of the order of several percent in the investigated SUSY scenarios unless for small CP-odd Higgs boson masses in the $\mbox{small}$-$\alpha_{\mbox{\small{eff}}}$ scenario, where a cancellation in the effective Born amplitude can occur and leads to a small lowest order result and a large relative correction. The corrections to the distributions are significant. Owing to the mixing between Higgs bosons, the potentially large corrections from the third generation fermions and sfermions to the coupling of the heavy CP-even Higgs boson and gauge bosons have been taken into account as well. Such corrections are comparable to other one-loop corrections in all investigated scenarios at small CP-odd Higgs boson masses, while decrease quite rapidly and become completely negligible when the CP-odd Higgs boson mass becomes large. In the evaluation of distributions, we also discussed the contribution of the higher order final state radiation, the effects of which are typically of the order of several percent and comparable to the contribution from the heavy CP-even Higgs boson-gauge bosons vertex correction at small values of the CP-odd Higgs boson mass.

The analytical expressions of the on-shell singular scalar integrals listed in Appendix~\ref{appenintegrals} can be useful for the computation of other processes involving unstable particles in the loop.

\section*{Acknowledgements}
We thank Stefan Dittmaier and Thomas Hahn for useful discussions.

\appendix
\section{Results of integrals}
\label{appenintegrals}
In this appendix we list the analytical expressions of the scalar integrals involving on-shell singularities discussed in section \ref{ossingvirtcorr}. The variables $k_i$, $S_{ij}$ and the mass parameters are also defined there. These integrals are obtained with the help of of refs.~\cite{Beenakker:1988jr,Denner:2005nn}. The expressions of integrals are given for zero gauge boson width. The real squared mass $M_V^2$ should be replaced by $M_V^2-iM_V\Gamma_V$ in the result, if singularities arise when $k_\pm^2$ approach $M_V^2$.
\begin{align}
B_0(k_-,0,M_V)&=\Delta+2+\ln\left(\frac{\mu^2}{M_V^2}\right)+\left(\frac{M_V^2}{k_-^2}-1\right)\ln\left(1-\frac{k_-^2}{M_V^2}-i\epsilon\right)\ , \non\\
C_0(-k_4,-k_-,0,m_\mu,M_V)&=\frac{1}{k_-^2}\Big\{\ln\frac{k_-^2}{m_\mu^2}\,\ln\Big(1-\frac{k_-^2}{M_V^2}-i\epsilon\Big)+\mbox{Li}_2\Big(1-\frac{k_-^2}{M_V^2}-i\epsilon\Big)\non\\
&-\frac{\pi^2}{6}\Big\}\ , \non\\
C_0(-k_2,k_-,0,m_e,M_V)&=\frac{1}{S_{24}+S_{25}}\Big\{\Big[\ln\Big( M_V^2m_e^2\Big)-2\ln\Big(-S_{24}-S_{25}-i\epsilon\Big)\Big]\non\\
&\times\ln\Big(M_V^2-k_-^2-i\epsilon\Big)-\ln\Big(M_V^2-k_-^2-S_{24}-S_{25}-i\epsilon\Big)\non\\
&\times\Big[\ln\Big( M_V^2m_e^2\Big)-\ln\Big(M_V^2-k_-^2-S_{24}-S_{25}-i\epsilon\Big)\Big]\non\\
&+\ln^2\Big(-S_{24}-S_{25}-i\epsilon\Big)+2\mbox{Li}_2\Big(\frac{M_V^2-k_-^2}{S_{24}+S_{25}}\Big)\non\\
&+\mbox{Li}_2\Big(\frac{k_-^2+S_{24}+S_{25}}{M_V^2}\Big)-\mbox{Li}_2\Big(\frac{k_-^2}{M_V^2}\Big)+\frac{\pi^2}{3}\Big\}\ , \non\\
D_0(-k_2,k_-,k_4,\lambda,m_e,M_V,m_\mu)&=\frac{1}{S_{24}(k_-^2-M_V^2)}\Big\{-\mbox{Li}_2\Big(-\frac{S_{25}+k_-^2-M_V^2}{S_{24}}\Big)\non\\
&+2\ln\Big(-\frac{S_{24}}{m_e m_\mu}-i\epsilon\Big)\ln\Big(\frac{M_V^2-k_-^2}{\lambda M_V}-i\epsilon\Big)\non\\
&-\ln^2\Big(\frac{m_\mu}{M_V}\Big)-\ln^2\Big(\frac{M_V^2-k_-^2-S_{24}-S_{25}}{m_e M_V}-i\epsilon\Big)\non\\
&-\frac{\pi^2}{3}\Big\}\ , \non
\end{align}
\begin{align}
D_0(-k_2,-k_+,k_4,\lambda,m_e,M_V,m_\mu)&=\frac{1}{S_{24}(k_+^2-M_V^2)}\Big\{-\mbox{Li}_2\Big(-\frac{S_{34}+k_+^2-M_V^2}{S_{24}}\Big)\non\\
&+2\ln\Big(-\frac{S_{24}}{m_e m_\mu}-i\epsilon\Big)\ln\Big(\frac{M_V^2-k_+^2}{\lambda M_V}-i\epsilon\Big)\non\\
&-\ln^2\Big(\frac{m_e}{M_V}\Big)-\ln^2\Big(\frac{M_V^2-k_+^2-S_{24}-S_{34}}{m_\mu M_V}-i\epsilon\Big)\non\\
&-\frac{\pi^2}{3}\Big\}\ ,
\end{align}
where $\Delta=\frac{2}{4-D}-\gamma_E+\ln(4\pi)$ and $\mu$ is the reference mass scale of dimensional regularization.

\section{Treatment of real soft and collinear photon emission}
\label{appensoftandcoll}
In this appendix we briefly describe the two approaches used to deal with the soft and collinear photon emission: the phase space slicing and the dipole subtraction method~\cite{Denner:2000bj,Dittmaier:1999mb}. In both approaches we regularize the soft and collinear singularities by the photon and fermion masses, respectively.
\subsection{Phase space slicing}
For the real photon Bremsstrahlung the phase space integral diverges in certain regions. One can divide the phase space into singular and non-singular regions. In the non-singular region the integral is finite and can be evaluated numerically without regulators. In the singular region the integral has to be evaluated analytically with regulators. The singular region consists of the soft region, where the photon energy is smaller than a given cutoff $\Delta E$; and the collinear region, in which the photon is emitted collinearly (but not soft) to a charged fermion, namely the angle between the emitted photon and the charged fermion is smaller than an angular cutoff $\Delta\theta$. The real corrections can be decomposed as follows
\beq
d\Gamma^{h^0\ra 4l\gamma}=d\Gamma_{soft}+d\Gamma_{coll}+d\Gamma_{finite}^{h^0\ra4l\gamma}\ .
\eeq
In the soft and collinear regions, the squared matrix element $|\mc M^{h^0\to 4l\gamma}|^2$ factorizes into the lowest order squared matrix element and a universal soft or collinear factor. The five particle phase space also factorizes into four particle phase space and a photon part, so that the photon momentum can be integrated over analytically. In the soft region, the soft photon approximation can be applied, in which the photon 4-momentum is omitted everywhere except in the IR singular propagators. Note that the fermion masses are kept only as regulators for the collinear singularities, the soft photon correction factor can then be written as~\cite{Denner:2000bj}
\begin{align}
d\Gamma_{soft}&=d\Gamma_{born}\;\frac{\alpha}{\pi}\;\sum_{i=2}^5\sum_{j=i+1}^5 Q_iQ_j\Bigg\{2\ln\left(\frac{2\Delta E}{\lambda}\right)\left[1-\ln\left(\frac{S_{ij}}{m_im_j}\right)\right] - \ln\left(\frac{4k_i^0k_j^0}{m_im_j}\right)\non\\
&+\ln^2\left(\frac{2k_i^0}{m_i}\right)+\ln^2\left(\frac{2k_j^0}{m_j}\right)+\frac{\pi^2}{3}+\mbox{Li}_2\left(1-\frac{4k_i^0k_j^0}{S_{ij}}\right)\Bigg\}\ .
\end{align}

In the collinear region, the squared matrix element and the phase space also factorize as in the soft region, and the collinear factor that describes the collinear final state radiation is given by
\beq\label{colinclu}
d\Gamma_{coll}=d\Gamma_{born}\;\frac{\alpha}{2\pi}\sum_{i=2}^5Q_i^2\Bigg\{\left[\frac{3}{2}+2\ln\left(\frac{\Delta E}{k_i^0}\right)\right]\left[1-2\ln\left(\frac{\Delta\theta k_i^0}{m_i}\right)\right]+3-\frac{2\pi^2}{3}\Bigg\}\ .
\eeq
The cutoff parameters $\Delta E$ and $\Delta\theta$ should be chosen sufficiently small so that the soft photon and leading-pole approximations apply. On the other hand, they should not be too small so that the instabilities of numerical integration can be avoided. Also note that this result assumes that a photon emitted collinearly to a charged fermion is treated inclusively, namely it is combined to the emitting charged fermion. As a result, all dependence on the photon and fermion masses will drop out in the final result. If the collinear photons are not treated inclusively, then in the collinear region one has~\cite{Bredenstein:2005zk}
\begin{align}\label{colexclu}
d\Gamma_{coll}&=\sum_{i=2}^5 \frac{\alpha}{2\pi}\, Q_i^2\, d\Gamma_{born}(\tilde k_i)\int_{0}^{1-\frac{\Delta E}{\tilde k_i^0}} dz_i \Bigg\{p_{ff}(z_i)\bigg[2\ln\Bigg(\frac{\Delta\theta\tilde k_i^0}{m_i}z_i\Bigg)-1\bigg] \non\\
&+(1-z_i)\Bigg\}\Theta(z_i)
\end{align}
with $z_i=k_i^0/\tilde k_i^0$ and the splitting function
\beq
P_{ff}(z_i)=\frac{1+z_i^2}{1-z_i}\ .
\eeq
Here $\tilde k_i^0$ and $k_i^0$ denote the energy of the charged fermion before and after emitting the collinear photon, the function $\Theta(z_i)$ summarizes the phase space cuts. The integration over $z_i$ in Eq.~(\ref{colexclu}) is constrained by the phase space cuts $\Theta(z_i)$ and cannot be performed analytically. Consequently the fermion mass singularities are not fully canceled in the combination of virtual and real corrections and thus become visible. If the photon is treated inclusively, the integration over $z_i$ will not be constrained by any phase space cut, and thus can be performed analytically, leading to Eq.~(\ref{colinclu}).

\subsection{Dipole subtraction}
In this approach \cite{Catani:1996jh,*Catani:1996vz,Dittmaier:1999mb,Roth:1999kk} one constructs an auxiliary function which contains the same singularities as the real bremsstrahlung integrand. Subtracting this auxiliary function from the bremsstrahlung integrand thus cancels all soft and collinear singularities and the difference can be integrated numerically, even in the singular region. In this numerical integration no regulators are needed for the soft and collinear singularities. The auxiliary function can then be integrated analytically (regulators required) and re-added to the original integral. Within the subtraction method there is no singular contribution involved in the numerical integration. Hence for computations within this method, the statistical uncertainty is smaller than that of the slicing method, in which the singular contributions are present in the numerical integration. The auxiliary function must possess the same asymptotic behavior as the original integrand in the soft and collinear limit, and has to be simple enough to be integrated over the singular regions analytically. In our case, soft and collinear singularities occur only in the final state. As the masses of the final state fermions can be consistently neglected, the expression of the auxiliary function is fairly simple \cite{Dittmaier:1999mb}
\beq
|\mc M_{sub}(\Phi_{4f\gamma})|^2=-\sum_{\stackrel{i,j=2}{i\not=j}}^5 Q_i\,Q_j\,g_{ij}^{sub}(k_i,k_j,k)|\mc M_{born}(\tilde\Phi_{4f,ij})|^2\ ,
\eeq
where $k_i$ and $k$ denote the respective momenta of final state fermions and photon, and the functions $g_{ij}^{sub}$ contain the soft and collinear singularities
\beq
g_{ij}^{sub}(k_i,k_j,k)=\frac{1}{(k_ik)(1-y_{ij})}\Big[\frac{2}{1-z_{ij}(1-y_{ij})}-1-z_{ij}\Big]
\eeq
with the variables
\beq
y_{ij}=\frac{k_ik}{k_ik_j+k_ik+k_jk}\ ,\;\;\;\;\;\;\;\;\;\;z_{ij}=\frac{k_ik_j}{k_ik_j+k_jk}\ .
\eeq
The mapping between the phase space of the radiative and non-radiative process,  $\Phi_{4f\gamma}$ and $\tilde\Phi_{4f}$ is defined as
\beq
\tilde k_i^\mu=k_i^\mu+k^\mu-\frac{y_{ij}}{1-y_{ij}}k_j^\mu\ ,\hspace{1cm}  \tilde k_j^\mu=\frac{1}{1-y_{ij}}k_j^\mu
\eeq
with all other momenta unchanged. The contribution of the auxiliary function should be computed analytically. After integrating over the photon momentum the result reads 
\beq
\int d\Phi_{4f\gamma}|\mc M_{sub}(\Phi_{4f\gamma})|^2=-\frac{\alpha}{2\pi}\sum_{\stackrel{i,j=2}{i\not=j}}^5 Q_i\,Q_j\int d\tilde\Phi_{4f,ij}\,G_{ij}^{sub}(S_{ij})|\mc M_{born}(\tilde\Phi_{4f,ij})|^2
\eeq
with the function
\beq
G_{ij}^{sub}(S_{ij})=L(S_{ij},m_i^2)-\frac{\pi^2}{3}+\frac{3}{2}\ ,
\eeq
where $L(S_{ij},m_i^2)$ has been defined in Eq. (\ref{Lfunction}).

\bibliographystyle{JHEP}
\bibliography{paper}

\providecommand{\href}[2]{#2}\begingroup\raggedright\begin{thebibliography}{10}

\bibitem{Heinemeyer:1998jw}
S.~Heinemeyer, W.~Hollik, and G.~Weiglein, {\it {{QCD} Corrections to the
  Masses of the Neutral CP-even Higgs Bosons in the MSSM}},  {\em Phys. Rev.}
  {\bf D58} (1998) 091701, [\href{http://xxx.lanl.gov/abs/hep-ph/9803277}{{\tt
  hep-ph/9803277}}].

\bibitem{Heinemeyer:1998kz}
S.~Heinemeyer, W.~Hollik, and G.~Weiglein, {\it {Precise Prediction for the
  Mass of the Lightest Higgs Boson in the MSSM}},  {\em Phys. Lett.} {\bf B440}
  (1998) 296--304, [\href{http://xxx.lanl.gov/abs/hep-ph/9807423}{{\tt
  hep-ph/9807423}}].

\bibitem{Heinemeyer:1998np}
S.~Heinemeyer, W.~Hollik, and G.~Weiglein, {\it {The Masses of the Neutral
  CP-even Higgs Bosons in the MSSM: Accurate Analysis at the Two-Loop Level}},
  {\em Eur. Phys. J.} {\bf C9} (1999) 343--366,
  [\href{http://xxx.lanl.gov/abs/hep-ph/9812472}{{\tt hep-ph/9812472}}].

\bibitem{Degrassi:2002fi}
G.~Degrassi, S.~Heinemeyer, W.~Hollik, P.~Slavich, and G.~Weiglein, {\it
  {Towards high-precision predictions for the MSSM Higgs sector}},  {\em Eur.
  Phys. J.} {\bf C28} (2003) 133--143,
  [\href{http://xxx.lanl.gov/abs/hep-ph/0212020}{{\tt hep-ph/0212020}}].

\bibitem{Allanach:2004rh}
B.~C. Allanach, A.~Djouadi, J.~L. Kneur, W.~Porod, and P.~Slavich, {\it
  {Precise determination of the neutral Higgs boson masses in the MSSM}},  {\em
  JHEP} {\bf 09} (2004) 044,
  [\href{http://xxx.lanl.gov/abs/hep-ph/0406166}{{\tt hep-ph/0406166}}].

\bibitem{Ball:2007zza}
{\bf CMS} Collaboration, G.~L. Bayatian {\em et~al.}, {\it {CMS Technical
  Design Report, Volume II: Physics performance}},  {\em J. Phys.} {\bf G34}
  (2007) 995--1579.

\bibitem{Atlas:1999fr}
{\it {ATLAS Detector and Physics Performance. Technical Design Report. Vol.
  2}}, . CERN-LHCC-99-15.

\bibitem{Dedes:2003cg}
A.~Dedes, S.~Heinemeyer, S.~Su, and G.~Weiglein, {\it {The Lightest Higgs Boson
  of mSUGRA, mGMSB and mAMSB at Present and Future Colliders: Observability and
  Precision Analyses}},  {\em Nucl. Phys.} {\bf B674} (2003) 271--305,
  [\href{http://xxx.lanl.gov/abs/hep-ph/0302174}{{\tt hep-ph/0302174}}].

\bibitem{Bredenstein:2006rh}
A.~Bredenstein, A.~Denner, S.~Dittmaier, and M.~M. Weber, {\it {Precise
  Predictions for the Higgs-boson Decay H $\to$ W W / Z Z $\to$ $4$ leptons}},
  {\em Phys. Rev.} {\bf D74} (2006) 013004,
  [\href{http://xxx.lanl.gov/abs/hep-ph/0604011}{{\tt hep-ph/0604011}}].

\bibitem{Bredenstein:2006ha}
A.~Bredenstein, A.~Denner, S.~Dittmaier, and M.~M. Weber, {\it {Radiative
  Corrections to the Semileptonic and Hadronic Higgs-boson Decays H $\to$ W W /
  Z Z $\to$ $4$ fermions}},  {\em JHEP} {\bf 02} (2007) 080,
  [\href{http://xxx.lanl.gov/abs/hep-ph/0611234}{{\tt hep-ph/0611234}}].

\bibitem{Carena:1999xa}
M.~S. Carena, S.~Heinemeyer, C.~E.~M. Wagner, and G.~Weiglein, {\it
  {Suggestions for Improved Benchmark Scenarios for Higgs- boson Searches at
  LEP2}},  \href{http://xxx.lanl.gov/abs/hep-ph/9912223}{{\tt hep-ph/9912223}}.

\bibitem{Carena:2002qg}
M.~S. Carena, S.~Heinemeyer, C.~E.~M. Wagner, and G.~Weiglein, {\it
  {Suggestions for Benchmark Scenarios for MSSM Higgs Boson Searches at Hadron
  Colliders}},  {\em Eur. Phys. J.} {\bf C26} (2003) 601--607,
  [\href{http://xxx.lanl.gov/abs/hep-ph/0202167}{{\tt hep-ph/0202167}}].

\bibitem{Gunion:2002zf}
J.~F. Gunion and H.~E. Haber, {\it {The CP-conserving two-Higgs-doublet model:
  The approach to the decoupling limit}},  {\em Phys. Rev.} {\bf D67} (2003)
  075019, [\href{http://xxx.lanl.gov/abs/hep-ph/0207010}{{\tt
  hep-ph/0207010}}].

\bibitem{Carena:2002es}
M.~S. Carena and H.~E. Haber, {\it {Higgs Boson Theory and Phenomenology.
  ((V))}},  {\em Prog. Part. Nucl. Phys.} {\bf 50} (2003) 63--152,
  [\href{http://xxx.lanl.gov/abs/hep-ph/0208209}{{\tt hep-ph/0208209}}].

\bibitem{Frank:2006yh}
M.~Frank {\em et~al.}, {\it {The Higgs Boson Masses and Mixings of the Complex
  MSSM in the Feynman-Diagrammatic Approach}},  {\em JHEP} {\bf 02} (2007) 047,
  [\href{http://xxx.lanl.gov/abs/hep-ph/0611326}{{\tt hep-ph/0611326}}].

\bibitem{Heinemeyer:2004ms}
S.~Heinemeyer, {\it {MSSM Higgs Physics at Higher Orders}},  {\em Int. J. Mod.
  Phys.} {\bf A21} (2006) 2659--2772,
  [\href{http://xxx.lanl.gov/abs/hep-ph/0407244}{{\tt hep-ph/0407244}}].

\bibitem{Freitas:2002um}
A.~Freitas and D.~Stockinger, {\it {Gauge Dependence and Renormalization of
  tan(beta) in the MSSM}},  {\em Phys. Rev.} {\bf D66} (2002) 095014,
  [\href{http://xxx.lanl.gov/abs/hep-ph/0205281}{{\tt hep-ph/0205281}}].

\bibitem{Brignole:1992uf}
A.~Brignole, {\it {Radiative Corrections to the Supersymmetric Neutral Higgs
  Boson Masses}},  {\em Phys. Lett.} {\bf B281} (1992) 284--294.

\bibitem{Frank:2002qf}
M.~Frank, S.~Heinemeyer, W.~Hollik, and G.~Weiglein, {\it {FeynHiggs1.2: Hybrid
  MS-bar / On-Shell Renormalization for the CP-even Higgs Boson Sector in the
  MSSM}},  \href{http://xxx.lanl.gov/abs/hep-ph/0202166}{{\tt hep-ph/0202166}}.

\bibitem{Denner:1991kt}
A.~Denner, {\it {Techniques for Calculation of Electroweak Radiative
  Corrections at the One Loop Level and Results for W Physics at LEP-200}},
  {\em Fortschr. Phys.} {\bf 41} (1993) 307--420,
  [\href{http://xxx.lanl.gov/abs/0709.1075}{{\tt 0709.1075}}].

\bibitem{Heinemeyer:2000fa}
S.~Heinemeyer, W.~Hollik, and G.~Weiglein, {\it {Decay Widths of the Neutral
  CP-even MSSM Higgs Bosons in the Feynman-Diagrammatic Approach}},  {\em Eur.
  Phys. J.} {\bf C16} (2000) 139--153,
  [\href{http://xxx.lanl.gov/abs/hep-ph/0003022}{{\tt hep-ph/0003022}}].

\bibitem{Heinemeyer:1998yj}
S.~Heinemeyer, W.~Hollik, and G.~Weiglein, {\it {FeynHiggs: A Program for the
  Calculation of the Masses of the Neutral CP-even Higgs Bosons in the MSSM}},
  {\em Comput. Phys. Commun.} {\bf 124} (2000) 76--89,
  [\href{http://xxx.lanl.gov/abs/hep-ph/9812320}{{\tt hep-ph/9812320}}].

\bibitem{Hahn:2005cu}
T.~Hahn, W.~Hollik, S.~Heinemeyer, and G.~Weiglein, {\it {Precision Higgs
  Masses with FeynHiggs 2.2}},
  \href{http://xxx.lanl.gov/abs/hep-ph/0507009}{{\tt hep-ph/0507009}}.

\bibitem{Hahn:2006np}
T.~Hahn {\em et~al.}, {\it {Higher-order Corrected Higgs Bosons in FeynHiggs
  2.5}},  {\em Pramana} {\bf 69} (2007) 861--870,
  [\href{http://xxx.lanl.gov/abs/hep-ph/0611373}{{\tt hep-ph/0611373}}].

\bibitem{Kublbeck:1990xc}
J.~Kublbeck, M.~Bohm, and A.~Denner, {\it {Feyn Arts: Computer Algebraic
  Generation of Feynman Graphs and Amplitudes}},  {\em Comput. Phys. Commun.}
  {\bf 60} (1990) 165--180.

\bibitem{Hahn:2000kx}
T.~Hahn, {\it {Generating Feynman Diagrams and Amplitudes with FeynArts 3}},
  {\em Comput. Phys. Commun.} {\bf 140} (2001) 418--431,
  [\href{http://xxx.lanl.gov/abs/hep-ph/0012260}{{\tt hep-ph/0012260}}].

\bibitem{Kublbeck:1992mt}
J.~Kublbeck, H.~Eck, and R.~Mertig, {\it {Computeralgebraic Generation and
  Calculation of Feynman Graphs Using FeynArts and FeynCalc}},  {\em Nucl.
  Phys. Proc. Suppl.} {\bf 29A} (1992) 204--208.

\bibitem{Hahn:2001rv}
T.~Hahn and C.~Schappacher, {\it {The Implementation of the Minimal
  Supersymmetric Standard Model in FeynArts and FormCalc}},  {\em Comput. Phys.
  Commun.} {\bf 143} (2002) 54--68,
  [\href{http://xxx.lanl.gov/abs/hep-ph/0105349}{{\tt hep-ph/0105349}}].

\bibitem{Hahn:2006qw}
T.~Hahn and M.~Rauch, {\it {News from FormCalc and LoopTools}},  {\em Nucl.
  Phys. Proc. Suppl.} {\bf 157} (2006) 236--240,
  [\href{http://xxx.lanl.gov/abs/hep-ph/0601248}{{\tt hep-ph/0601248}}].

\bibitem{Hahn:2007px}
T.~Hahn and J.~I. Illana, {\it {Extensions in FormCalc 5.3}},
  \href{http://xxx.lanl.gov/abs/0708.3652}{{\tt 0708.3652}}.

\bibitem{Hahn:1999wr}
T.~Hahn, {\it {Generating and Calculating One-loop Feynman Diagrams with
  FeynArts, FormCalc, and LoopTools}},
  \href{http://xxx.lanl.gov/abs/hep-ph/9905354}{{\tt hep-ph/9905354}}.

\bibitem{Denner:1999gp}
A.~Denner, S.~Dittmaier, M.~Roth, and D.~Wackeroth, {\it {Predictions for All
  Processes e+ e- $\to$ $4$ fermions + gamma}},  {\em Nucl. Phys.} {\bf B560}
  (1999) 33--65, [\href{http://xxx.lanl.gov/abs/hep-ph/9904472}{{\tt
  hep-ph/9904472}}].

\bibitem{Denner:2005fg}
A.~Denner, S.~Dittmaier, M.~Roth, and L.~H. Wieders, {\it {Electroweak
  Corrections to Charged-Current e+ e- $\to$ $4$ fermion Processes: Technical
  Details and Further Results}},  {\em Nucl. Phys.} {\bf B724} (2005) 247--294,
  [\href{http://xxx.lanl.gov/abs/hep-ph/0505042}{{\tt hep-ph/0505042}}].

\bibitem{Stuart:1991xk}
R.~G. Stuart, {\it {Gauge Invariance, Analyticity and Physical Observables at
  the Z0 Resonance}},  {\em Phys. Lett.} {\bf B262} (1991) 113--119.

\bibitem{Aeppli:1993cb}
A.~Aeppli, F.~Cuypers, and G.~J. van Oldenborgh, {\it {O(Gamma) Corrections to
  W Pair Production in e+ e- and Gamma Gamma Collisions}},  {\em Phys. Lett.}
  {\bf B314} (1993) 413--420,
  [\href{http://xxx.lanl.gov/abs/hep-ph/9303236}{{\tt hep-ph/9303236}}].

\bibitem{Aeppli:1993rs}
A.~Aeppli, G.~J. van Oldenborgh, and D.~Wyler, {\it {Unstable Particles in One
  Loop Calculations}},  {\em Nucl. Phys.} {\bf B428} (1994) 126--146,
  [\href{http://xxx.lanl.gov/abs/hep-ph/9312212}{{\tt hep-ph/9312212}}].

\bibitem{Baur:1991pp}
U.~Baur, J.~A.~M. Vermaseren, and D.~Zeppenfeld, {\it {Electroweak Vector Boson
  Production in High-energy e p Collisions}},  {\em Nucl. Phys.} {\bf B375}
  (1992) 3--44.

\bibitem{Kurihara:1994fz}
Y.~Kurihara, D.~Perret-Gallix, and Y.~Shimizu, {\it {e+ e- $\to$ e-
  anti-electron-neutrino u anti-d from LEP to Linear Collider Energies}},  {\em
  Phys. Lett.} {\bf B349} (1995) 367--374,
  [\href{http://xxx.lanl.gov/abs/hep-ph/9412215}{{\tt hep-ph/9412215}}].

\bibitem{Denner:2003iy}
A.~Denner, S.~Dittmaier, M.~Roth, and M.~M. Weber, {\it {Electroweak radiative
  corrections to e+ e- $\to$ nu anti-nu H}},  {\em Nucl. Phys.} {\bf B660}
  (2003) 289--321, [\href{http://xxx.lanl.gov/abs/hep-ph/0302198}{{\tt
  hep-ph/0302198}}].

\bibitem{Hahn:2002gm}
T.~Hahn, S.~Heinemeyer, and G.~Weiglein, {\it {MSSM Higgs-boson Production at
  the Linear Collider: Dominant Corrections to the W W Fusion Channel}},  {\em
  Nucl. Phys.} {\bf B652} (2003) 229--258,
  [\href{http://xxx.lanl.gov/abs/hep-ph/0211204}{{\tt hep-ph/0211204}}].

\bibitem{Denner:1997ia}
A.~Denner, S.~Dittmaier, and M.~Roth, {\it {Non-factorizable Photonic
  Corrections to e+ e- $\to$ W W $\to$ $4$ fermions}},  {\em Nucl. Phys.} {\bf
  B519} (1998) 39--84, [\href{http://xxx.lanl.gov/abs/hep-ph/9710521}{{\tt
  hep-ph/9710521}}].

\bibitem{Denner:2002ii}
A.~Denner and S.~Dittmaier, {\it {Reduction of One-loop Tensor 5-point
  Integrals}},  {\em Nucl. Phys.} {\bf B658} (2003) 175--202,
  [\href{http://xxx.lanl.gov/abs/hep-ph/0212259}{{\tt hep-ph/0212259}}].

\bibitem{Dittmaier:2001ay}
S.~Dittmaier and M.~Kramer, {\it {Electroweak Radiative Corrections to W-boson
  Production at Hadron Colliders}},  {\em Phys. Rev.} {\bf D65} (2002) 073007,
  [\href{http://xxx.lanl.gov/abs/hep-ph/0109062}{{\tt hep-ph/0109062}}].

\bibitem{Kinoshita:1962ur}
T.~Kinoshita, {\it {Mass Singularities of Feynman Amplitudes}},  {\em J. Math.
  Phys.} {\bf 3} (1962) 650--677.

\bibitem{Lee:1964is}
T.~D. Lee and M.~Nauenberg, {\it {Degenerate Systems and Mass Singularities}},
  {\em Phys. Rev.} {\bf 133} (1964) B1549--B1562.

\bibitem{Catani:1996jh}
S.~Catani and M.~H. Seymour, {\it {The Dipole Formalism for the Calculation of
  QCD Jet Cross Sections at Next-to-Leading Order}},  {\em Phys. Lett.} {\bf
  B378} (1996) 287--301, [\href{http://xxx.lanl.gov/abs/hep-ph/9602277}{{\tt
  hep-ph/9602277}}].

\bibitem{Catani:1996vz}
S.~Catani and M.~H. Seymour, {\it {A General Algorithm for Calculating Jet
  Cross Sections in NLO QCD}},  {\em Nucl. Phys.} {\bf B485} (1997) 291--419,
  [\href{http://xxx.lanl.gov/abs/hep-ph/9605323}{{\tt hep-ph/9605323}}].

\bibitem{Dittmaier:1999mb}
S.~Dittmaier, {\it {A General Approach to Photon Radiation off Fermions}},
  {\em Nucl. Phys.} {\bf B565} (2000) 69--122,
  [\href{http://xxx.lanl.gov/abs/hep-ph/9904440}{{\tt hep-ph/9904440}}].

\bibitem{Roth:1999kk}
M.~Roth, {\it {Precise Predictions for Four-fermion Production in Electron
  Positron Annihilation}},  \href{http://xxx.lanl.gov/abs/hep-ph/0008033}{{\tt
  hep-ph/0008033}}.

\bibitem{Denner:2000bj}
A.~Denner, S.~Dittmaier, M.~Roth, and D.~Wackeroth, {\it {Electroweak Radiative
  Corrections to e+ e- $\to$ W W $\to$ $4$ fermions in Double-pole
  Approximation: The RACOONWW Approach}},  {\em Nucl. Phys.} {\bf B587} (2000)
  67--117, [\href{http://xxx.lanl.gov/abs/hep-ph/0006307}{{\tt
  hep-ph/0006307}}].

\bibitem{Yennie:1961ad}
D.~R. Yennie, S.~C. Frautschi, and H.~Suura, {\it {The Infrared Divergence
  Phenomena and High-energy Processes}},  {\em Ann. Phys.} {\bf 13} (1961)
  379--452.

\bibitem{Beenakker:1996kt}
W.~Beenakker {\em et~al.}, {\it {WW Cross-sections and Distributions}},
  \href{http://xxx.lanl.gov/abs/hep-ph/9602351}{{\tt hep-ph/9602351}}.

\bibitem{Kuraev:1985hb}
E.~A. Kuraev and V.~S. Fadin, {\it {On Radiative Corrections to e+ e- Single
  Photon Annihilation at High-Energy}},  {\em Sov. J. Nucl. Phys.} {\bf 41}
  (1985) 466--472.

\bibitem{Altarelli:1986kq}
G.~Altarelli and G.~Martinelli, {\it {Radiative Corrections to the Z0 Line
  Shape at LEP}}, . In *Ellis, J. ( Ed.), Peccei, R.d. ( Ed.): Physics At LEP,
  Vol. 1*, 47-57.

\bibitem{Nicrosini:1986sm}
O.~Nicrosini and L.~Trentadue, {\it {Soft Photons and Second Order Radiative
  Corrections to e+ e- $\to$ Z0}},  {\em Phys. Lett.} {\bf B196} (1987) 551.

\bibitem{Nicrosini:1987sw}
O.~Nicrosini and L.~Trentadue, {\it {Second Order Electromagnetic Radiative
  Corrections to e+ e- $\to$ gamma*, Z0 $\to$ mu+ mu-}},  {\em Z. Phys.} {\bf
  C39} (1988) 479.

\bibitem{Berends:1987ab}
F.~A. Berends, W.~L. van Neerven, and G.~J.~H. Burgers, {\it {Higher Order
  Radiative Corrections at LEP Energies}},  {\em Nucl. Phys.} {\bf B297} (1988)
  429.

\bibitem{Yao:2006px}
{\bf Particle Data Group} Collaboration, W.~M. Yao {\em et~al.}, {\it {Review
  of Particle Physics}},  {\em J. Phys.} {\bf G33} (2006) 1--1232.

\bibitem{Arguin:2005cc}
{\bf CDF} Collaboration, J.~F. Arguin {\em et~al.}, {\it {Combination of CDF
  and D0 Results on the Top-quark Mass}},
  \href{http://xxx.lanl.gov/abs/hep-ex/0507091}{{\tt hep-ex/0507091}}.

\bibitem{Brein:2007da}
O.~Brein and W.~Hollik, {\it {Distributions for MSSM Higgs Boson + Jet
  Production at Hadron Colliders}},  {\em Phys. Rev.} {\bf D76} (2007) 035002,
  [\href{http://xxx.lanl.gov/abs/0705.2744}{{\tt 0705.2744}}].

\bibitem{Barate:2003sz}
{\bf LEP Working Group for Higgs Boson Searches} Collaboration, R.~Barate {\em
  et~al.}, {\it {Search for the Standard Model Higgs Boson at LEP}},  {\em
  Phys. Lett.} {\bf B565} (2003) 61--75,
  [\href{http://xxx.lanl.gov/abs/hep-ex/0306033}{{\tt hep-ex/0306033}}].

\bibitem{Schael:2006cr}
{\bf ALEPH} Collaboration, S.~Schael {\em et~al.}, {\it {Search for Neutral
  MSSM Higgs Bosons at LEP}},  {\em Eur. Phys. J.} {\bf C47} (2006) 547--587,
  [\href{http://xxx.lanl.gov/abs/hep-ex/0602042}{{\tt hep-ex/0602042}}].

\bibitem{Beenakker:1988jr}
W.~Beenakker and A.~Denner, {\it {Infrared Divergent Scalar Box Integrals with
  Applications in the Electroweak Standard Model}},  {\em Nucl. Phys.} {\bf
  B338} (1990) 349--370.

\bibitem{Denner:2005nn}
A.~Denner and S.~Dittmaier, {\it {Reduction Schemes for One-loop Tensor
  Integrals}},  {\em Nucl. Phys.} {\bf B734} (2006) 62--115,
  [\href{http://xxx.lanl.gov/abs/hep-ph/0509141}{{\tt hep-ph/0509141}}].

\bibitem{Bredenstein:2005zk}
A.~Bredenstein, S.~Dittmaier, and M.~Roth, {\it {Four-fermion production at
  gamma gamma colliders. II: Radiative corrections in double-pole
  approximation}},  {\em Eur. Phys. J.} {\bf C44} (2005) 27--49,
  [\href{http://xxx.lanl.gov/abs/hep-ph/0506005}{{\tt hep-ph/0506005}}].

\end{thebibliography}\endgroup

\begin{figure}[htbp]
\begin{center}
$m_h^{\mbox{\small{max}}}\;\mbox{scenario}$ 
\end{center}
\vskip -0.5cm
\includegraphics[width=0.42\textwidth]{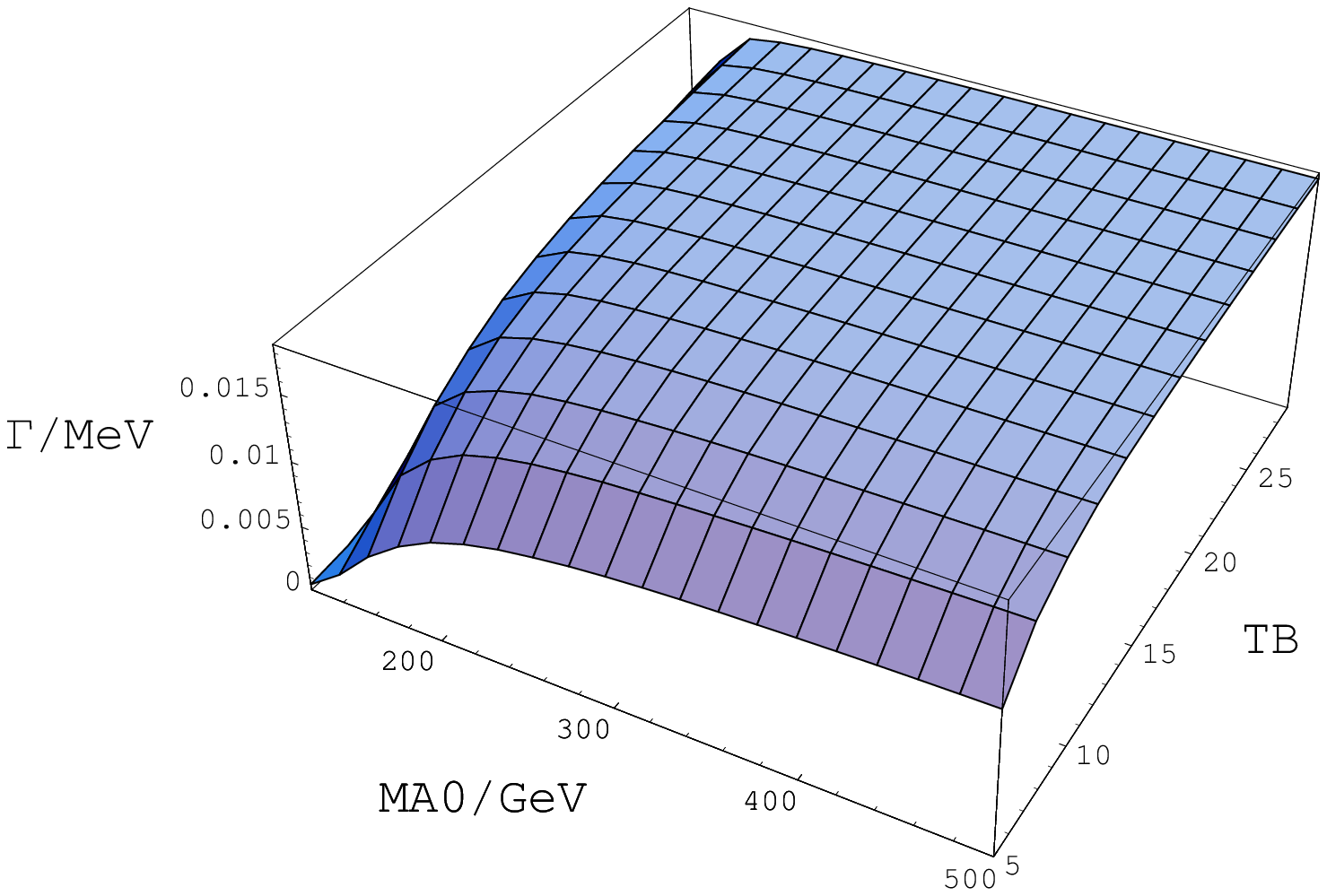}
\hspace{2em}
\includegraphics[width=0.42\textwidth]{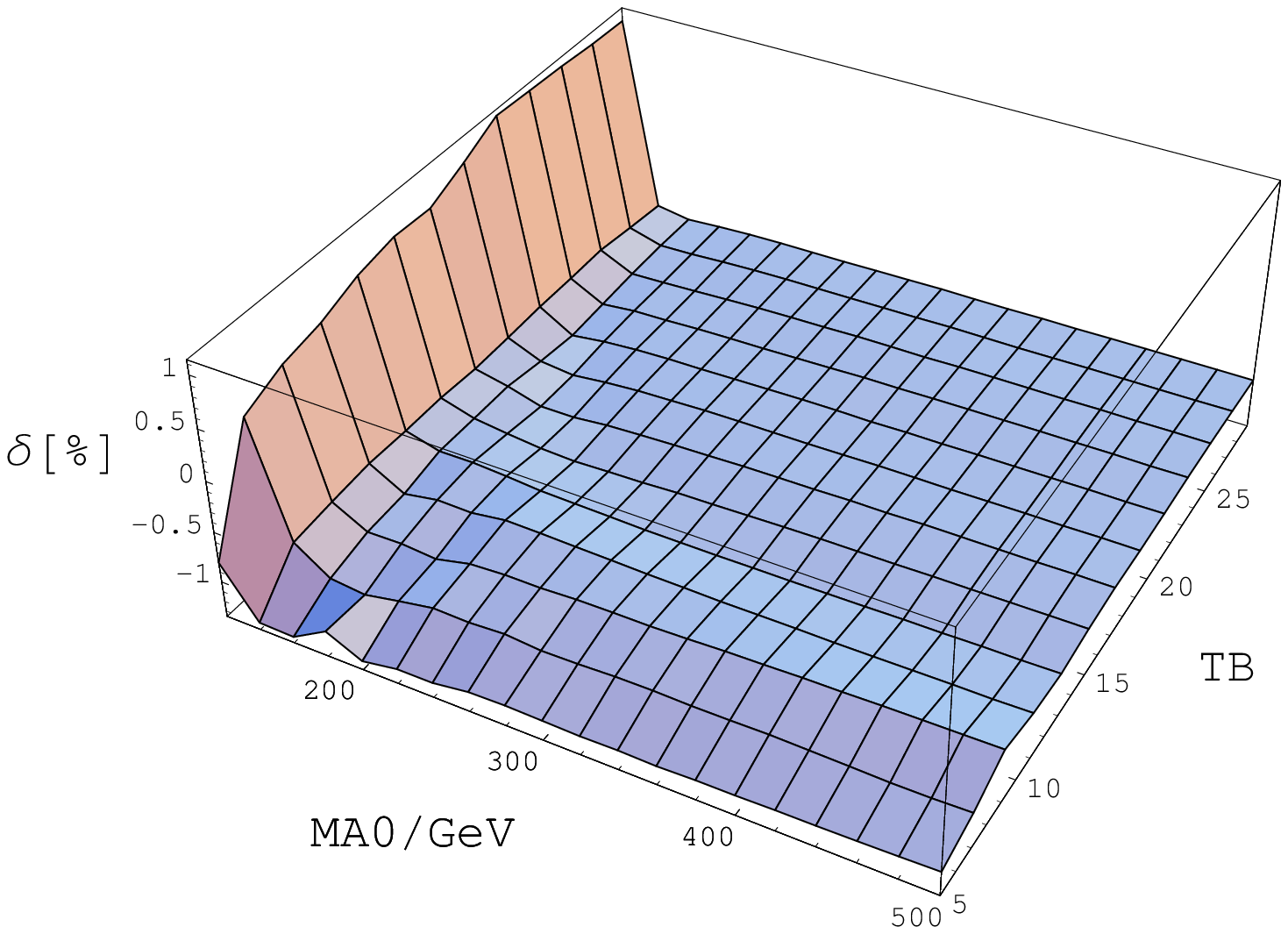}

\begin{center}
no-mixing scenario
\end{center}
\vskip -0.5cm
\includegraphics[width=0.42\textwidth]{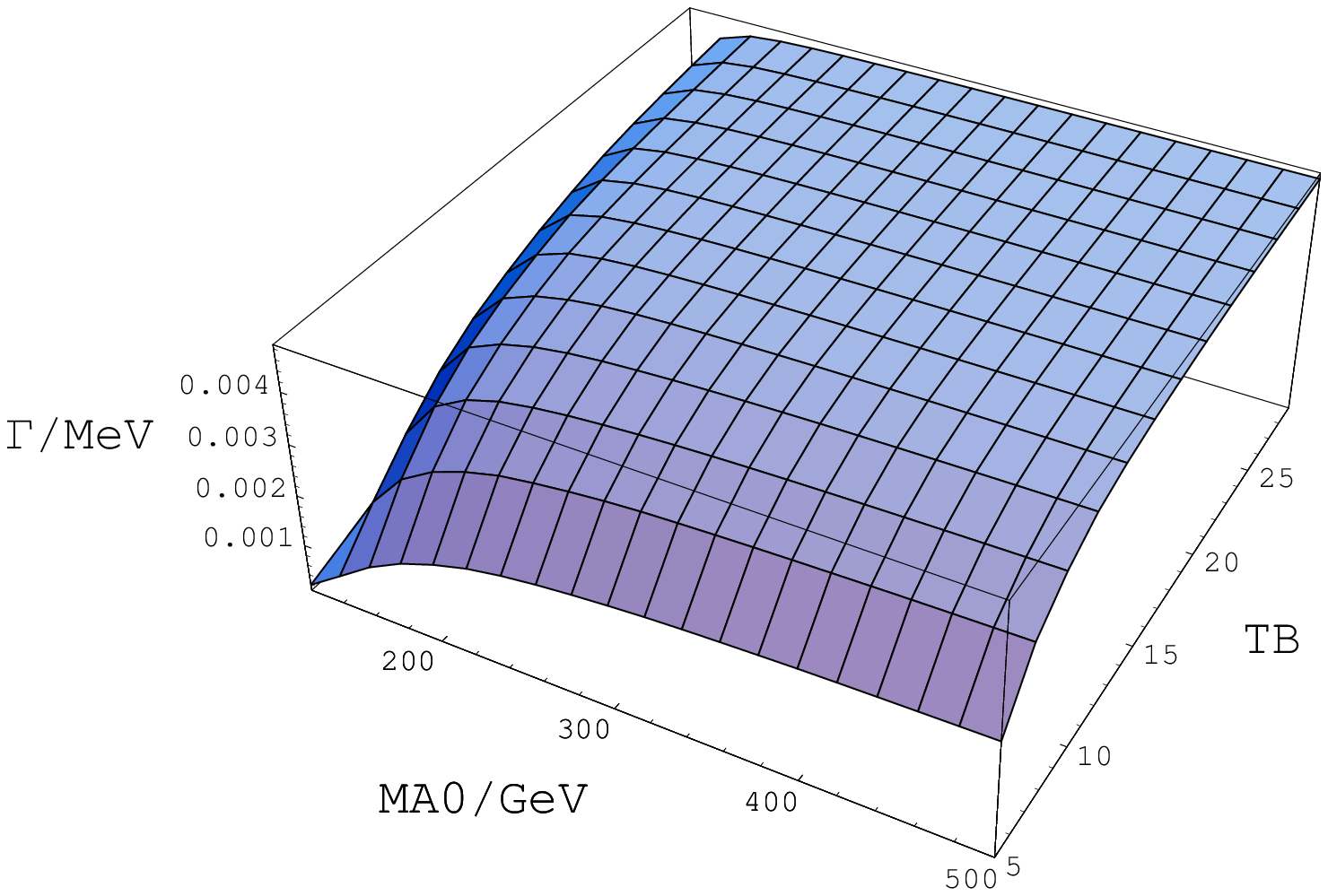}
\hspace{2em}
\includegraphics[width=0.42\textwidth]{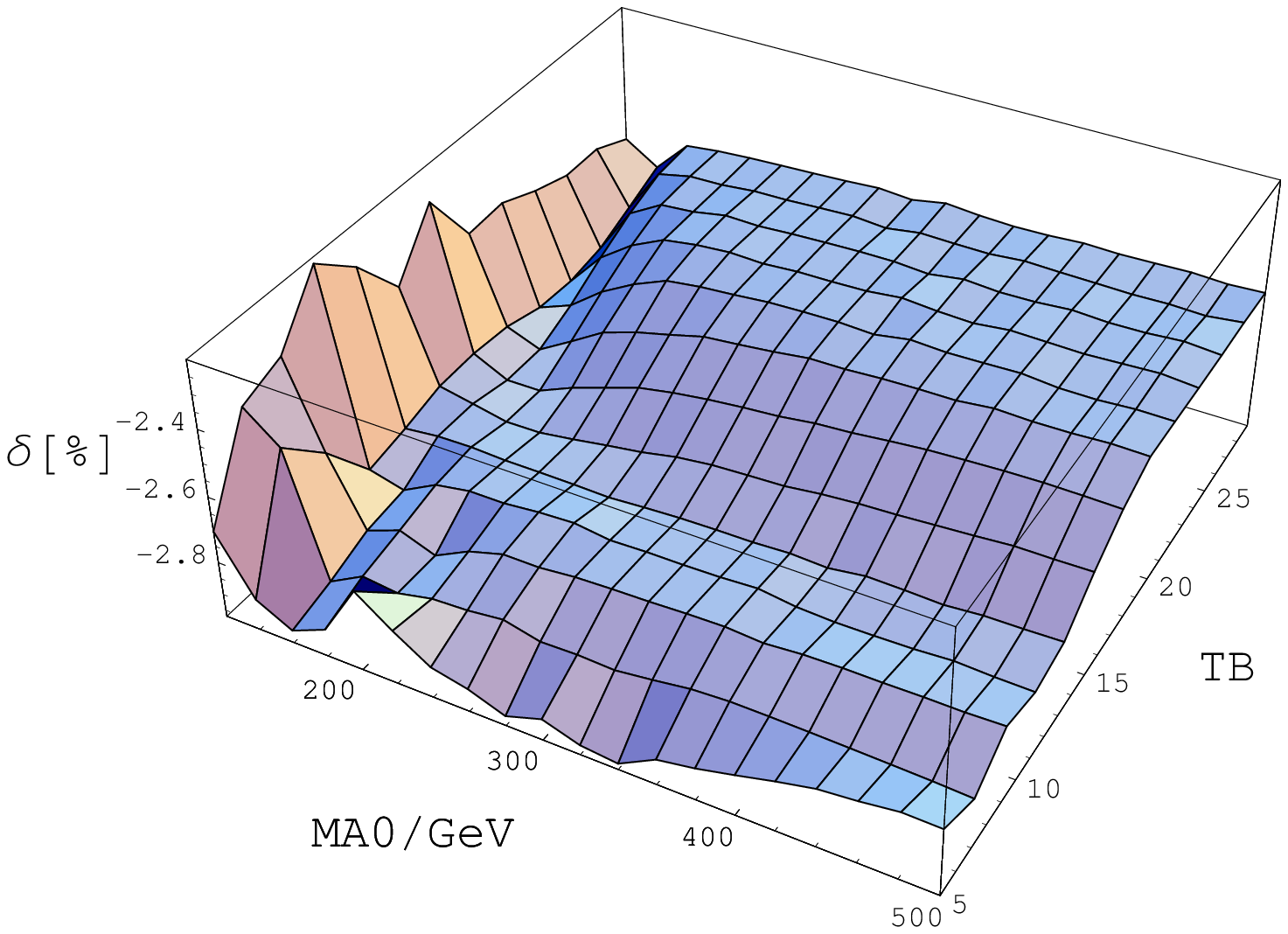}

\begin{center}
small-$\alpha_{\mbox{\small{eff}}}$ scenario
\end{center}
\vskip -0.5cm
\includegraphics[width=0.42\textwidth]{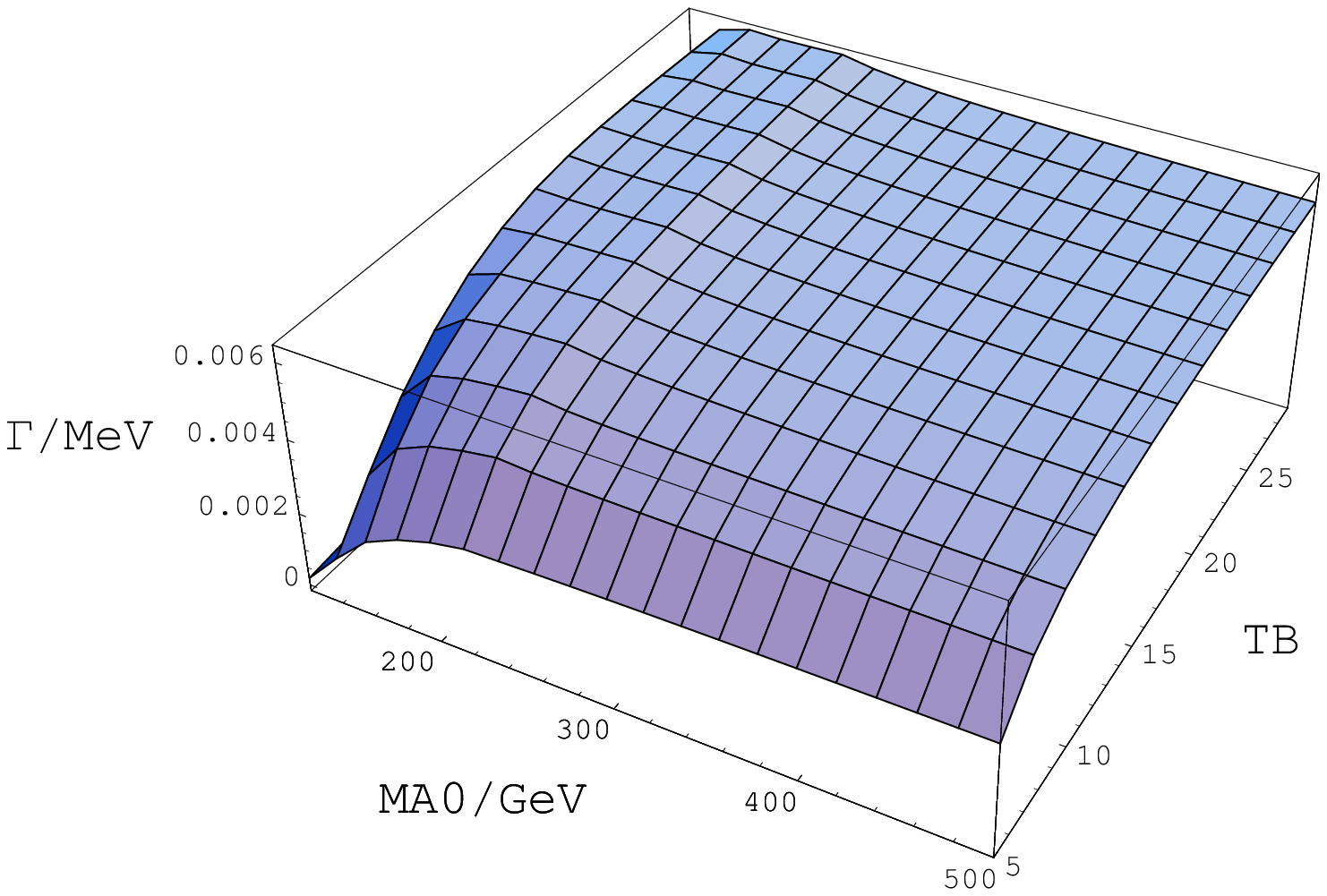}
\hspace{2em}
\includegraphics[width=0.42\textwidth]{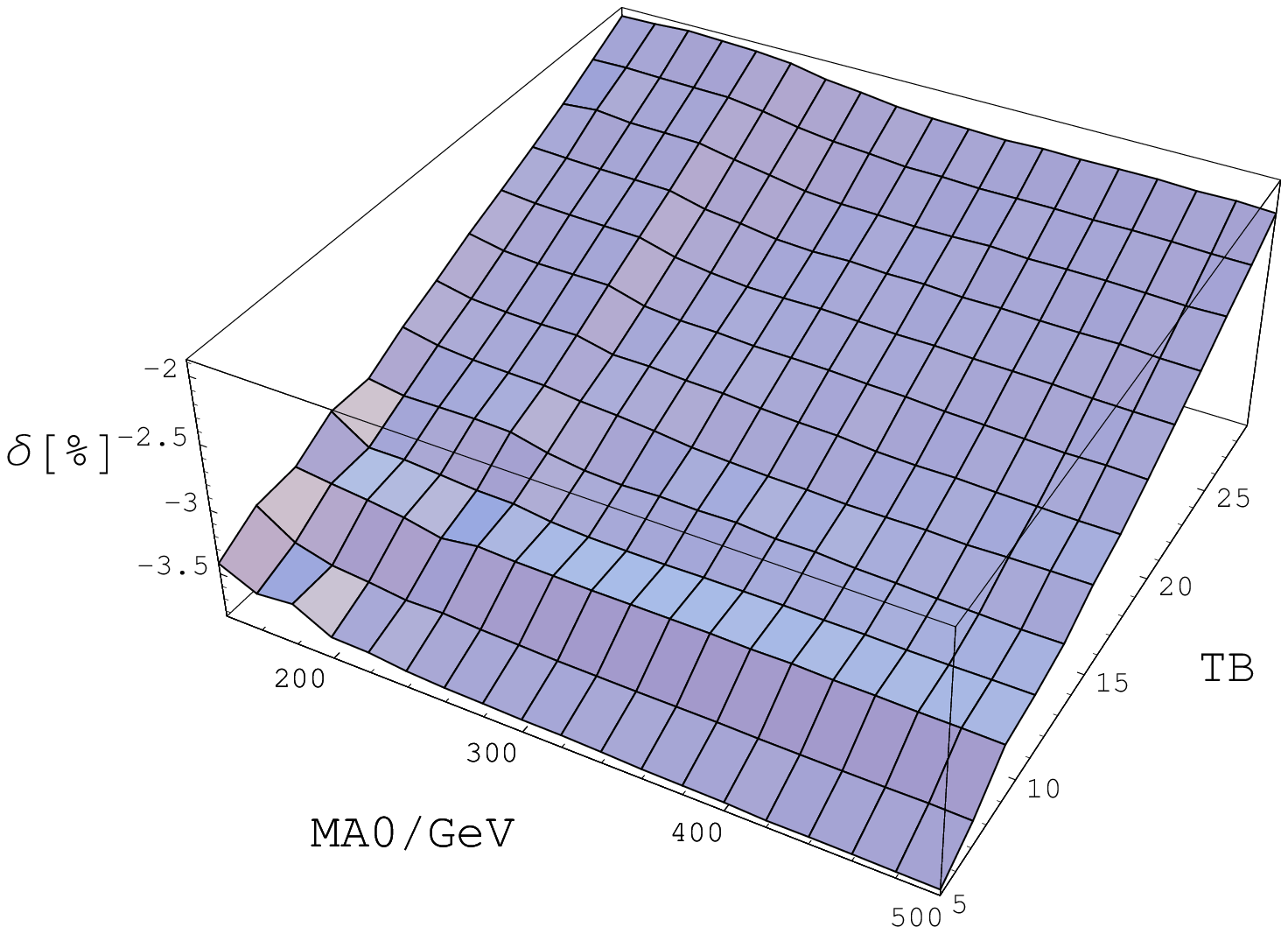}

\caption{Results for the partial decay width of $h^0\ra WW^*\ra e^-\bar\nu_e\mu^+\nu_\mu$ in three different benchmark scenarios (the contribution due to the third generation fermion and sfermion loop corrections to the $H^0WW$ coupling is not included. TB denotes $\tan\beta$). The left column shows the corrected partial decay width, while the right column shows the relative corrections.}
\label{hWWwidth}
\end{figure}

\begin{figure}[htbp]
\begin{center}
$m_h^{\mbox{\small{max}}}\;\mbox{scenario}$ 
\end{center}
\vskip -0.5cm
\includegraphics[width=0.42\textwidth]{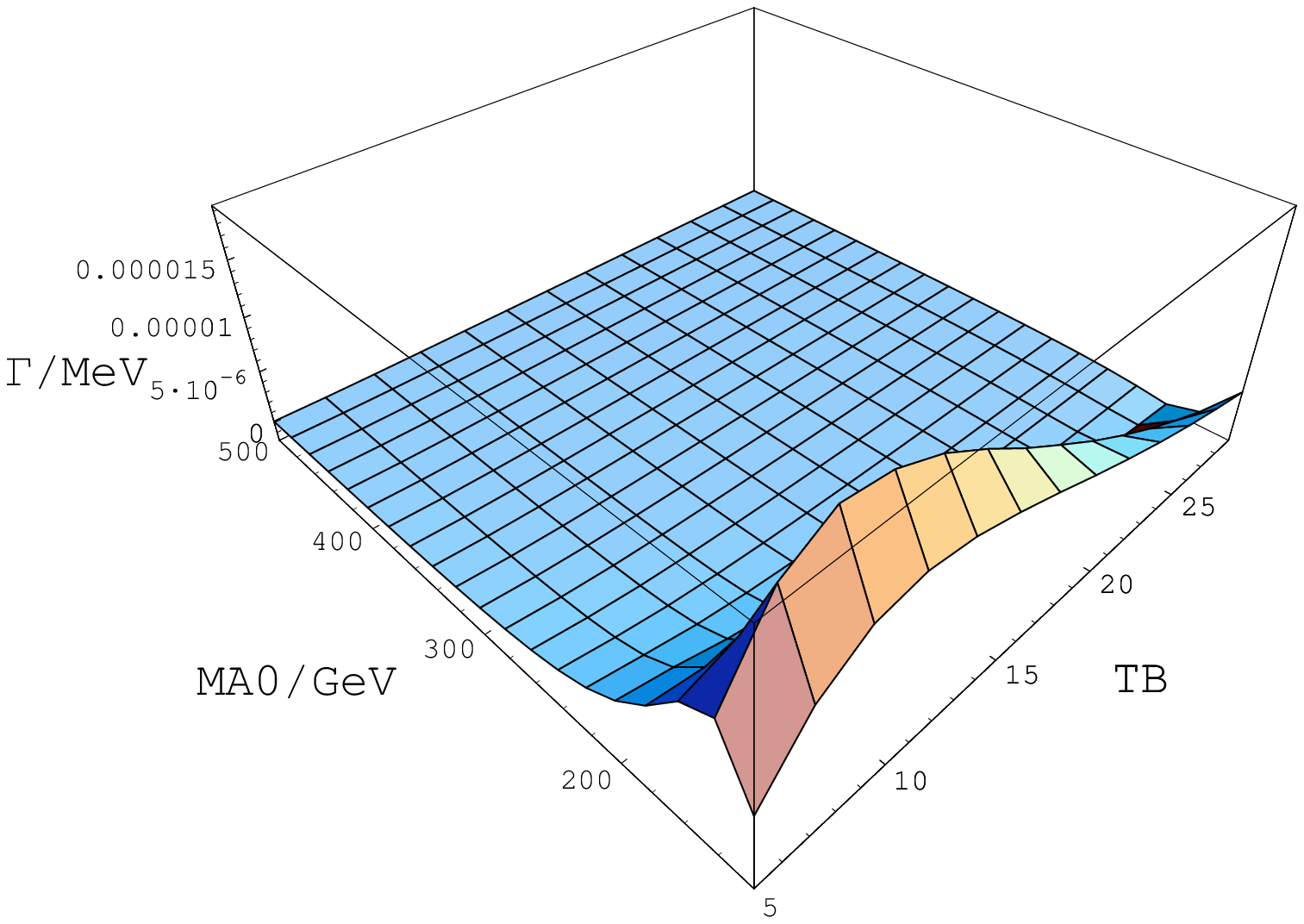}
\hspace{2em}
\includegraphics[width=0.42\textwidth]{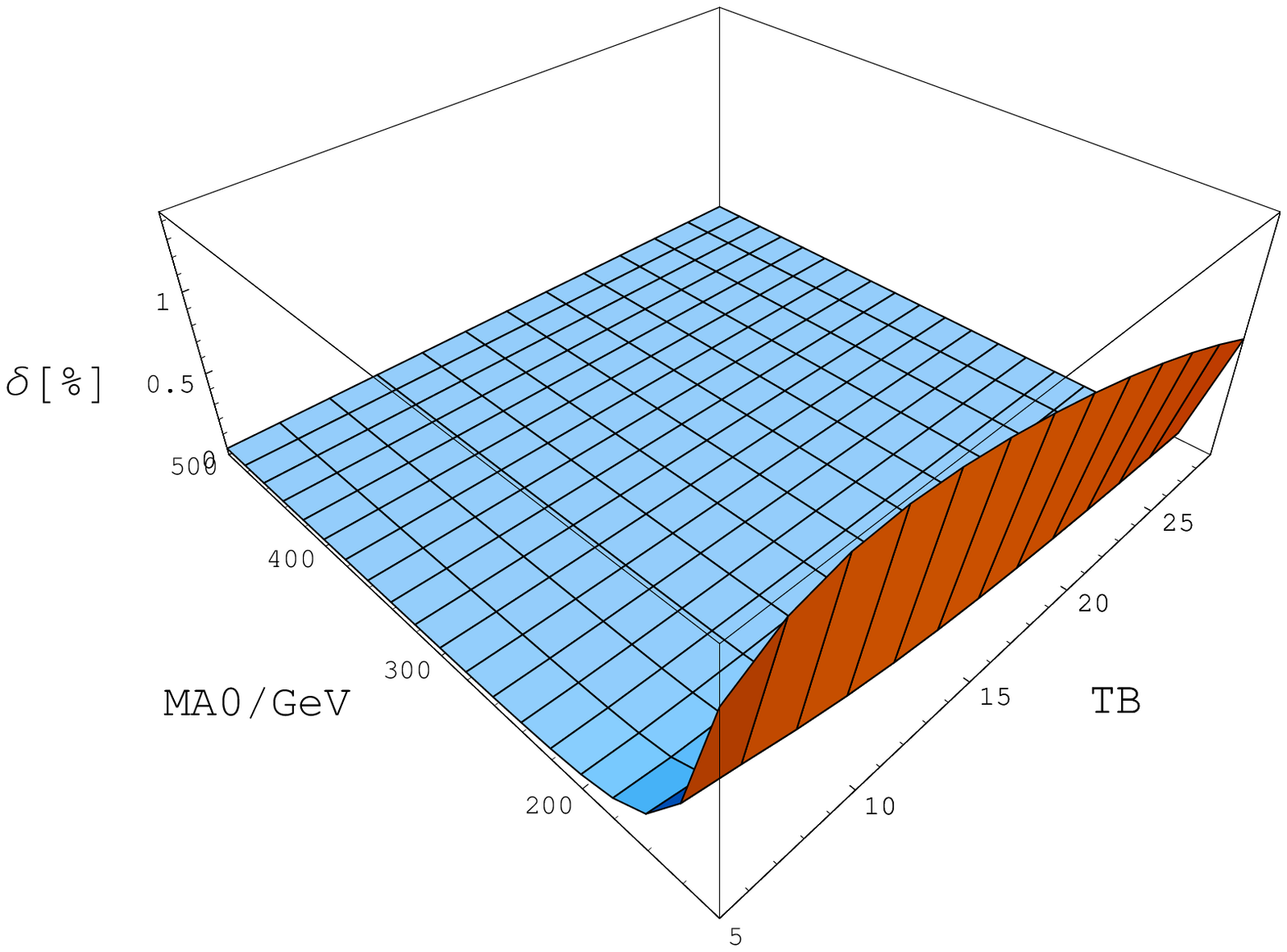}

\begin{center}
no-mixing scenario
\end{center}
\vskip -0.5cm
\includegraphics[width=0.42\textwidth]{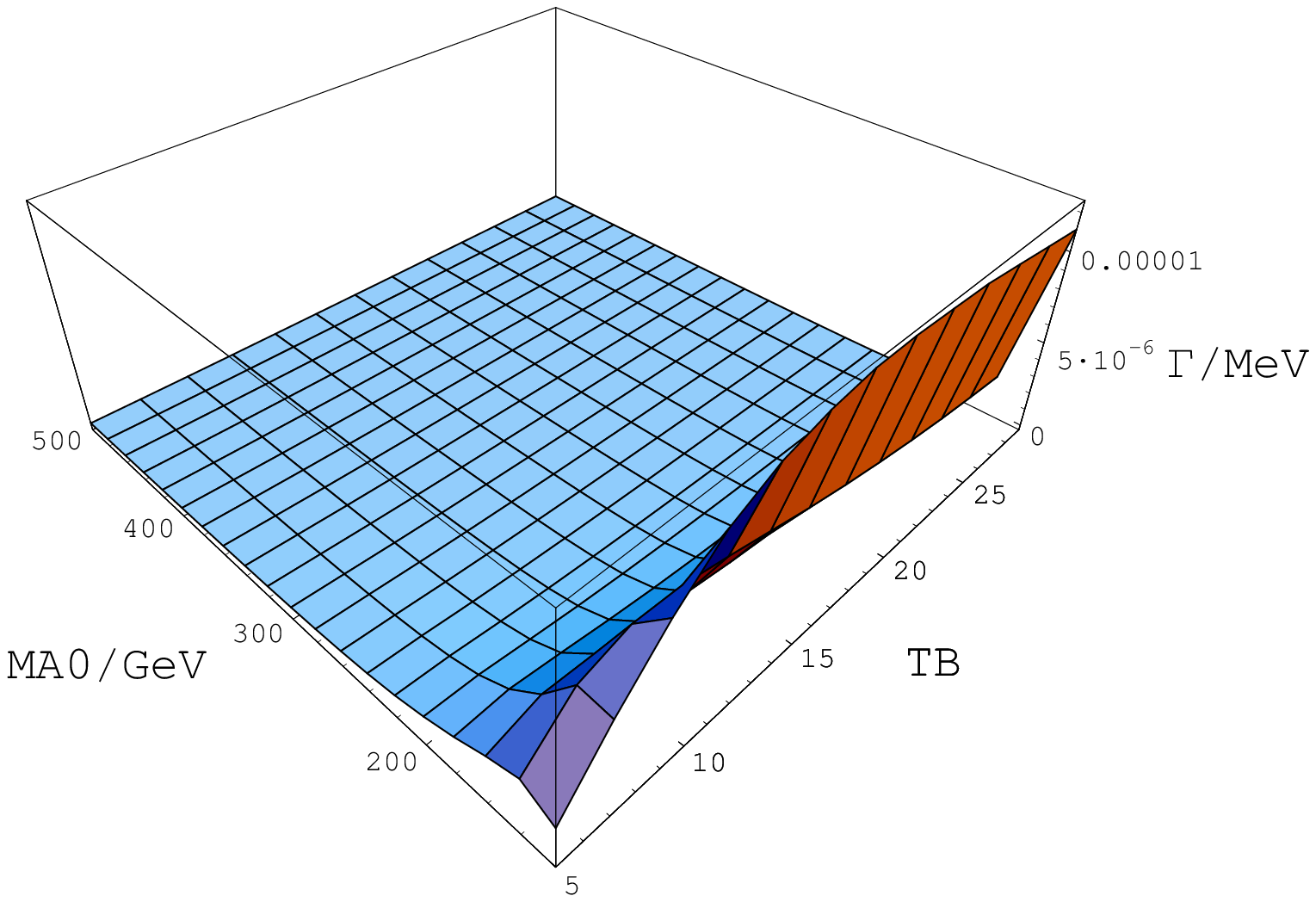}
\hspace{2em}
\includegraphics[width=0.42\textwidth]{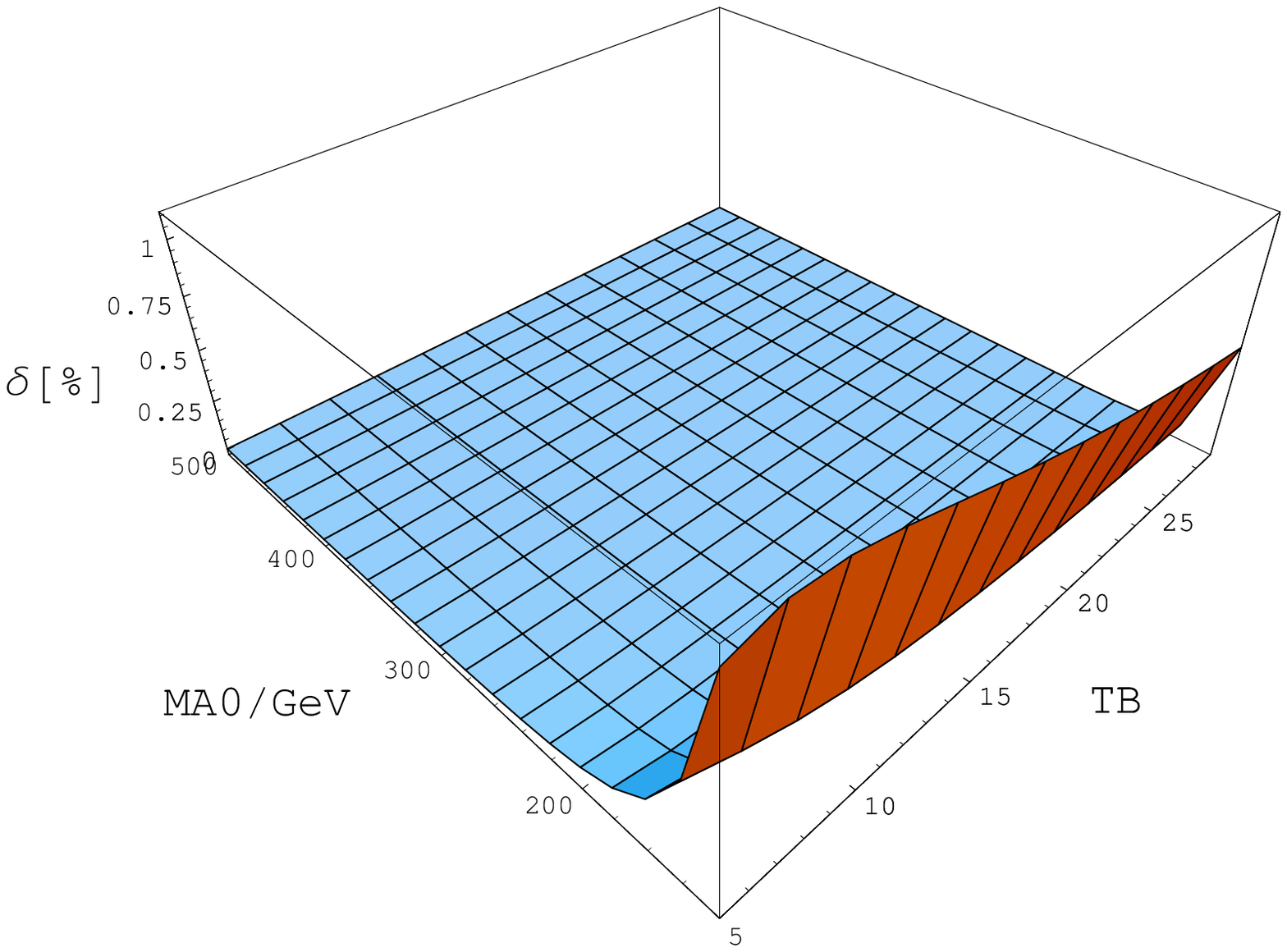}

\begin{center}
small-$\alpha_{\mbox{\small{eff}}}$ scenario
\end{center}
\vskip -0.5cm
\includegraphics[width=0.42\textwidth]{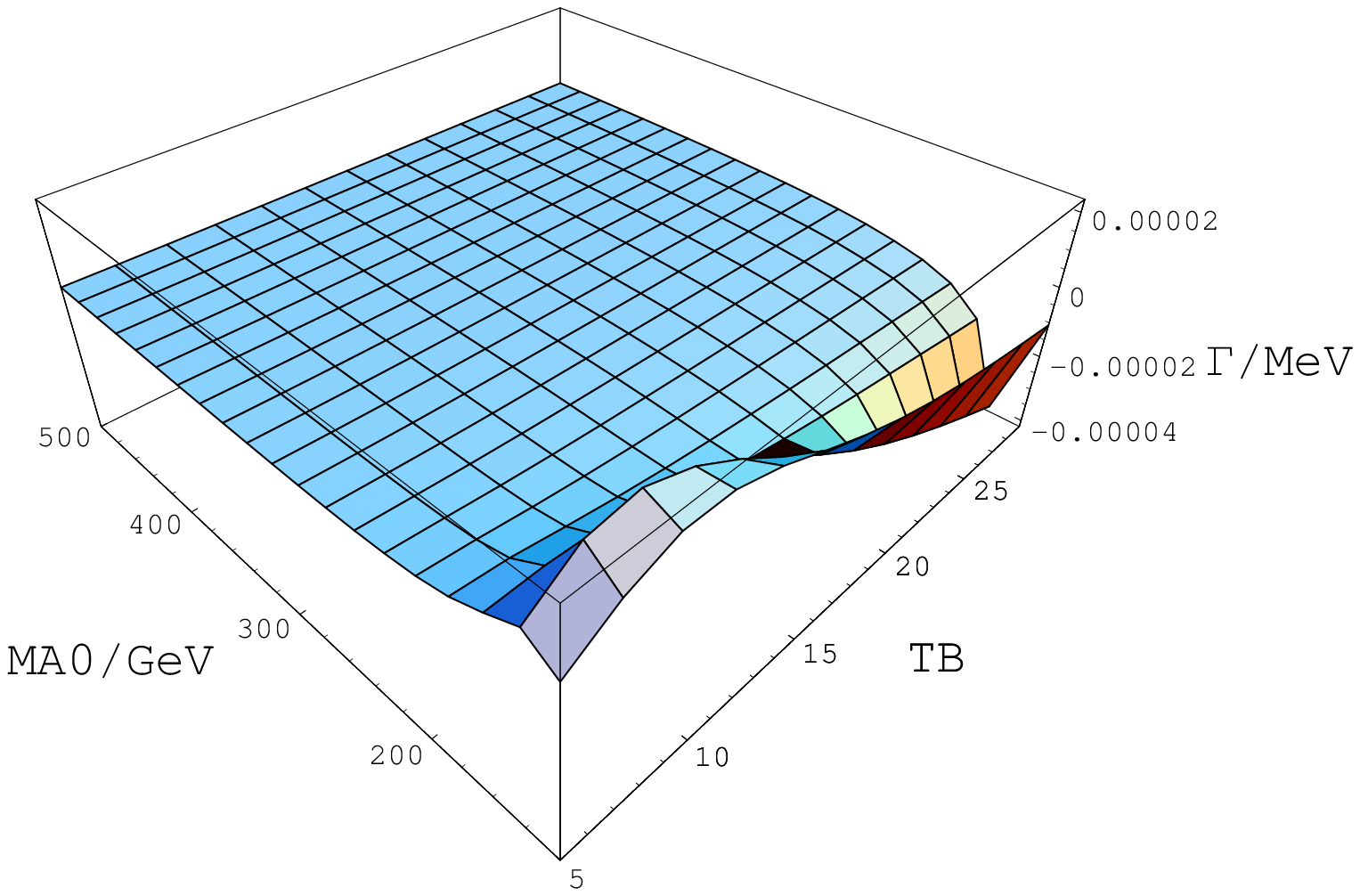}
\hspace{2em}
\includegraphics[width=0.42\textwidth]{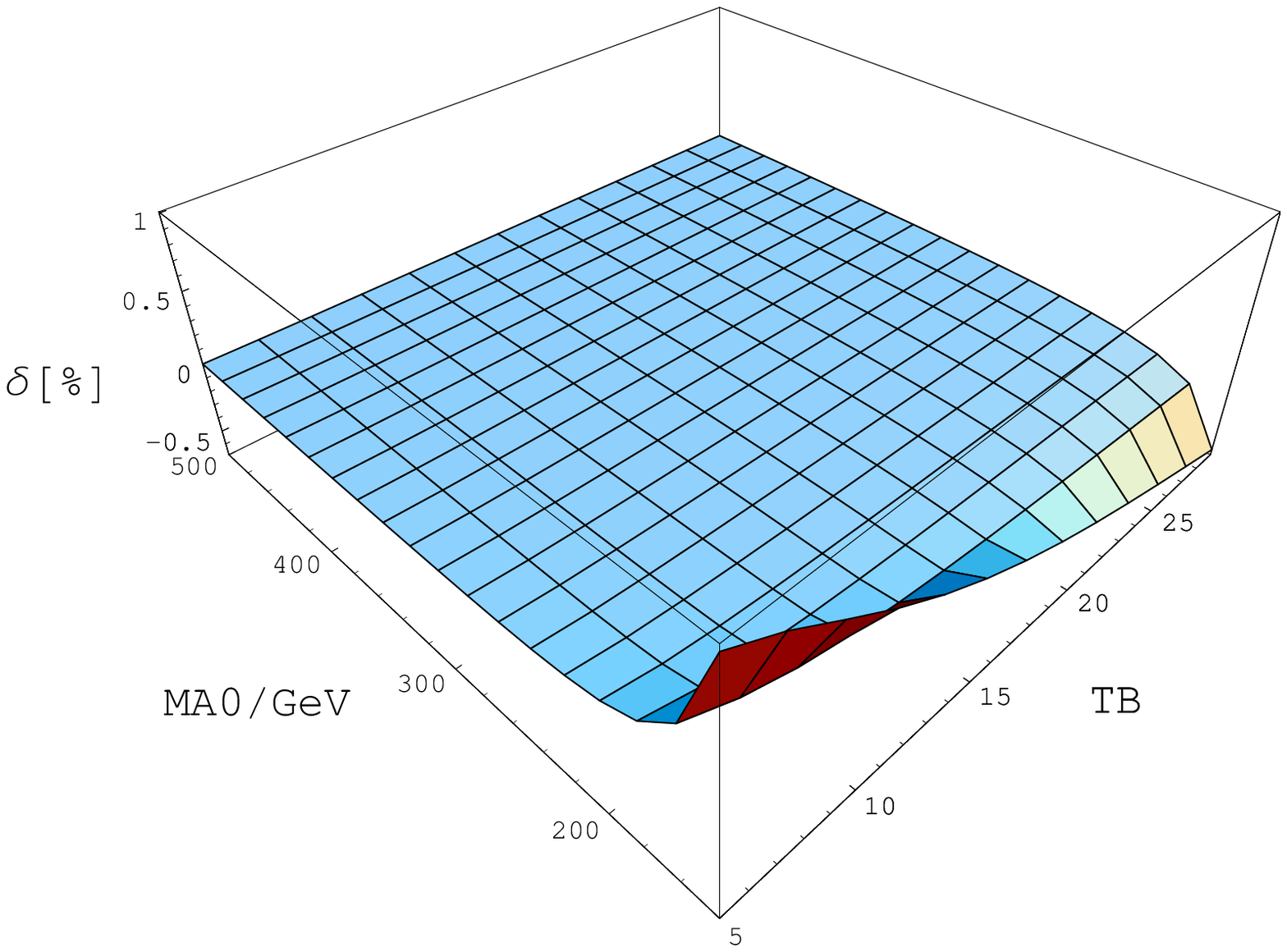}

\caption{Contribution to the partial decay width of $h^0\ra WW^*\ra e^-\bar\nu_e\mu^+\nu_\mu$ due to the correction to the $H^0WW$ coupling from the third generation fermions and sfermions in three different benchmark scenarios (TB denotes $\tan\beta$). The left column shows the corrections to the partial decay width, while the right column shows their relative size.}
\label{HWWwidth}
\end{figure}

\begin{figure}[htbp]
\begin{center}
$m_h^{\mbox{\small{max}}}\;\mbox{scenario}$ 
\end{center}
\vskip -0.5cm
\includegraphics[width=0.42\textwidth]{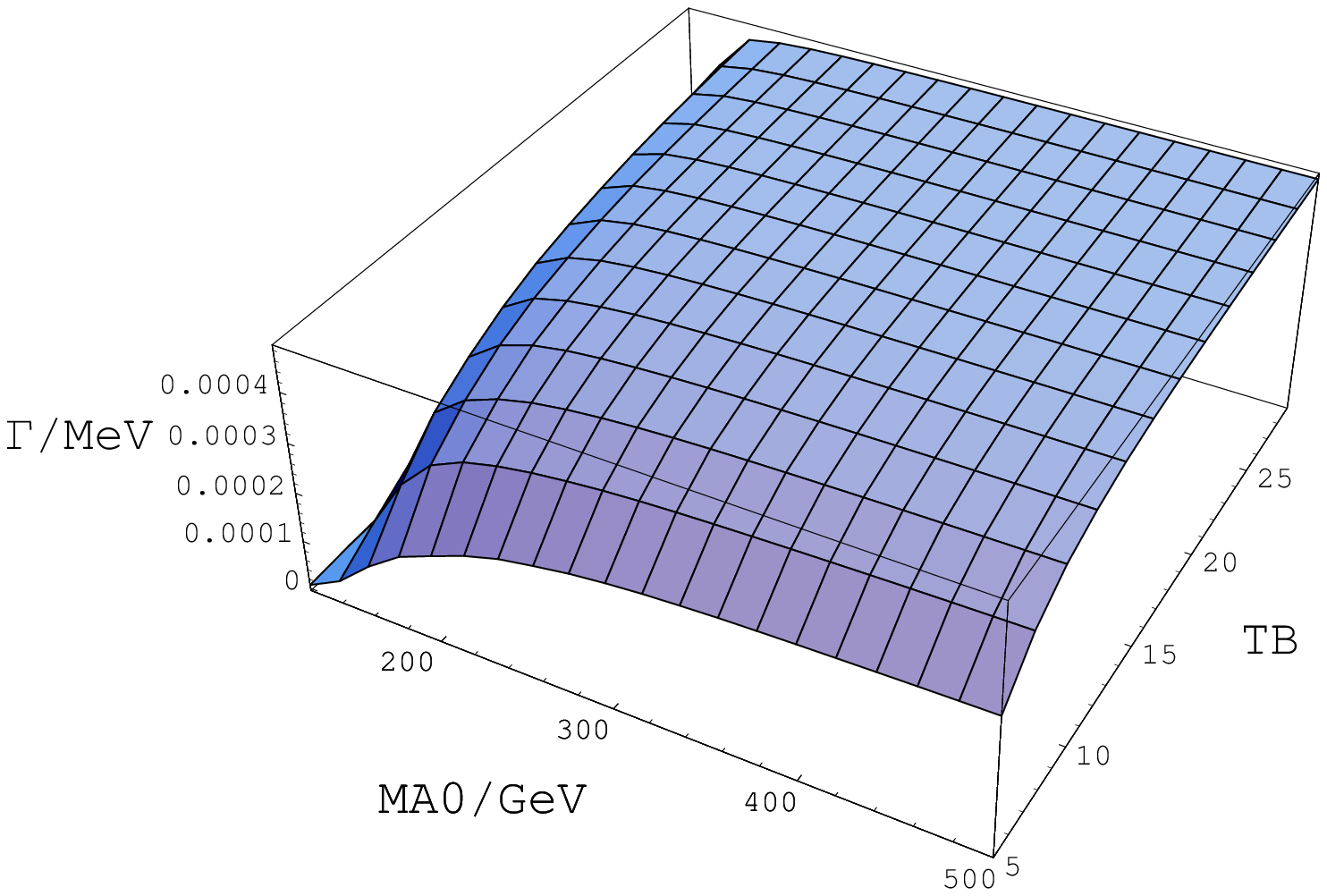}
\hspace{2em}
\includegraphics[width=0.42\textwidth]{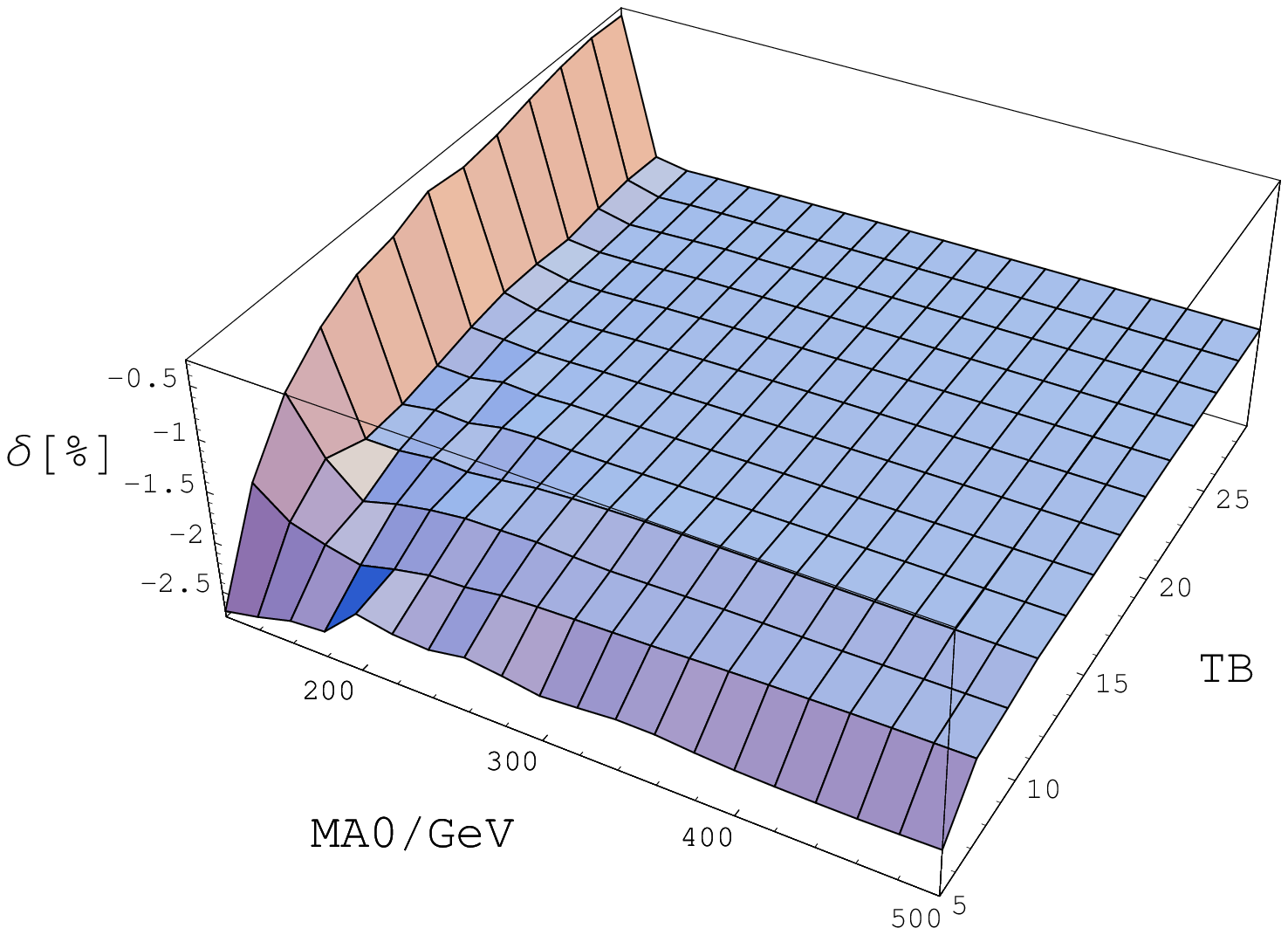}

\begin{center}
no-mixing scenario
\end{center}
\vskip -0.5cm
\includegraphics[width=0.42\textwidth]{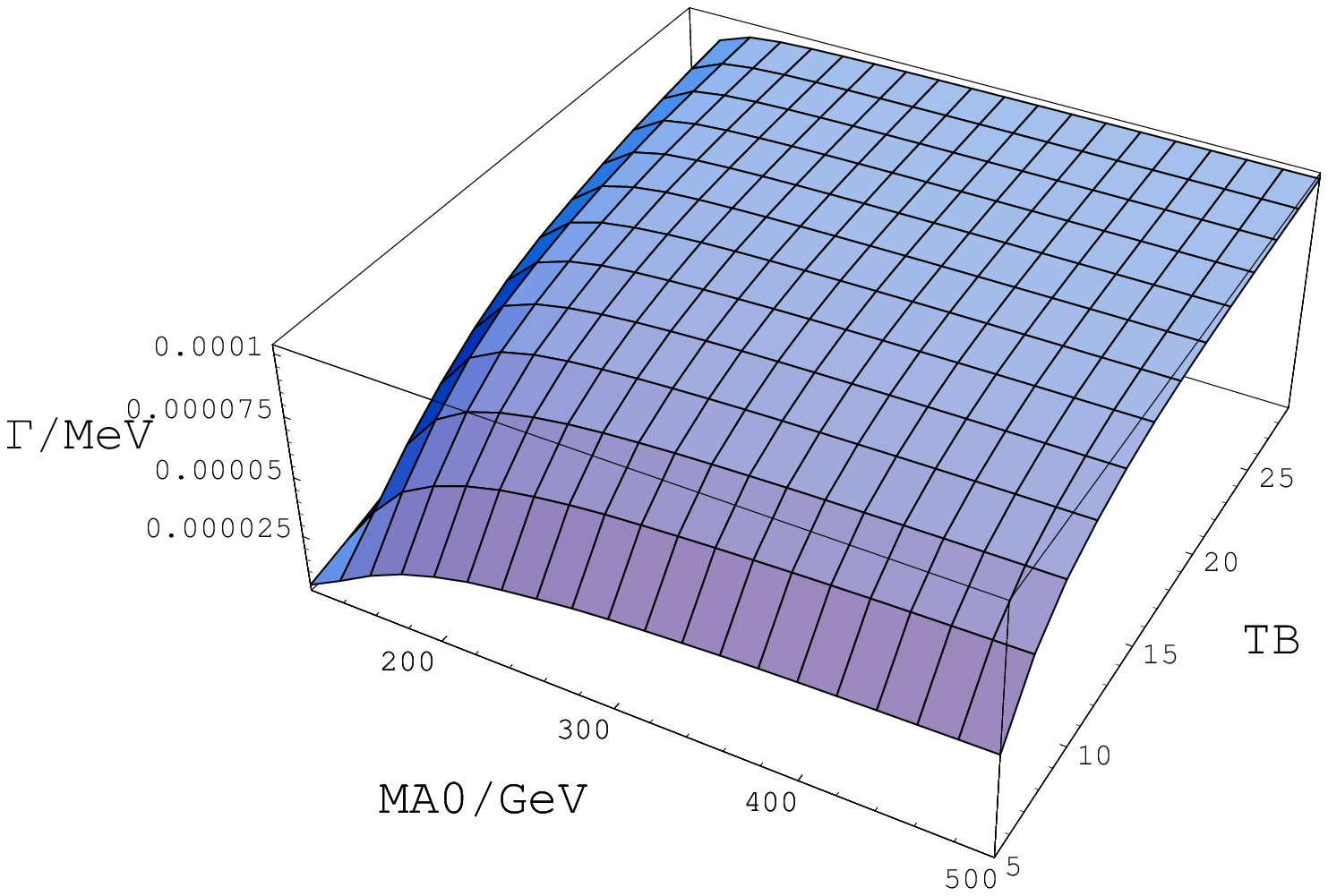}
\hspace{2em}
\includegraphics[width=0.42\textwidth]{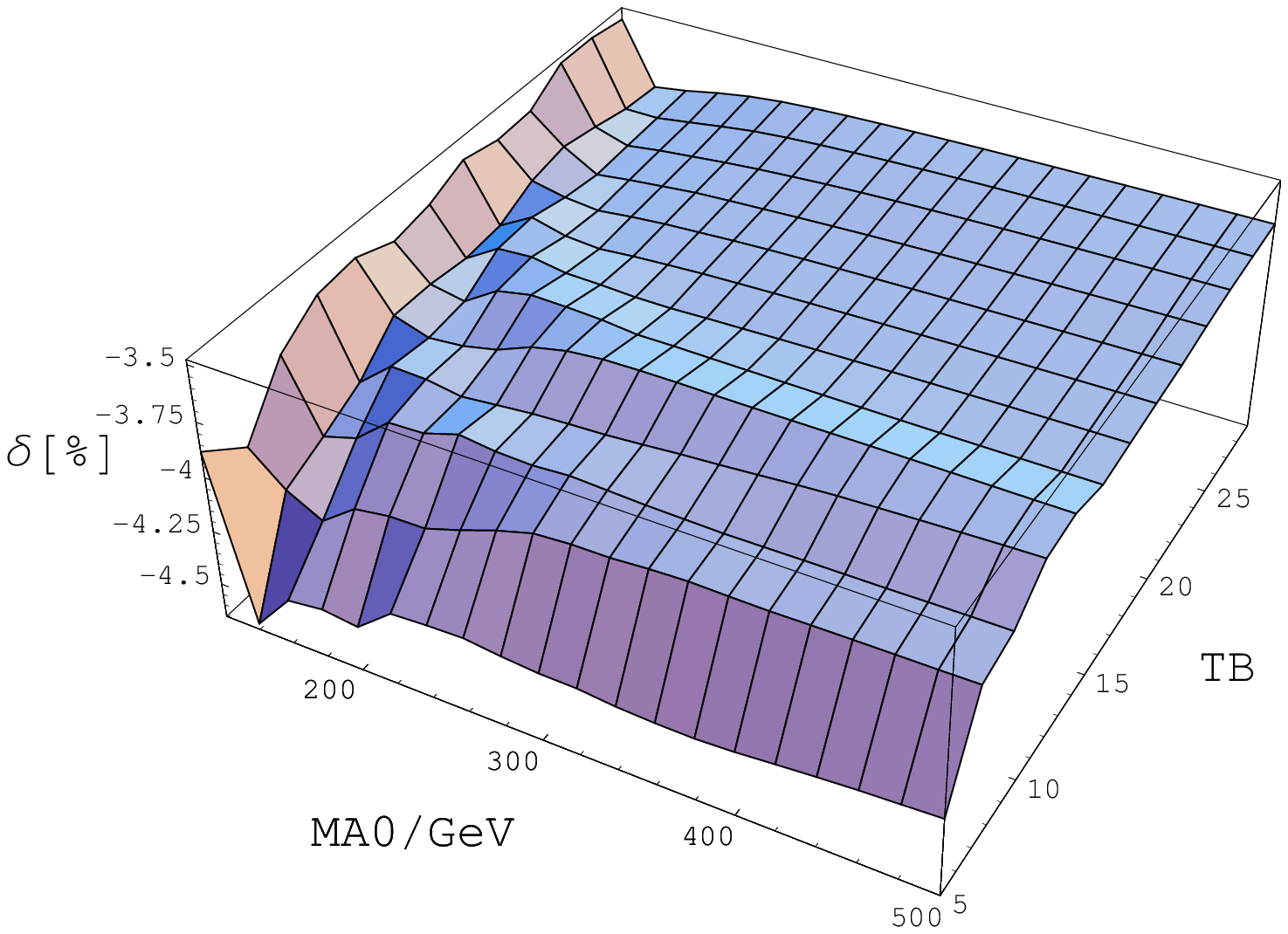}

\begin{center}
small-$\alpha_{\mbox{\small{eff}}}$ scenario
\end{center}
\vskip -0.5cm
\includegraphics[width=0.42\textwidth]{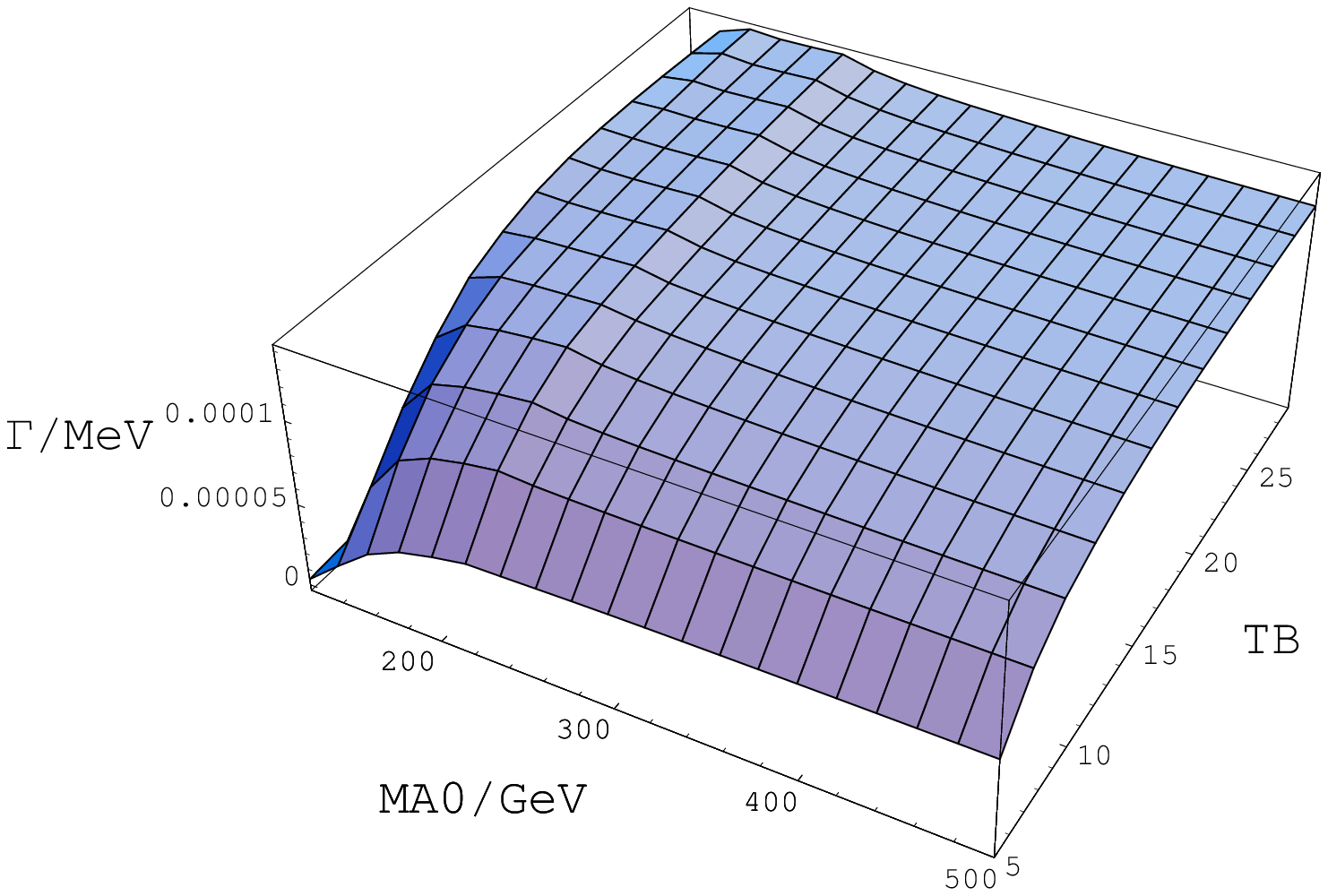}
\hspace{2em}
\includegraphics[width=0.42\textwidth]{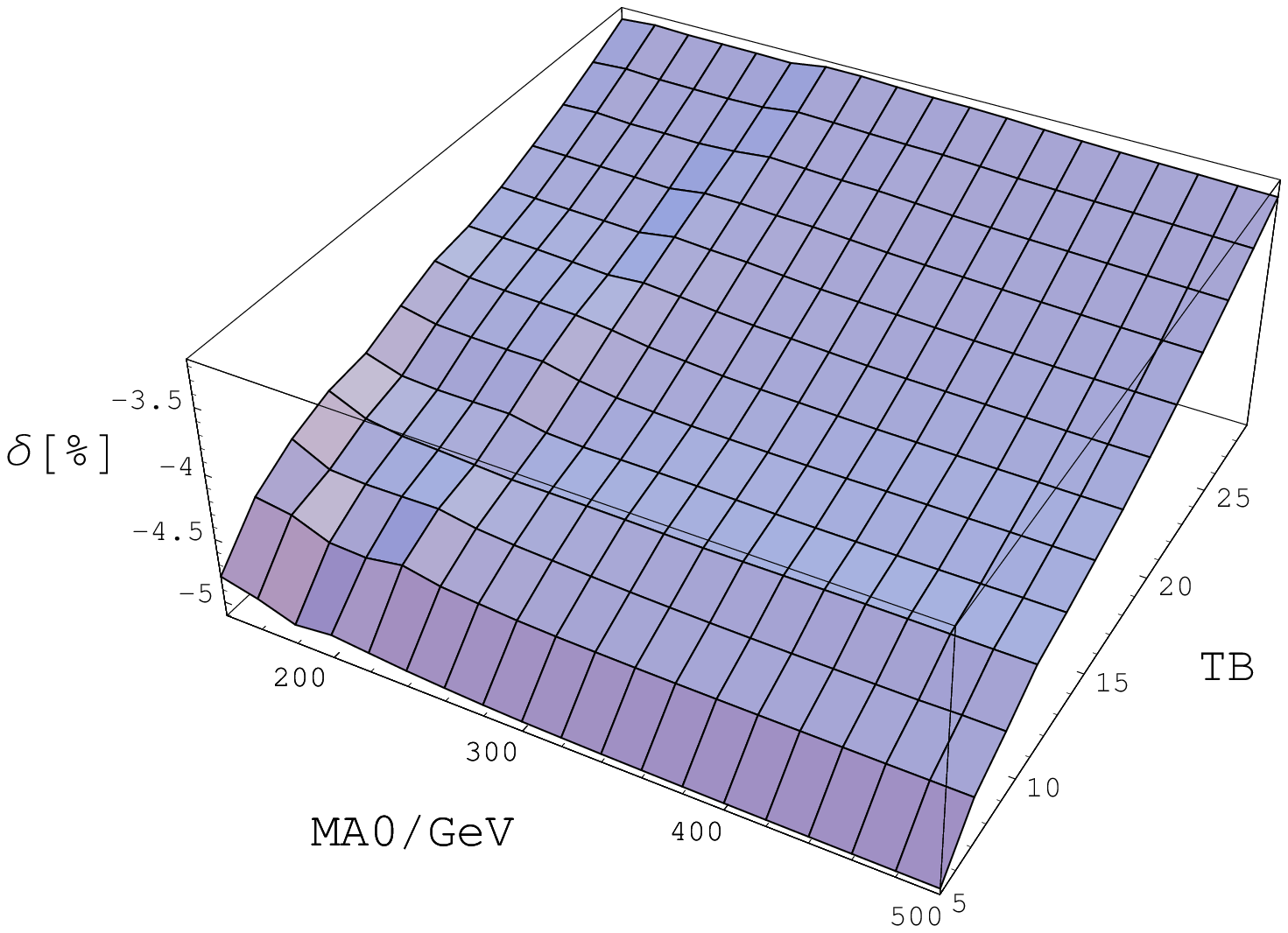}

\caption{Results for the partial decay width of $h^0\ra ZZ^*\ra e^-e^+\mu^+\mu^-$ in three different benchmark scenarios (the contribution due to the third generation fermion and sfermion loop corrections to the $H^0ZZ$ coupling is not included. TB denotes $\tan\beta$). The left column shows the corrected partial decay width, while the right column shows the relative corrections.}
\label{hZZwidth}
\end{figure}

\begin{figure}[htbp]
\begin{center}
$m_h^{\mbox{\small{max}}}\;\mbox{scenario}$ 
\end{center}
\vskip -0.5cm
\includegraphics[width=0.42\textwidth]{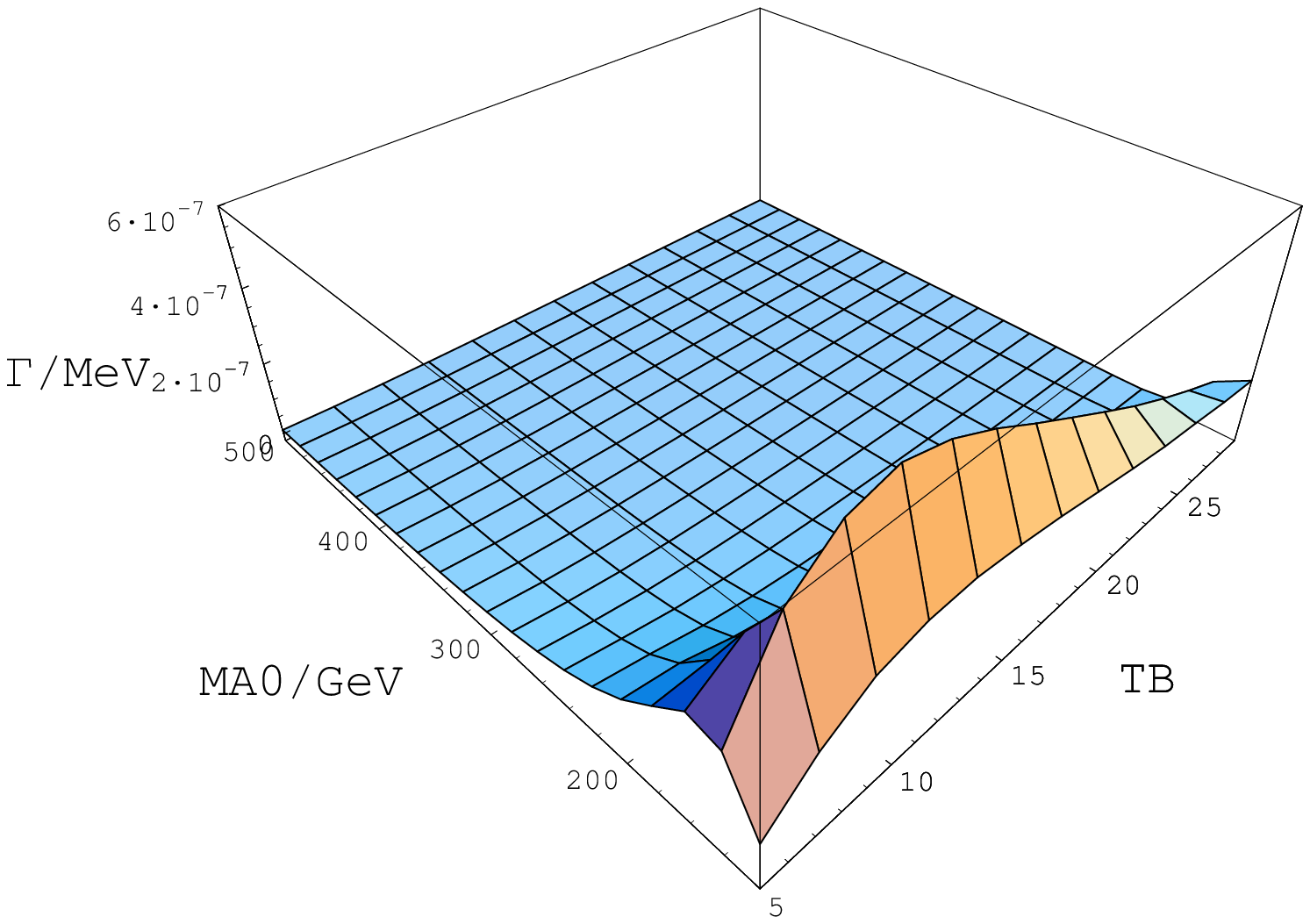}
\hspace{2em}
\includegraphics[width=0.42\textwidth]{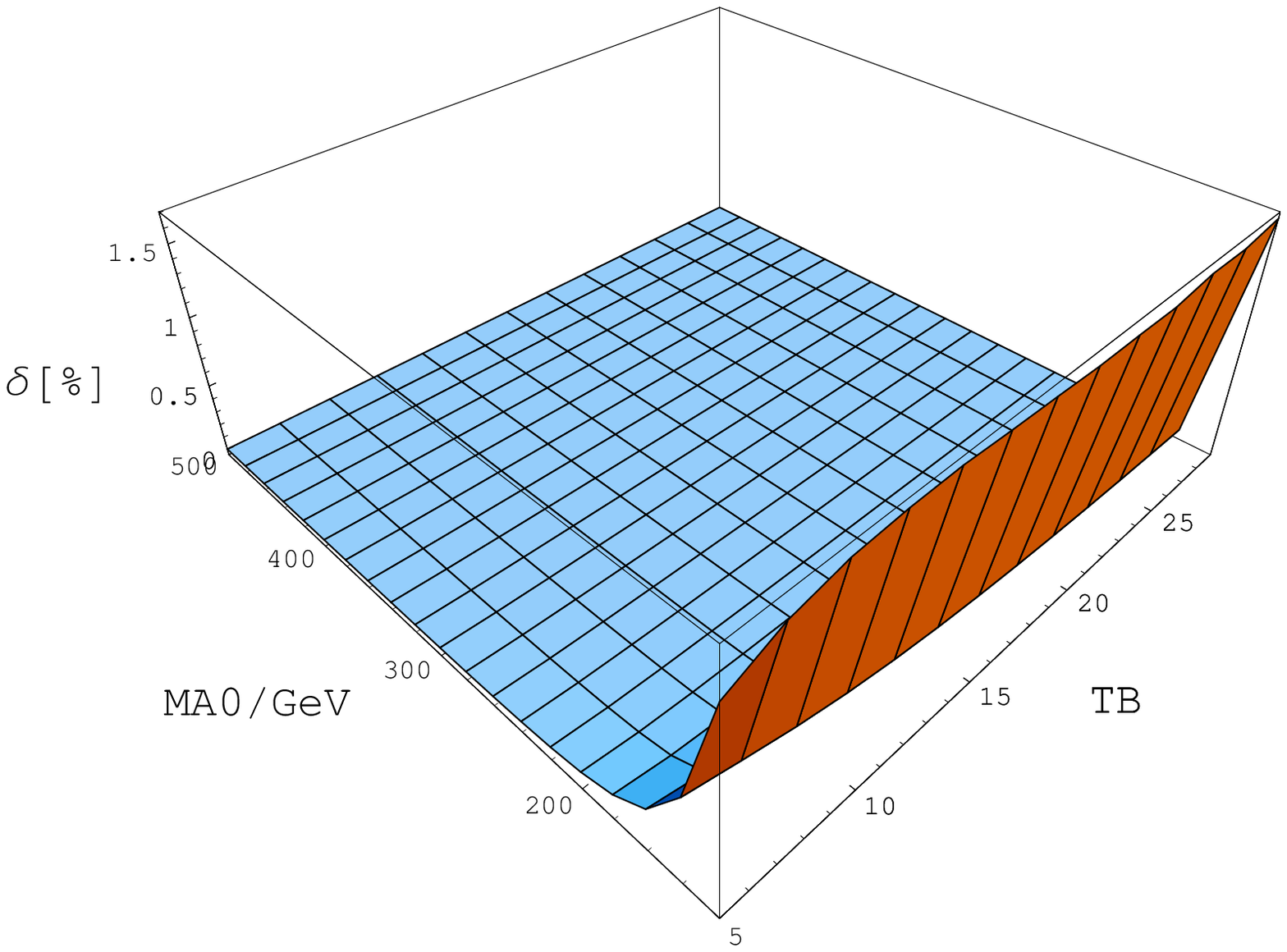}

\begin{center}
no-mixing scenario
\end{center}
\vskip -0.5cm
\includegraphics[width=0.42\textwidth]{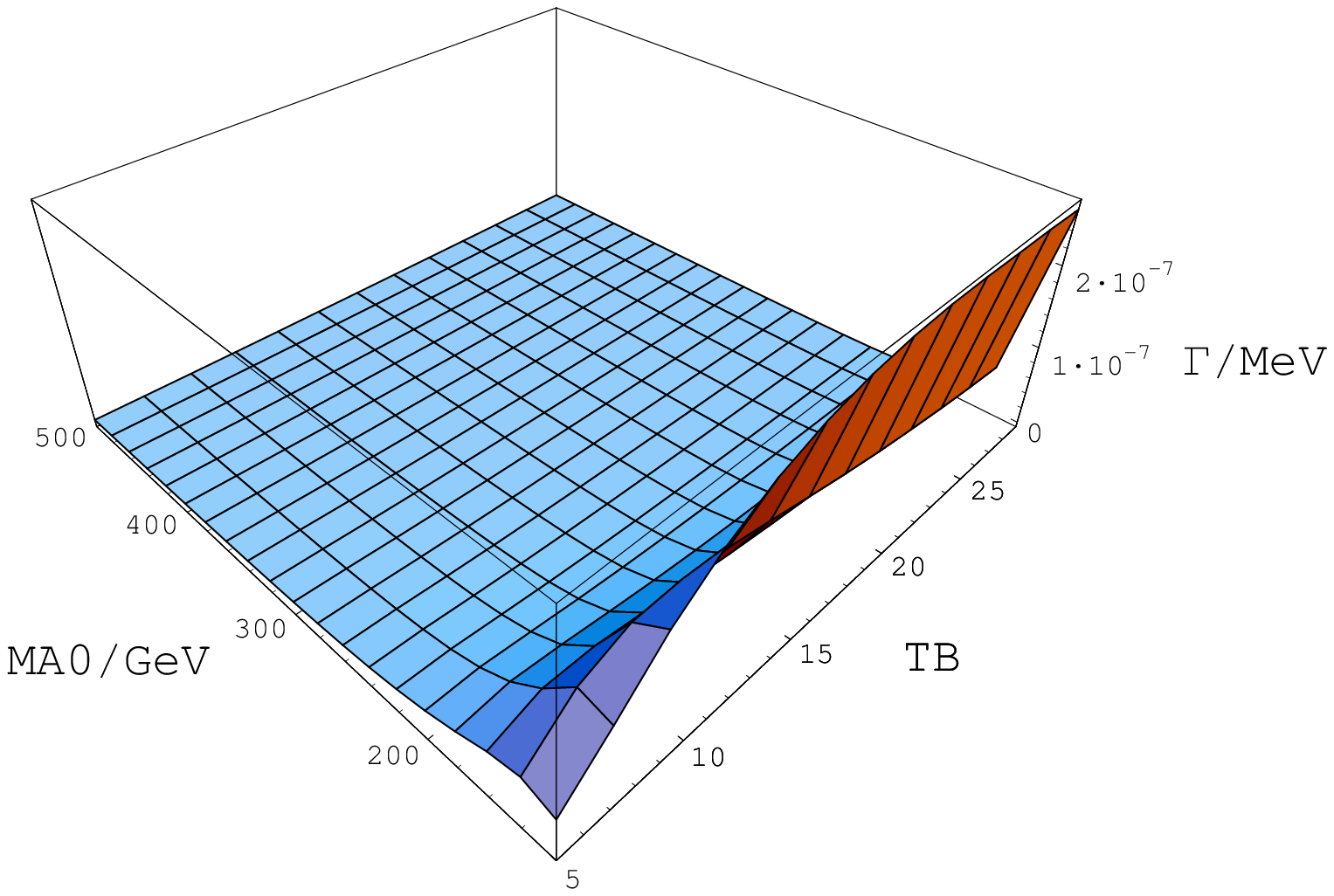}
\hspace{2em}
\includegraphics[width=0.42\textwidth]{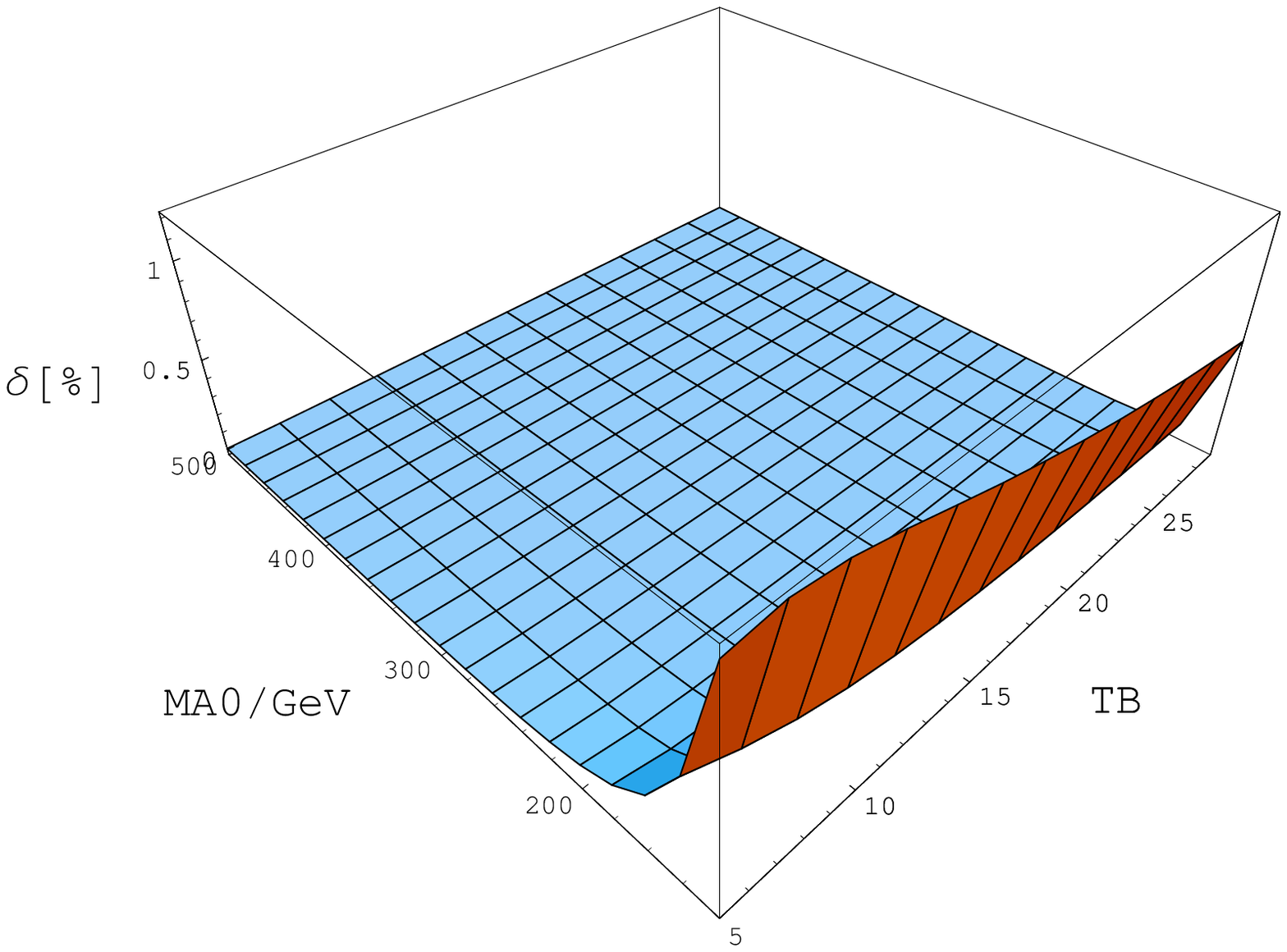}

\begin{center}
small-$\alpha_{\mbox{\small{eff}}}$ scenario
\end{center}
\vskip -0.5cm
\includegraphics[width=0.42\textwidth]{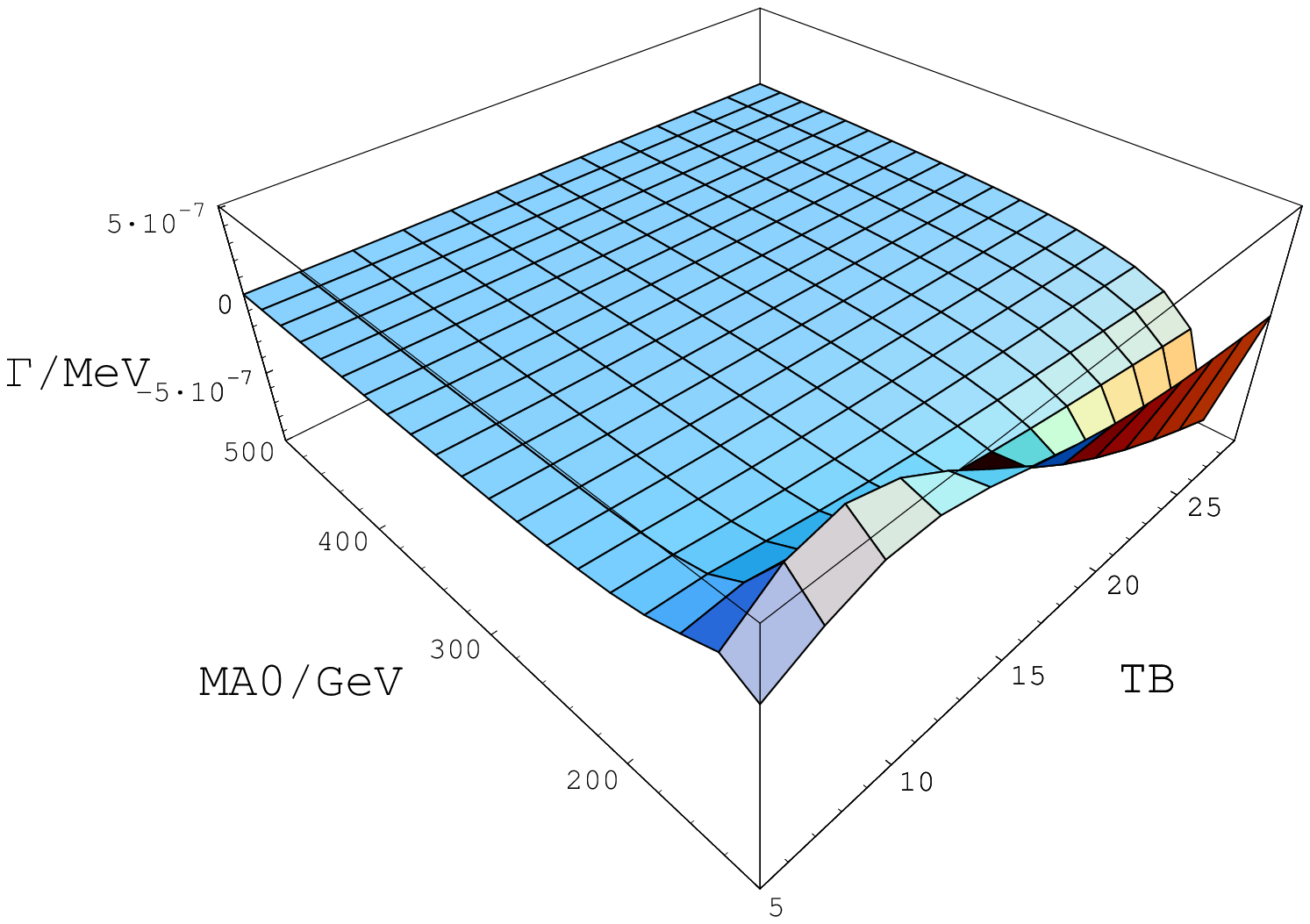}
\hspace{2em}
\includegraphics[width=0.42\textwidth]{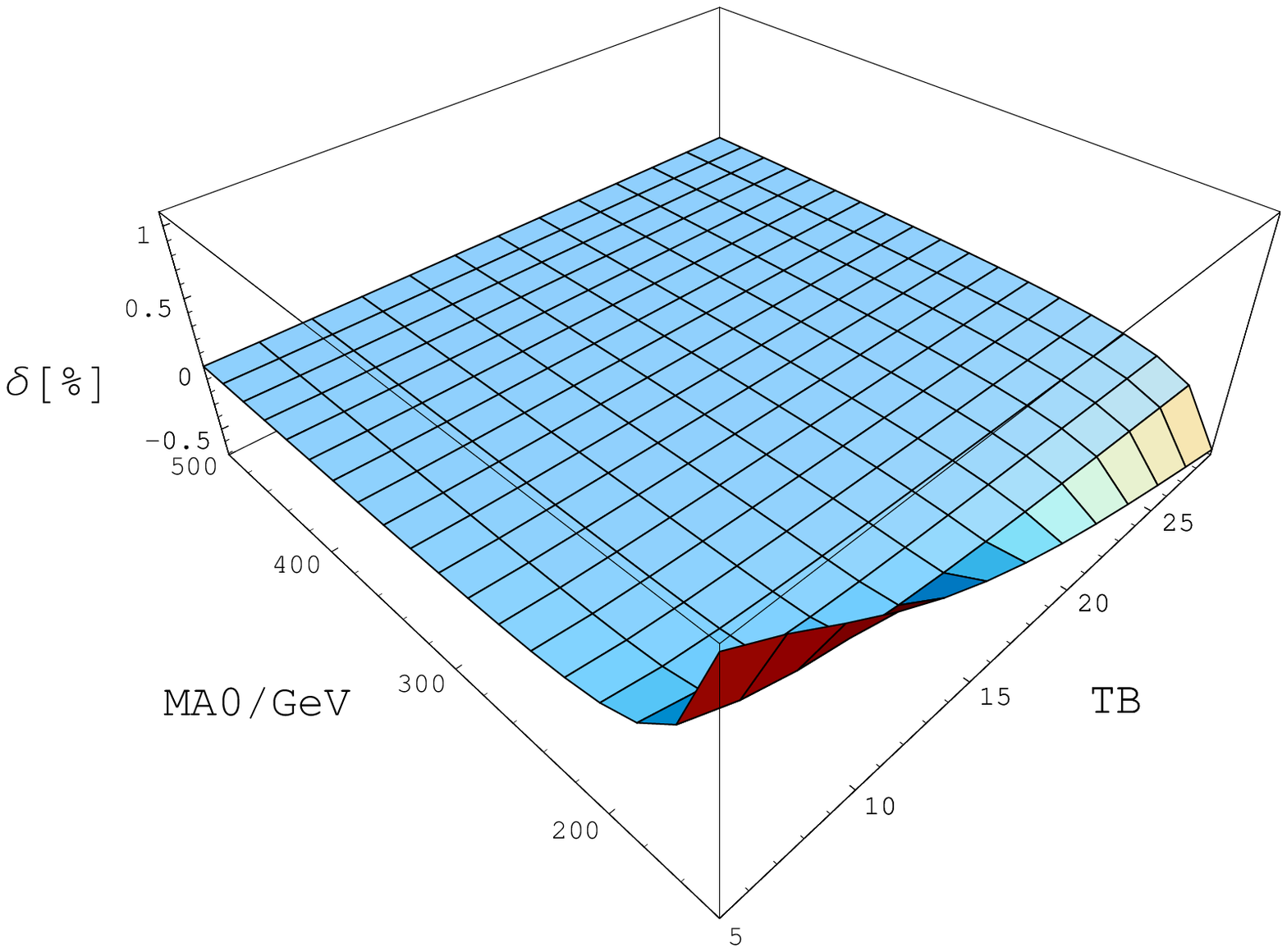}

\caption{Contribution to the partial decay width of $h^0\ra ZZ^*\ra e^-e^+\mu^+\mu^-$ due to the correction to the $H^0ZZ$ coupling from the third generation fermions and sfermions in three different benchmark scenarios (TB denotes $\tan\beta$). The left column shows the corrections to the partial decay width, while the right column shows their relative size.}
\label{HZZwidth}
\end{figure}

\begin{figure}[htbp]
\includegraphics[width=0.47\textwidth]{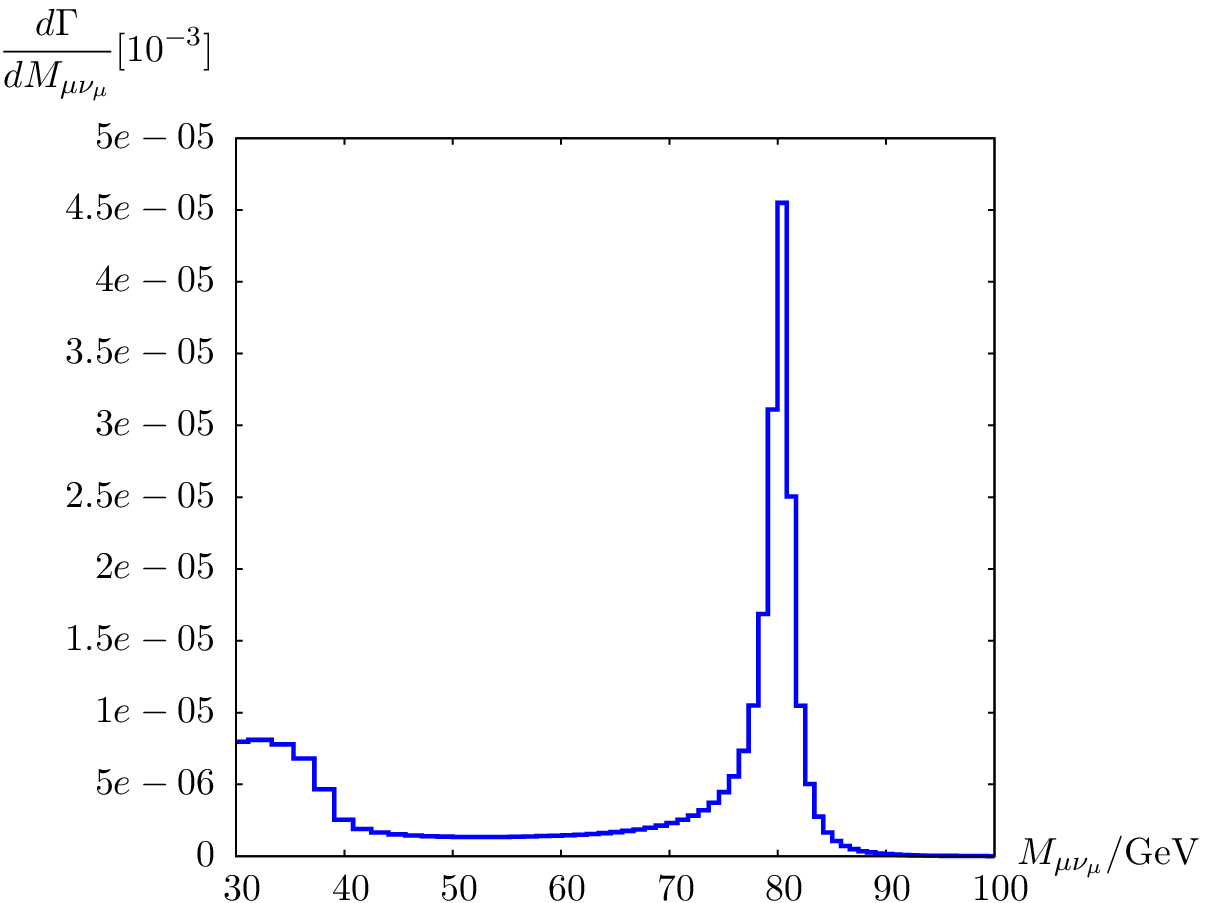}
\hspace{2em}
\includegraphics[width=0.47\textwidth]{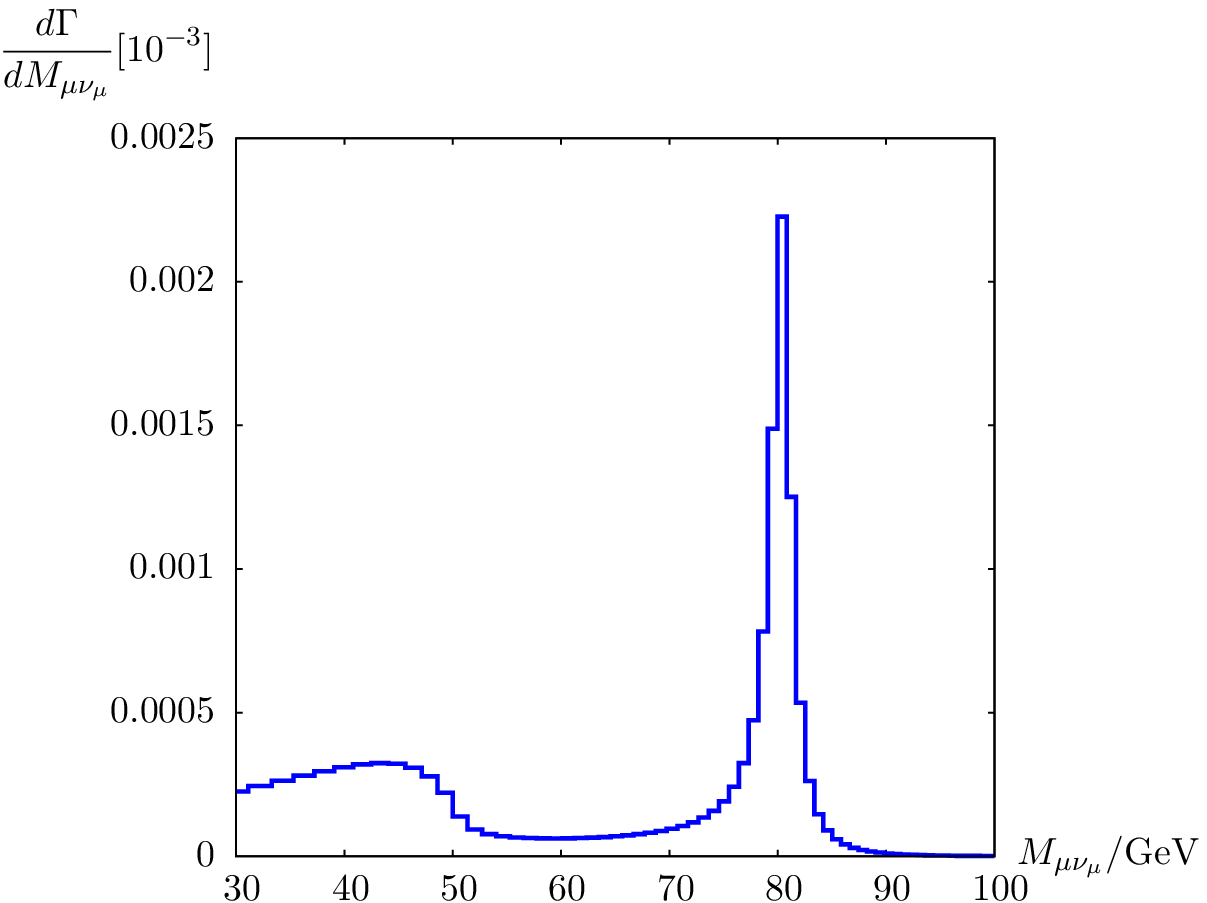}
\caption{Invariant mass distribution of $\mu^+\nu_\mu$ in the decay $h^0\ra e^-\bar\nu_e\mu^+\nu_\mu$ (no photon combination, the contribution from the $H^0WW$ vertex correction is not included) in the $m_h^{\mbox{\small{max}}}$ scenario, with $\tan\beta=30,\,M_{A^0}=120\,\mbox{GeV}$ (left) and $\tan\beta=30,\,M_{A^0}=400\,\mbox{GeV}$ (right).}
\label{HWWimdmax120and400}
\end{figure}

\begin{figure}[htbp]
\hspace {2em}
\includegraphics[width=0.42\textwidth]{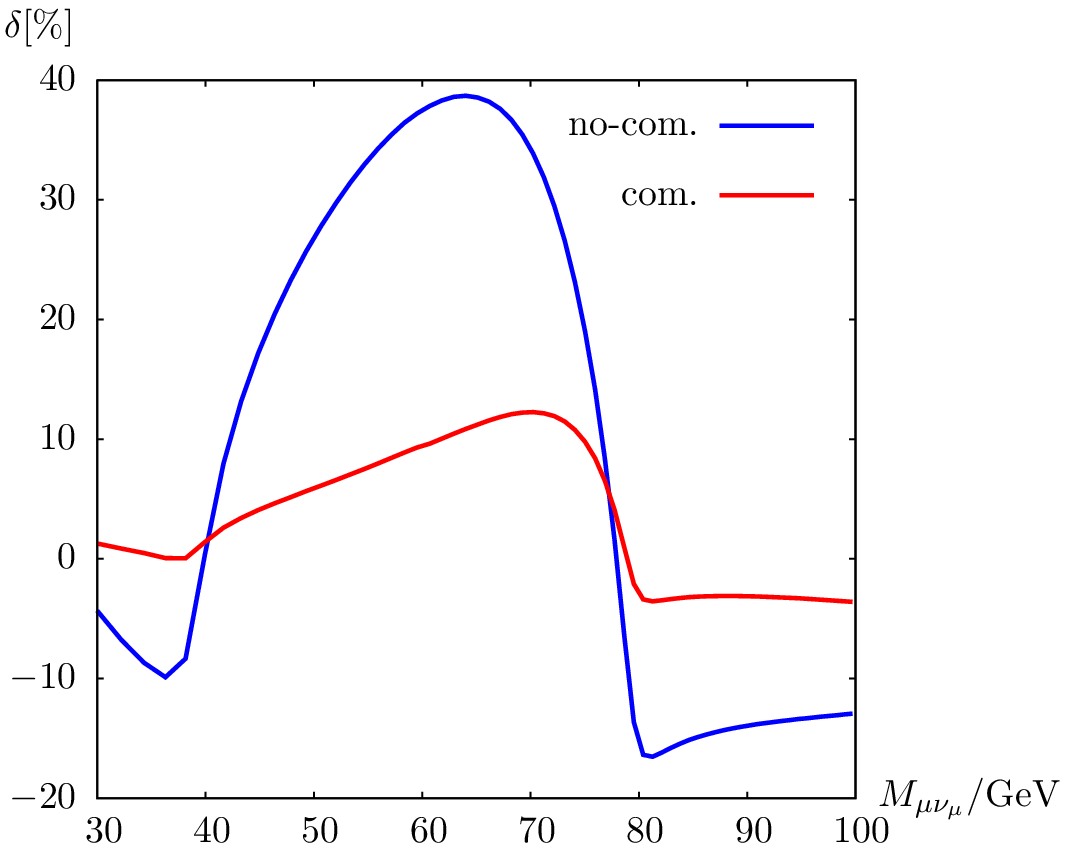}
\hspace {4em}
\includegraphics[width=0.42\textwidth]{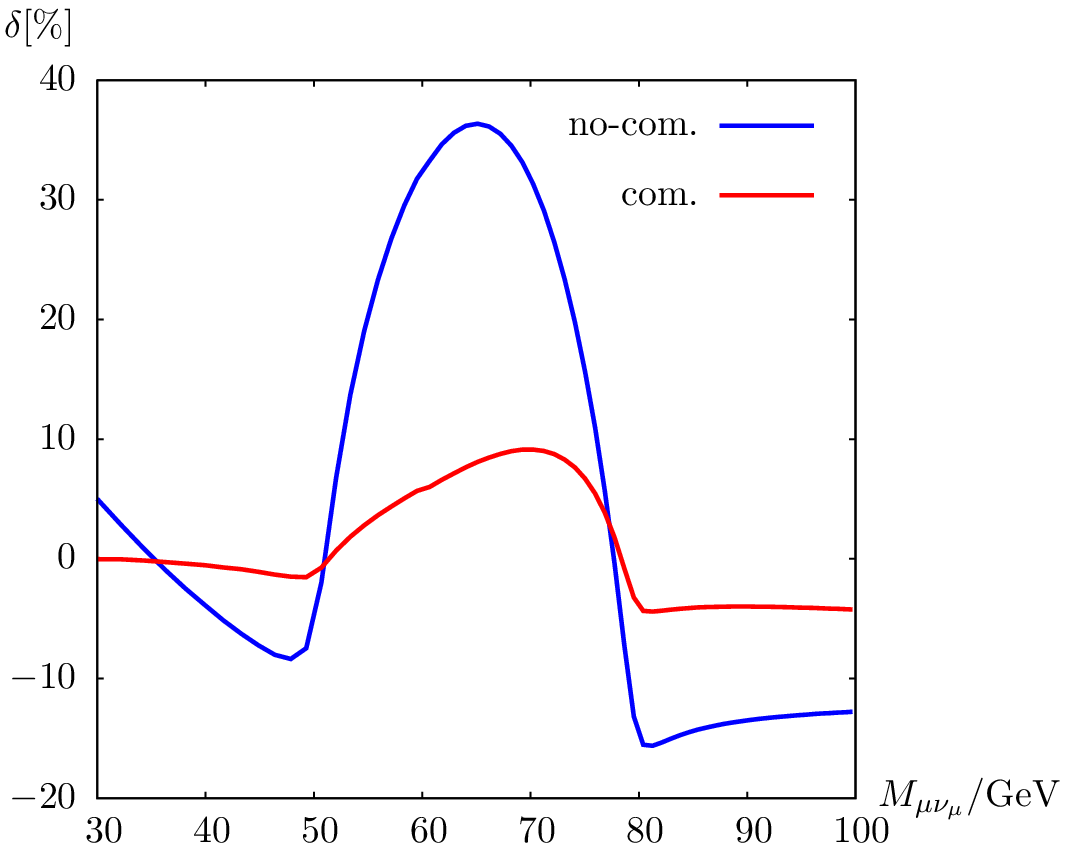}
\caption{Relative correction to the invariant mass distribution of $\mu^+\nu_\mu$ in the decay $h^0\ra e^-\bar\nu_e\mu^+\nu_\mu$ (the contribution from the $H^0WW$ vertex correction is not included) in the $m_h^{\mbox{\small{max}}}$ scenario, with $\tan\beta=30,\,M_{A^0}=120\,\mbox{GeV}$ (left) and $\tan\beta=30,\,M_{A^0}=400\,\mbox{GeV}$ (right). "com." and "no-com." indicate the results with and without photon combination, respectively. }
\label{HWWimdmax120and400relative1}
\vspace*{1cm}
\end{figure}

\begin{figure}[t]
\hspace {2em}
\includegraphics[width=0.42\textwidth]{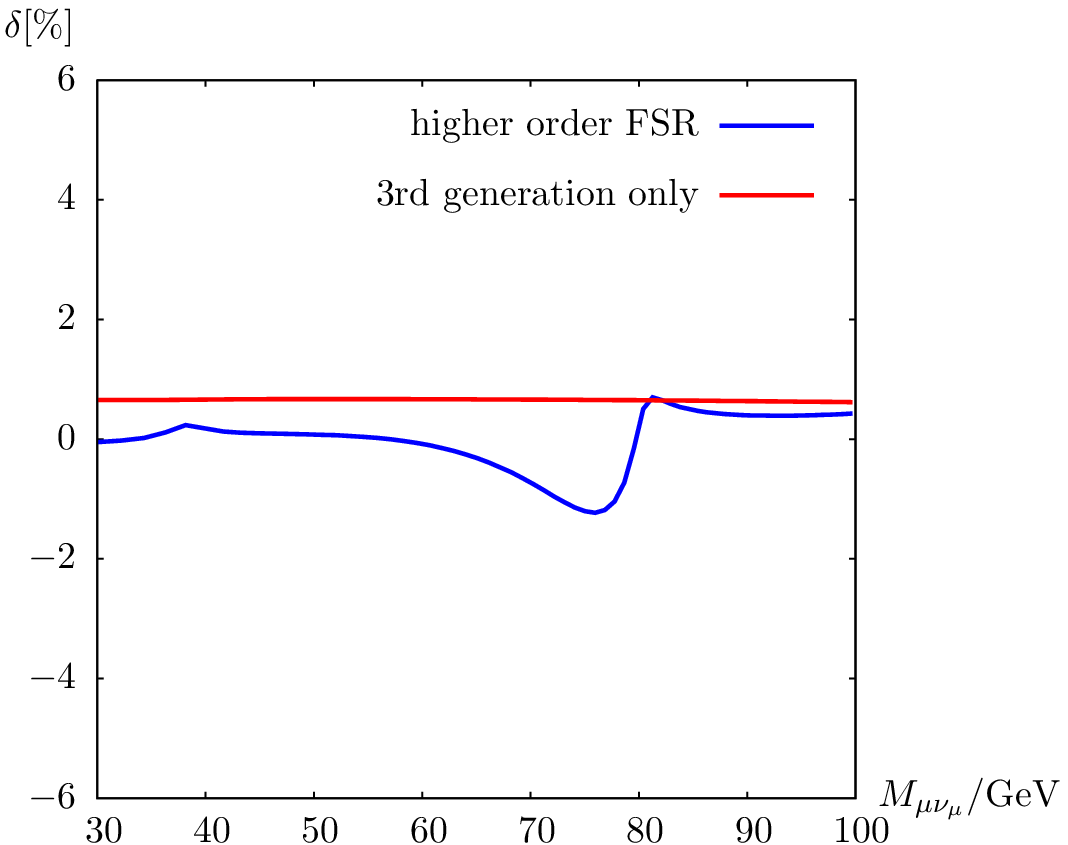}
\hspace {4em}
\includegraphics[width=0.42\textwidth]{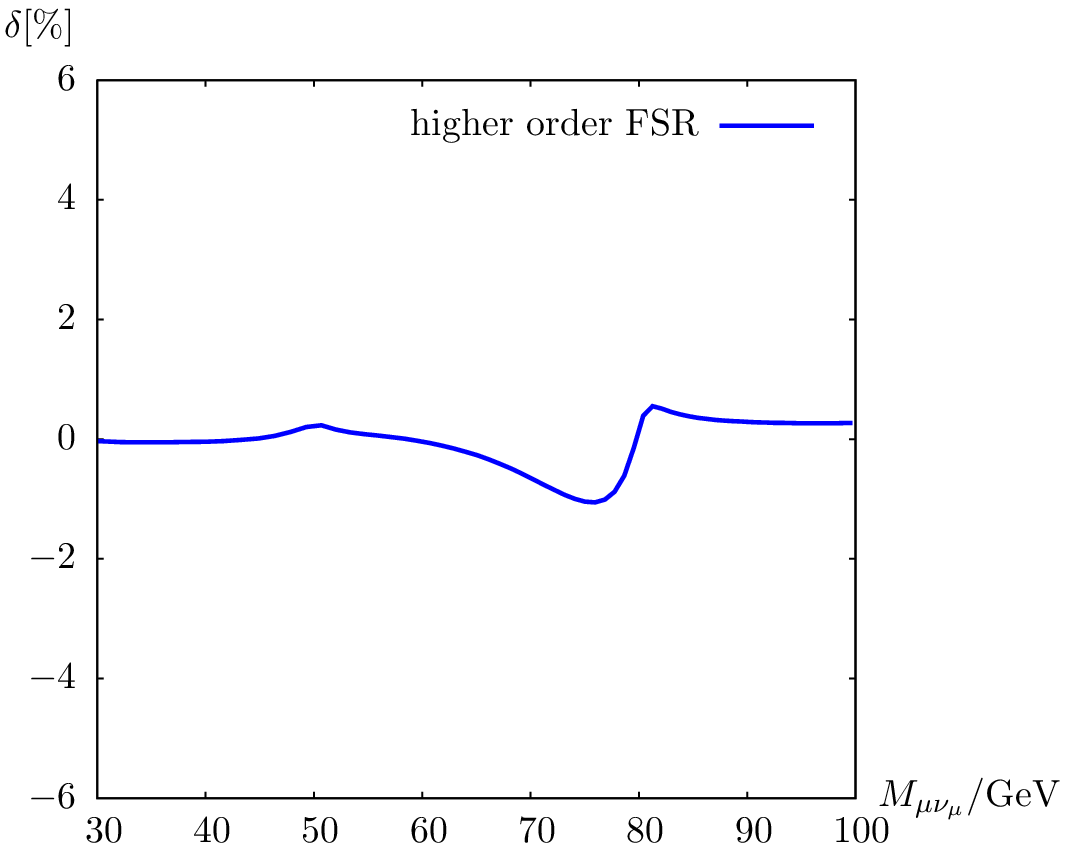}
\caption{Relative contribution to the invariant mass distribution of $\mu^+\nu_\mu$ in the decay $h^0\ra e^-\bar\nu_e\mu^+\nu_\mu$ from the higher order final state radiation ("higher order FSR") and from the $H^0WW$ vertex correction ("3rd generation only") in the $m_h^{\mbox{\small{max}}}$ scenario, with $\tan\beta=30,\,M_{A^0}=120\,\mbox{GeV}$ (left) and $\tan\beta=30,\,M_{A^0}=400\,\mbox{GeV}$ (right). }
\label{HWWimdmax120and400relative2}
\end{figure}

\begin{figure}[htbp]
\includegraphics[width=0.47\textwidth]{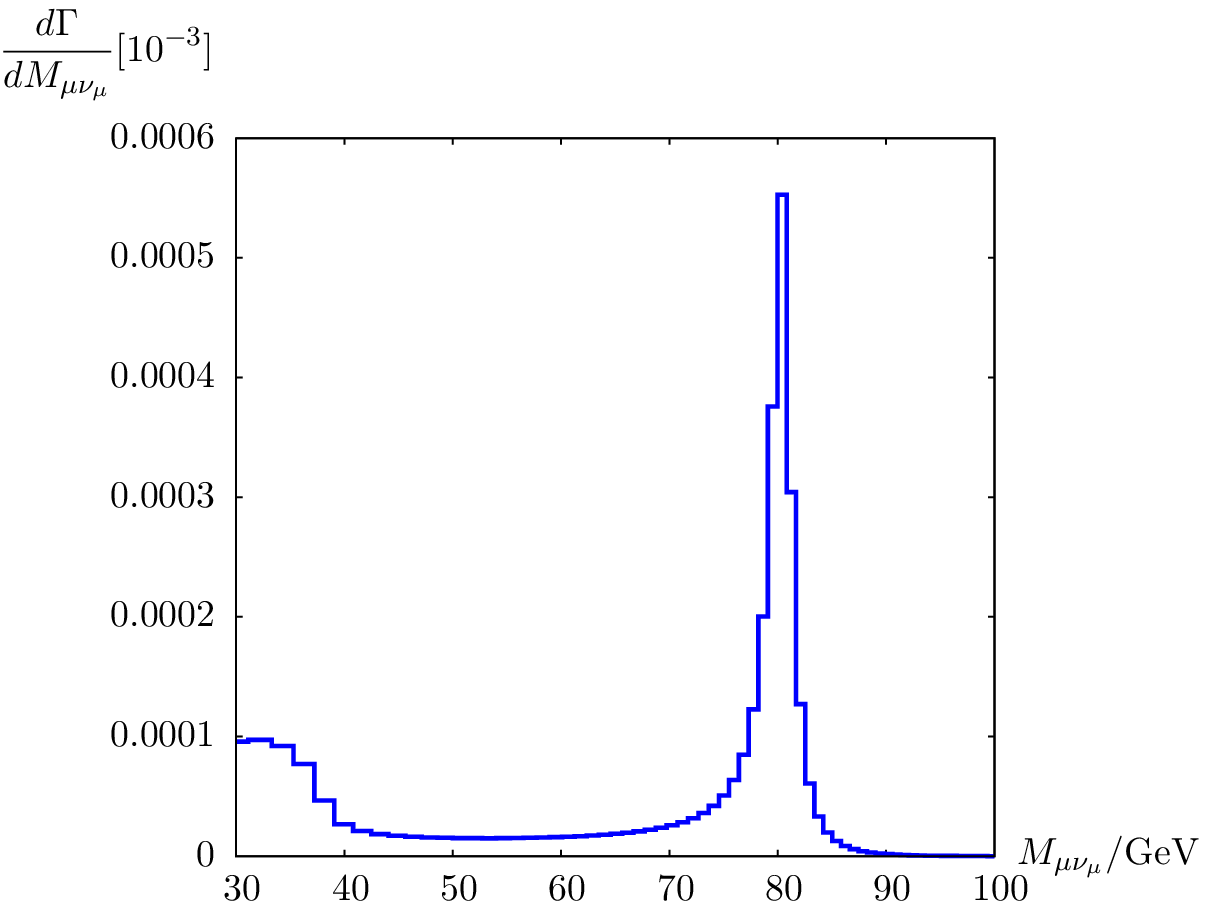}
\hspace{3.8em}
\includegraphics[width=0.42\textwidth]{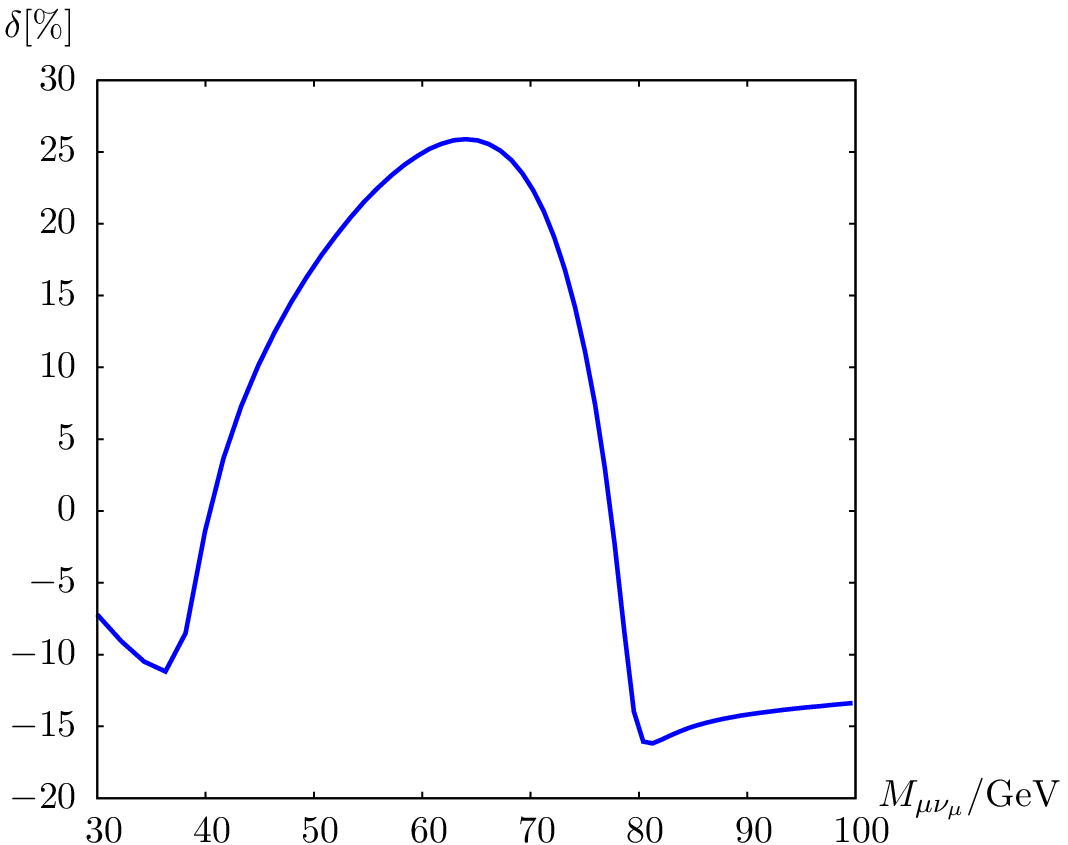}
\includegraphics[width=0.47\textwidth]{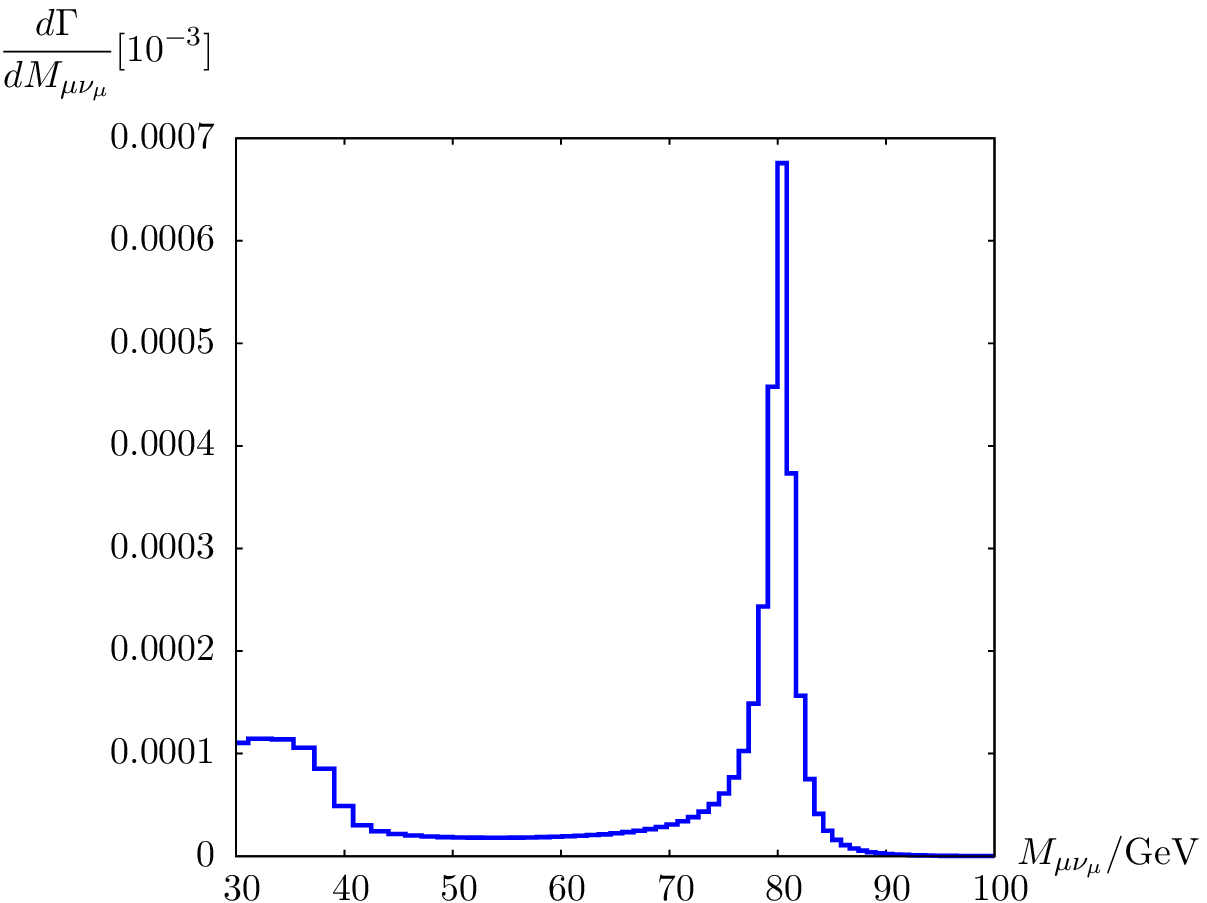}
\hspace{3.8em}
\includegraphics[width=0.42\textwidth]{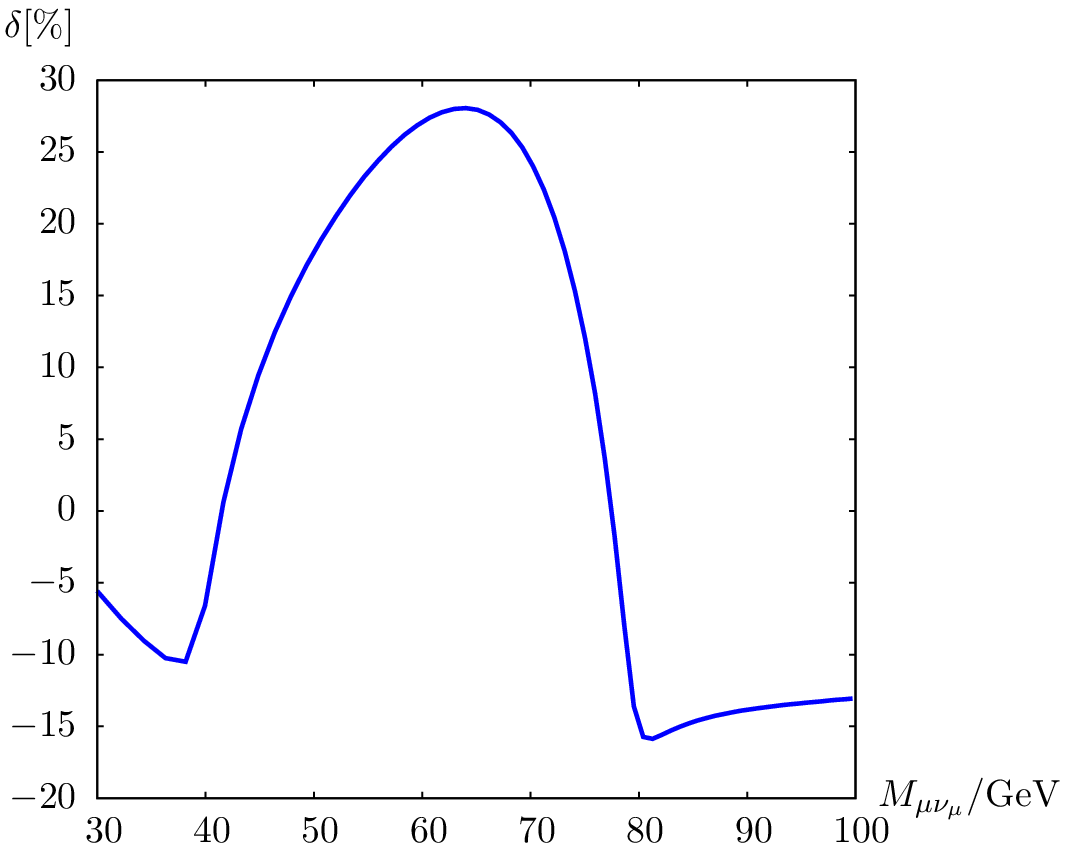}
\caption{Invariant mass distribution and relative correction to the invariant mass distribution of $\mu^+\nu_\mu$ in the decay $h^0\ra e^-\bar\nu_e\mu^+\nu_\mu$ (no photon combination) in the no-mixing scenario (upper) and the small-$\alpha_{\mbox{\small{eff}}}$ scenario (lower), with $\tan\beta=30,\,M_{A^0}=400\,\mbox{GeV}$.}
\label{HWWimdnoandsmall400}
\end{figure}

\begin{figure}[htbp]
\includegraphics[width=0.47\textwidth]{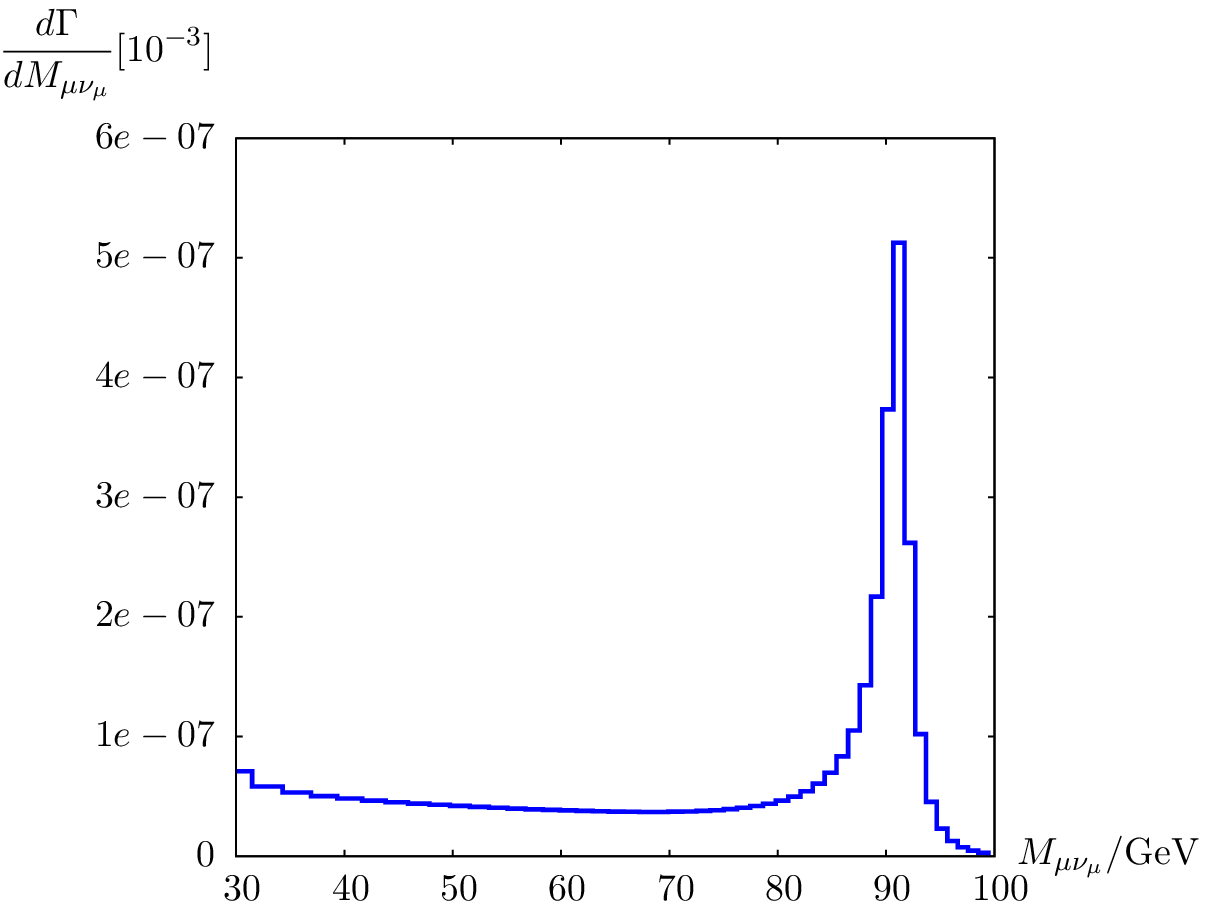}
\hspace{2em}
\includegraphics[width=0.47\textwidth]{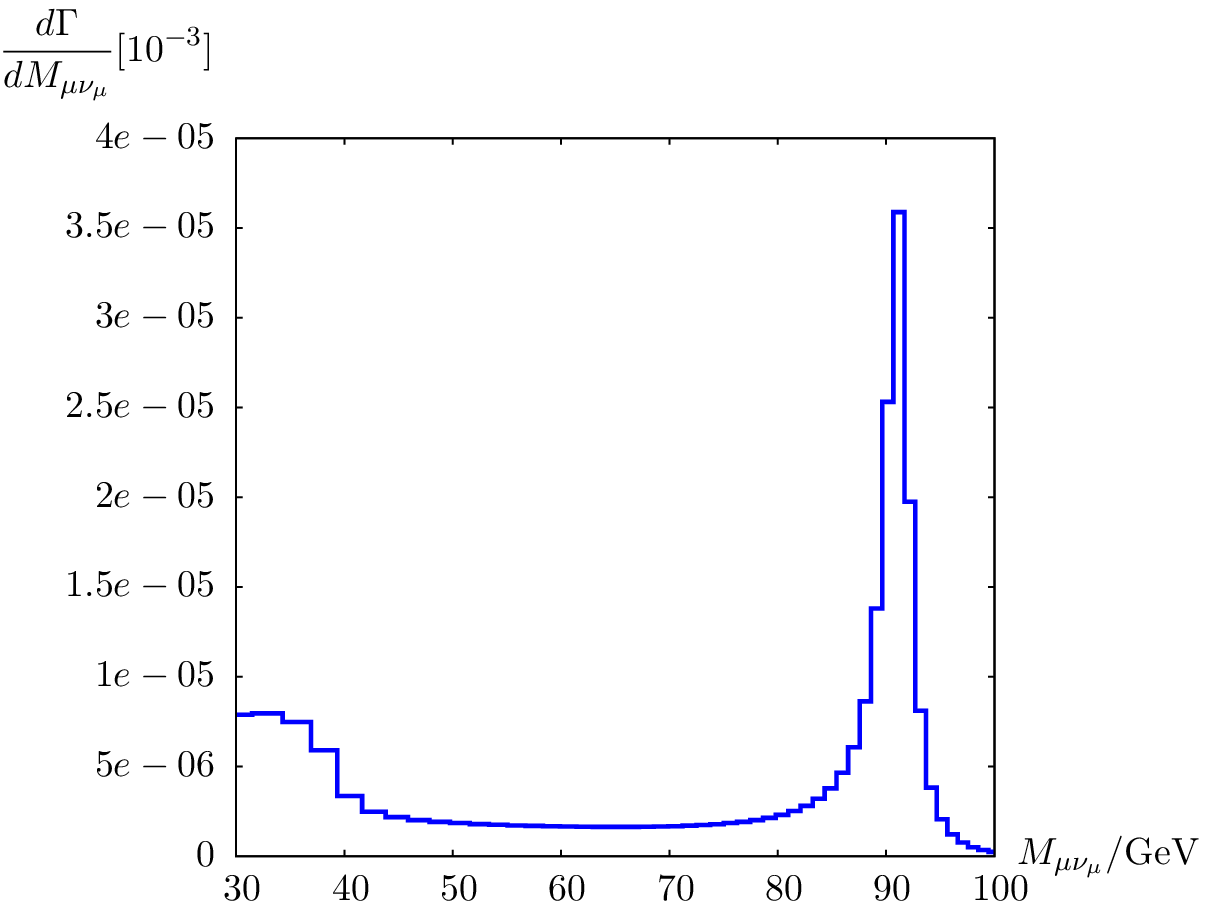}
\caption{Invariant mass distribution of $\mu^+\mu^-$ in the decay $h^0\ra e^-e^+\mu^+\mu^-$ (no photon combination, the contribution from the $H^0ZZ$ vertex correction is not included) in the $m_h^{\mbox{\small{max}}}$ scenario, with $\tan\beta=30,\,M_{A^0}=120\,\mbox{GeV}$ (left) and $\tan\beta=30,\,M_{A^0}=400\,\mbox{GeV}$ (right).}
\label{HZZimdmax120and400}
\end{figure}

\begin{figure}[htbp]
\hspace {2em}
\includegraphics[width=0.42\textwidth]{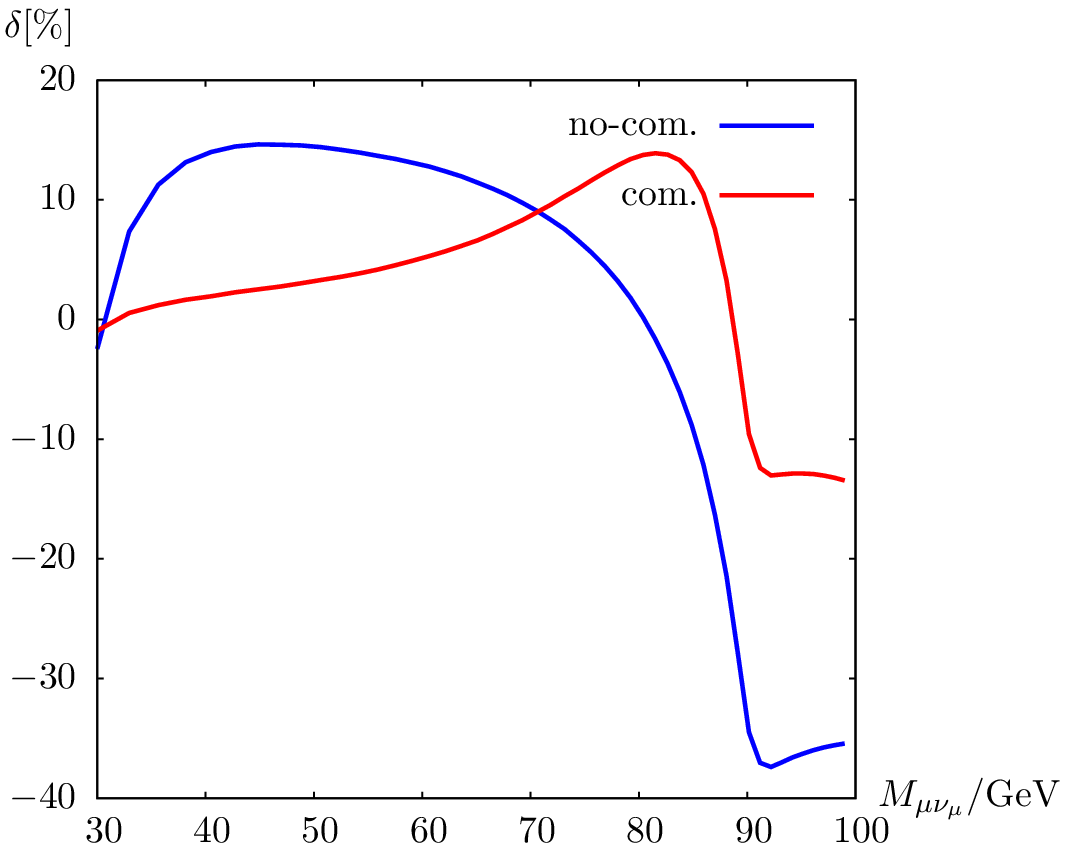}
\hspace {4em}
\includegraphics[width=0.42\textwidth]{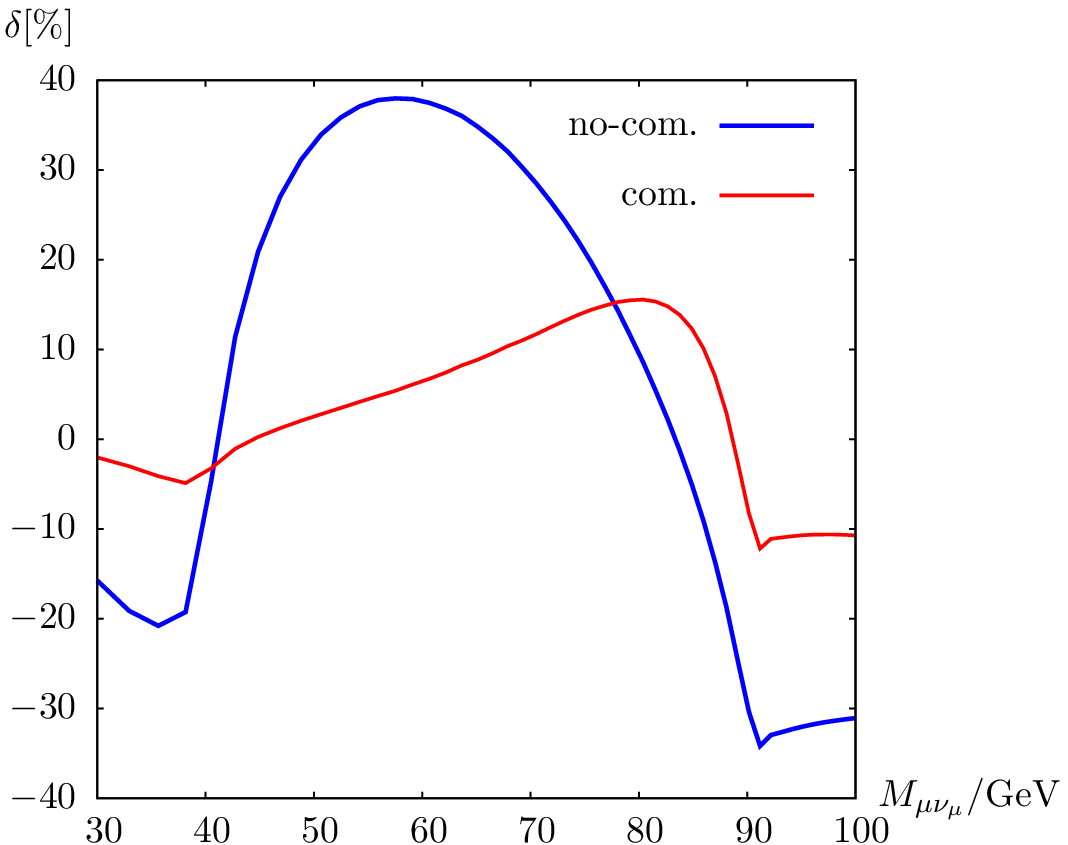}
\caption{Relative correction to the invariant mass distribution of $\mu^+\mu^-$ in the decay $h^0\ra e^-e^+\mu^+\mu^-$ (the contribution from the $H^0ZZ$ vertex correction is not included) in the $m_h^{\mbox{\small{max}}}$ scenario, with $\tan\beta=30,\,M_{A^0}=120\,\mbox{GeV}$ (left) and $\tan\beta=30,\,M_{A^0}=400\,\mbox{GeV}$ (right).  "com." and "no-com." indicate the results with and without photon combination, respectively. }
\label{HZZimdmax120and400relative1}
\vspace*{1cm}
\end{figure}

\begin{figure}[t]
\hspace {2em}
\includegraphics[width=0.42\textwidth]{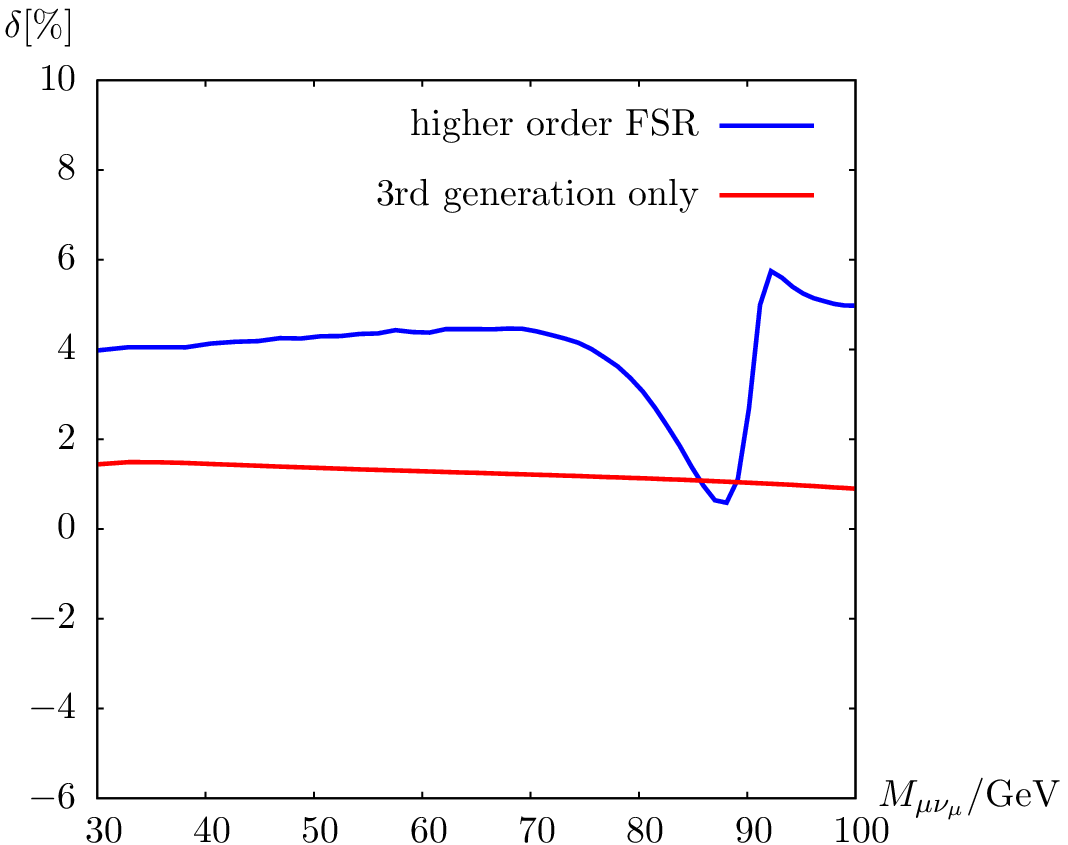}
\hspace {4em}
\includegraphics[width=0.42\textwidth]{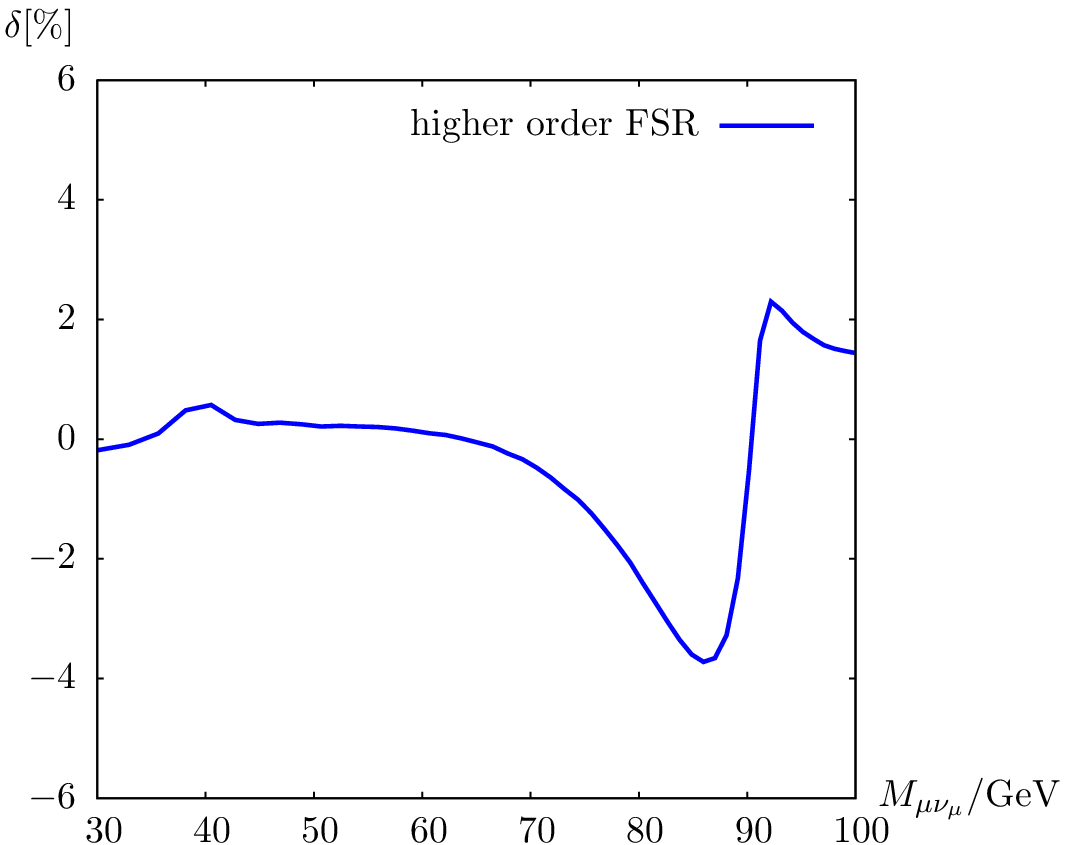}
\caption{Relative contribution to the invariant mass distribution of $\mu^+\mu^-$ in the decay $h^0\ra e^-e^+\mu^+\mu^-$ from the higher order final state radiation ("higher order FSR") and from the $H^0ZZ$ vertex correction ("3rd generation only") in the $m_h^{\mbox{\small{max}}}$ scenario, with $\tan\beta=30,\,M_{A^0}=120\,\mbox{GeV}$ (left) and $\tan\beta=30,\,M_{A^0}=400\,\mbox{GeV}$ (right). }
\label{HZZimdmax120and400relative2}
\end{figure}

\begin{figure}[htbp]
\includegraphics[width=0.47\textwidth]{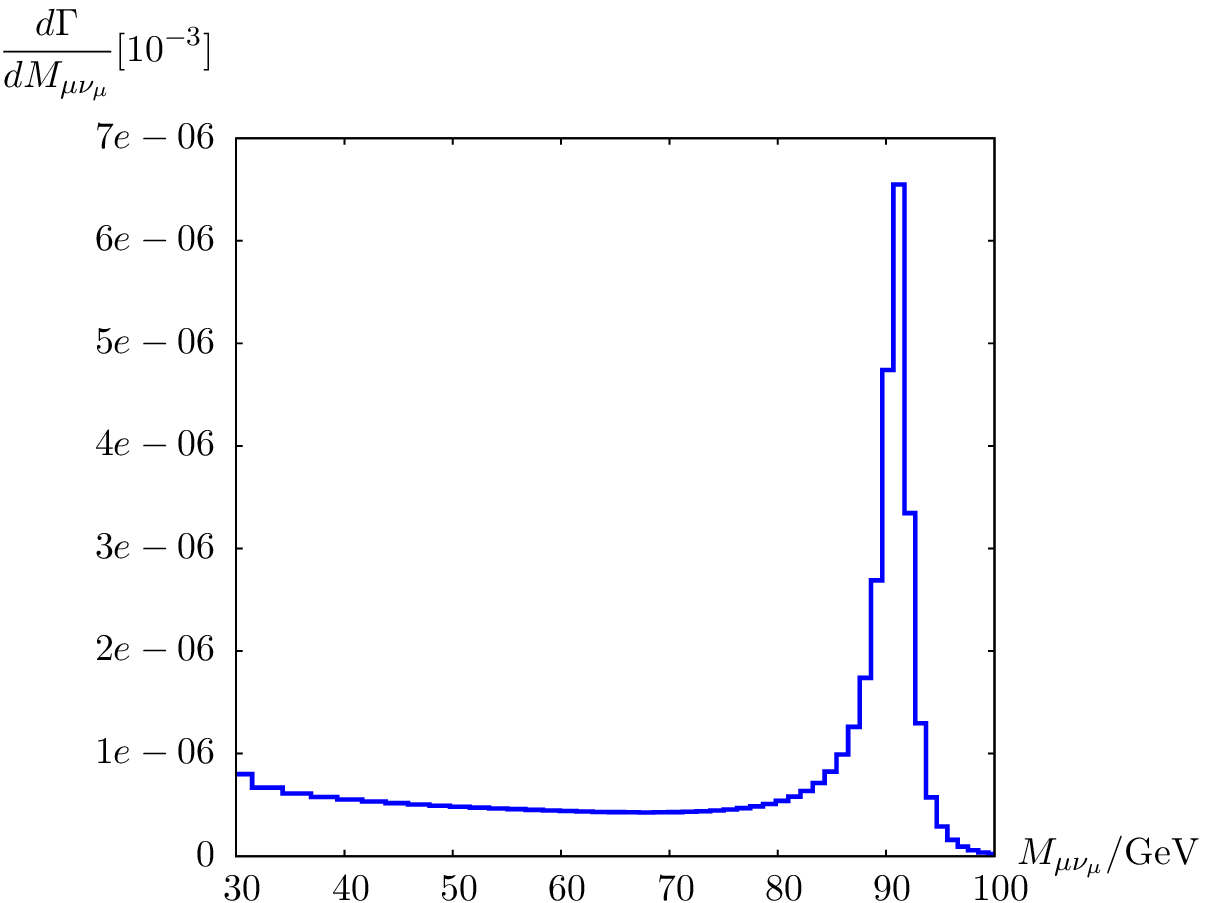}
\hspace{3.8em}
\includegraphics[width=0.42\textwidth]{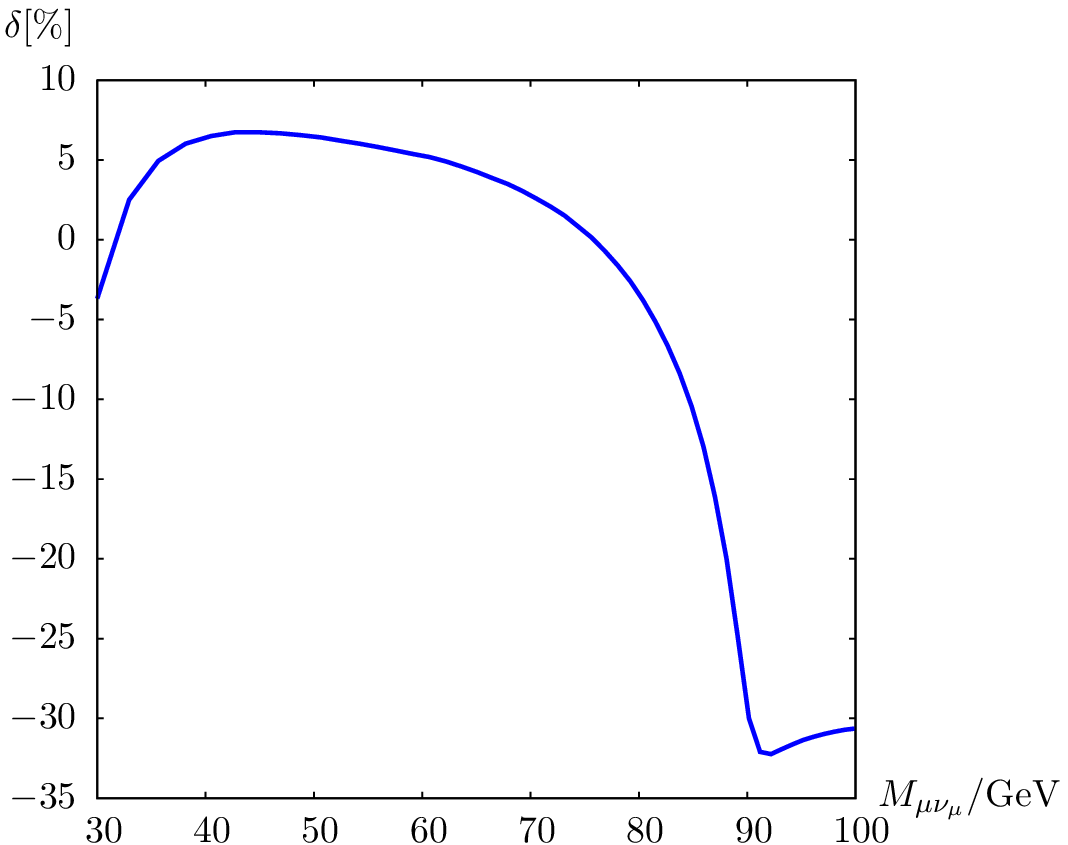}
\includegraphics[width=0.47\textwidth]{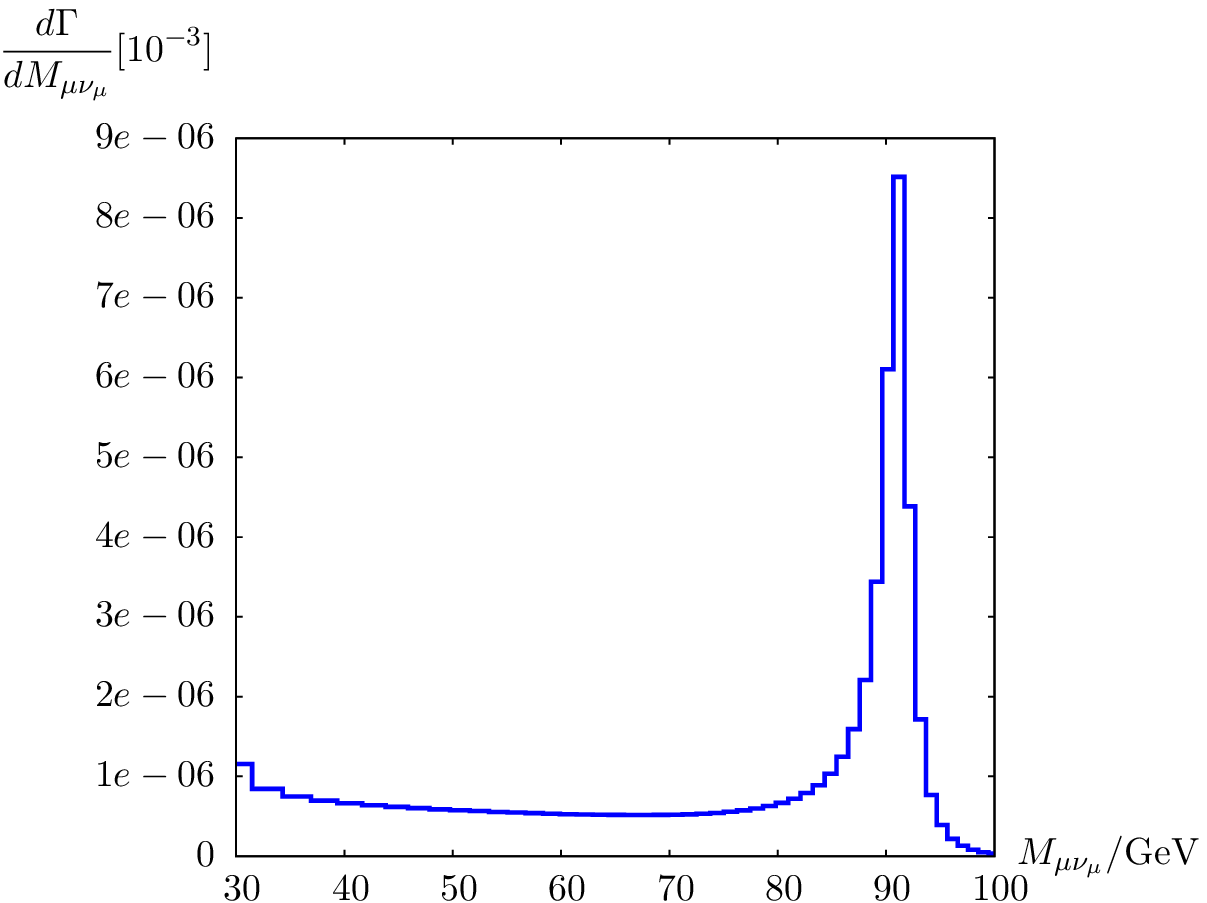}
\hspace{3.8em}
\includegraphics[width=0.42\textwidth]{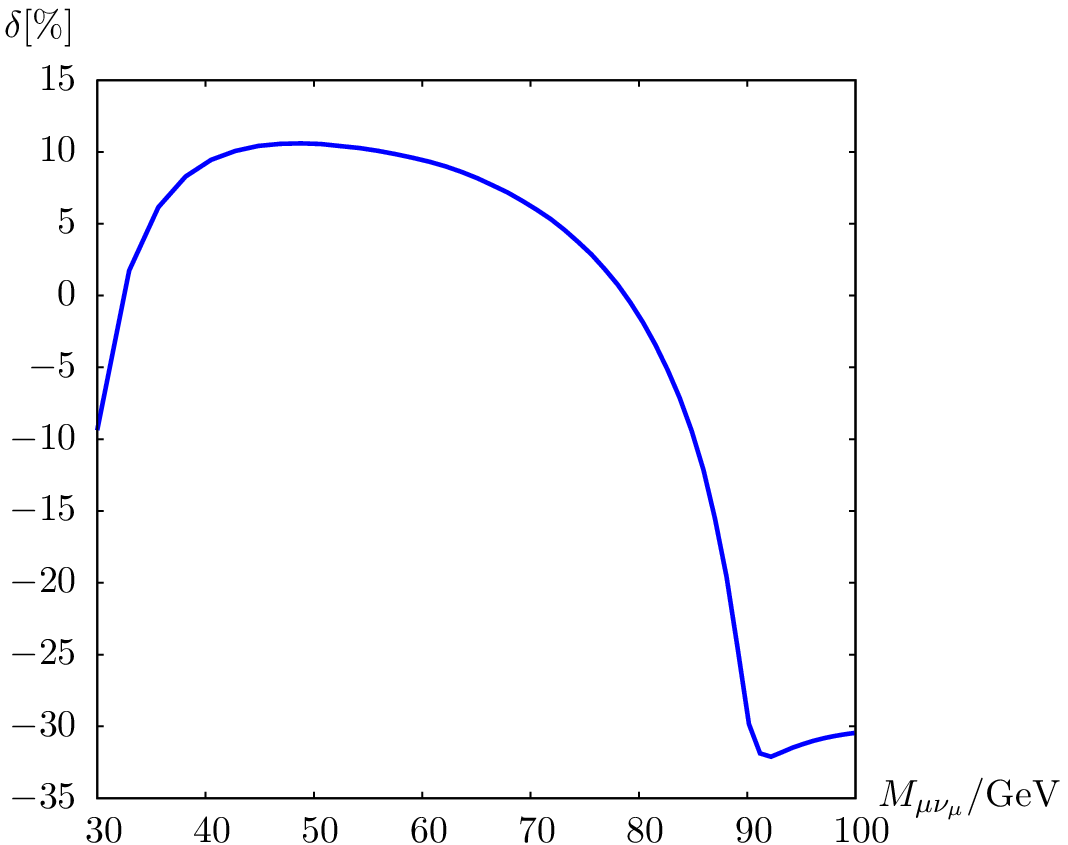}
\caption{Invariant mass distribution and relative correction to the invariant mass distribution of $\mu^+\mu^-$ in the decay $h^0\ra e^-e^+\mu^+\mu^-$ (no photon combination) in the no-mixing scenario (upper) and the small-$\alpha_{\mbox{\small{eff}}}$ scenario (lower), with $\tan\beta=30,\,M_{A^0}=400\,\mbox{GeV}$.}
\label{HZZimdnoandsmall400}
\end{figure}

\end{document}